\documentclass[published]{JHEP3} 
\JHEP{03(2010)031}
\usepackage{epsfig,multicol,bbm}
\newcommand\fverb{\setbox\fverbbox=\hbox\bgroup\verb}
\newcommand\fverbdo{\egroup\medskip\noindent%
			\fbox{\unhbox\fverbbox}\ }
\newcommand\fverbit{\egroup\item[\fbox{\unhbox\fverbbox}]}
\newbox\fverbbox

\newcommand{\be}{\begin{equation}}
\newcommand{\ee}{\end{equation}}
\newcommand{\bea}{\begin{eqnarray}}
\newcommand{\eea}{\end{eqnarray}}
\def\eps{\epsilon}
\def\veps{\varepsilon}
\def\g{\gamma}
\def\p{\prime}
\title{Flavor conversion of cosmic neutrinos from hidden jets}
\author{{
Soebur Razzaque$^{a,b}$\thanks{email: \tt srazzaque@ssd5.nrl.navy.mil}
~and
A. Yu. Smirnov$^{c,d}$\thanks{email: \tt smirnov@ictp.it}} \\ \\
{$^a$Space Science Division, U.S. Naval Research Laboratory,} \\ 
{\, 4555 Overlook Ave. SW,  Washington, DC 20375, USA } \\
{$^{b}$National Research Council Research Associate}\\ \\
{$^{c}$The Abdus Salam International Centre for Theoretical Physics,} \\
{\, Strada Costiera 11, I-34014 Trieste, Italy} \\
{$^{d}$Institute for Nuclear Research, Russian Academy of Sciences, 
Moscow, Russia}}
\abstract{ High energy cosmic neutrino fluxes can be produced inside
relativistic jets under the envelopes of collapsing stars. In the
energy range $E \sim (0.3 - 10^5)$ GeV, flavor conversion of these
neutrinos is modified by various matter effects inside the star and
the Earth.  We present a comprehensive (both analytic and numerical)
description of the flavor conversion of these neutrinos which
includes: (i) oscillations inside jets, (ii) flavor-to-mass state
transitions in an envelope, (iii) loss of coherence on the way to
observer, and (iv) oscillations of the mass states inside the
Earth. We show that conversion has several new features which are not
realized in other objects, in particular interference effects (``L-
and H- wiggles'') induced by the adiabaticity violation.  The $\nu
-\nu$ scattering inside jet and inelastic neutrino interactions in the
envelope may produce some additional features at $E \gtrsim 10^{4}$
GeV.  We study dependence of the probabilities and flavor ratios in
the matter-affected region on angles $\theta_{13}$ and $\theta_{23}$,
on the CP-phase $\delta$, as well as on the initial flavor content and
density profile of the star. We show that measurements of the energy
dependence of the flavor ratios will, in principle, allow to determine
independently the neutrino and astrophysical parameters.  }

\received{January 9, 2010}
\accepted{February 8, 2010}


\begin{document}

\section{Introduction}

It is difficult to overestimate importance of future detection of high
energy cosmic neutrinos which will open new window to the Universe
(for reviews see e.g. \cite{nu_astro}).  This detection will bring unique information on
astrophysical sources of the neutrinos, as well as on neutrinos
themselves: their propagation, interactions and flavor conversion.  A
number of different high-energy, $\gtrsim 1$~GeV, neutrino sources
have been proposed in literature, which includes active galactic
nuclei (AGNs)~\cite{agn_nu}, gamma ray bursts (GRBs)~\cite{grb_nu},
core collapse supernovae (SNe)~\cite{sn_nu}, supernova
remnants~\cite{snr_nu}, etc.  Properties of neutrino fluxes, energy
range, shape of the energy spectra and flavor content depend on
physical conditions in the sources, as well as on effects of
propagation between the source and the Earth.  In particular, the
flavor content of fluxes is modified by the vacuum oscillations
between the production region and the Earth-based detectors.  For
known neutrino mass squared differences the oscillations are averaged
due to large distances (baselines), and the oscillation effects are
described by the averaged vacuum oscillation probabilities which do
not depend on neutrino energy.

The sources listed above are detected by their electromagnetic (EM)
radiation with space- and ground-based detectors.  No astrophysical
source has been detected so far by its high energy neutrino signal.
Apart from the sources visible in EM radiation, it is expected that
there are various hidden sources of neutrinos: the sources in which
the EM radiation produced in the inner part of the object is absorbed
in the surrounding dense material and only neutrinos can pass through
and be detected at the Earth. One such optically thick source is a
core collapse SN with mildly relativistic jets emitted by a central
engine, a black hole or a highly magnetized neutron star.  These jets
form promptly after the collapse~\cite{Razzaque04} and may not break
the envelope unlike highly-relativistic GRB jets.  Therefore they may
not lead to emission of high energy non-thermal EM
radiation. Observation of late time radio afterglow and explosion
geometry provide evidences of a plausible hidden jet in some
SNe~\cite{hidden_jet}.  These hidden jets, however, can be the sources
of high-energy neutrino fluxes generated by interactions of the shock
accelerated protons with surrounding matter and radiation
\cite{Razzaque04,Ando05,Razzaque05}.

The number of hidden sources can be much larger than the number of
observable ones, limited only by the ratio of type Ib/c and type II
SNe to GRB rates. The combined SN rate from roughly 4000 galaxies
within 20 Mpc is larger than one per year and the SN rate in nearby
starburst galaxies such as M82 and NGC253 is much higher than in the
Milky way \cite{Razzaque04,Ando05b}. Recent $\gamma$-ray observation
of M82 and NGC253 with the Fermi Large Area Space Telescope also
supports an increased SN activity~\cite{Fermi_starburst}.  The
prospect of high-energy neutrino detection from SNe in these galaxies
by neutrino detectors such as IceCube\cite{IceCube}
ANTARES~\cite{ANTARES} and KM3NeT~\cite{KM3NeT} is rather high, if a
significant fraction of the SNe are endowed with mildly-relativistic
jets. Optical follow-up triggered by neutrino events in IceCube along
the direction of the neutrino trajectories will further enhance
detection prospect by identifying the SN and extrapolating its light
curve back to the explosion time~\cite{Kowalski07}.

Neutrinos produced in the hidden jets cross an envelope and in
\cite{Mena:2006eq} it was shown that in general matter effects on
neutrino oscillations are not small. In particular, the minimal width
condition (the lower bound on the column density of electrons on the
way of neutrinos which is required for strong matter
effect)~\cite{Luna} can be satisfied. It was shown that the matter
effects inside a star can substantially modify the average vacuum
oscillation probabilies in the range from $10^2$ GeV (the IceCube
threshold) up to $10^{5}$ GeV.  Neutrino fluxes at the surface of the
Earth have been computed.  As an observable in the forthcoming
experiments, the ratio of shower events (induced mainly by the
electron and tau neutrinos) and the muon track events (produced mainly
by interactions of muon neutrinos and antineutrinos) has been
proposed~\cite{Mena:2006eq}.  The ratio as a function of energy
depends on values of the mixing angles $\theta_{13}$, $\theta_{23}$,
the CP-violation phase $\delta$ as well as on density profiles of the
stellar envelope.  It was concluded, in particular, that with about
$10^3$ events one can explore the neutrino properties such as the type
of mass hierarchy and CP violating phase, provided that the mixing
angles will be measured at future reactor and accelerator experiments
with high accuracy.  Also certain information about properties of the
source can be extracted.  The rate of detection of individual source
from nearby galaxies is one in 5--10 years.  If the neutrino
telescopes are able to detect the diffuse flux of neutrinos from all
the sources, still one can see some deviation of the flavor ratio from
that produced by the vacuum oscillations. Inversely, observation of
such a deviation in certain energy range can be explained by matter
induced transformations and large population of the hidden
sources. With high statistics this feature in the energy spectrum can
be extracted from large background of atmospheric neutrinos.

In this paper we have reconsidered the flavor conversion of neutrinos
from hidden jets. Our results differ from those in
Ref.~\cite{Mena:2006eq}. The difference originates from treatment of
averaging and the coherence loss.  Neutrinos produced by pions, muons
and kaons in the strong magnetic field have very short wave packets
and lose coherence due to separation of the wave
packets~\cite{Farzan}.  So, neutrinos arrive at the surface of the
Earth as incoherent fluxes of the mass eigenstates and coherence is
not restored in detector.  Therefore one should compute probabilities
of the flavor-to-mass transitions ($\nu_\alpha \rightarrow \nu_i$)
inside the star.  In contrast, in Ref.~\cite{Mena:2006eq} the
flavor-to-flavor transitions have been computed from the production
point to the surface of the star and then the oscillations on the way
from the surface of the star to the earth have been averaged.  This
leads to different results of numerical computations.  In particular,
fast oscillatory behavior of probabilities with energy appears
according to~\cite{Mena:2006eq}.  We have included in consideration
also additional effects not considered in~\cite{Mena:2006eq}, such as
energy-dependent particle to antiparticle ratio and flavor ratio in
the initial neutrino fluxes, oscillations in jets, inelastic
interactions of neutrinos and neutrino-neutrino scattering.  Detailed
and comprehensive study of the conversion inside the star and on the
way between the star and the Earth is performed.  We show that the
main effect is due to the adiabatic and partially adiabatic conversion
(the MSW-effect)~\cite{MSW} inside an envelope.  Both numerical and
analytical results for probabilities, neutrino fluxes and flavor
ratios at the Earth are presented.  Computation of number of events in
specific detectors is beyond the scope of this paper and will be given
elsewhere.

The paper is organized as follows.  In Sec.\ 2 we describe the model
of hidden source, summarize physical conditions at the neutrino
production site, and present properties of the generated neutrino
fluxes.  We describe neutrino conversion inside the star in Sec.\ 3
that includes the adiabatic conversion at low energies, adiabaticity
violation at energies above the 1-3 resonance, interference effects in
the range of adiabaticity violation which lead to H- and L- wiggles.
We describe properties of the conversion probabilities in specific
channels, their dependence on neutrino parameters and characteristic
of density profiles in Sec.\ 4.  In Sec.\ 5 we present neutrino fluxes
and flavor ratios at the surface of the Earth.  We consider their
dependence on neutrino parameters, original flavor content as well as
on the density profile of the stellar envelope. Conclusions are
presented in Sec.\ 6.  Some details of neutrino flux calculation,
explanation of the difference of results of~\cite{Mena:2006eq} and
this paper, and details of estimations of the Earth matter effects are
presented in Appendices A, B and C correspondingly.

\section{Neutrinos from jets: production, physical conditions, fluxes}

Here we discuss generic properties of source, conditions of neutrino
production and characteristics of neutrino fluxes.

\subsection{Properties of relativistic jets}

Hidden neutrino sources are associated to the core collapses of stars
with masses $M_\star \lesssim 28 M_{\odot}$. Models of these sources
are based on extrapolation of the observed properties of GRB and
models of observed jets.  Recall that stars with mass $M_\star \gtrsim
28 M_{\odot}$ and a fast-rotating core are widely believed to be the
progenitors of the long-duration GRBs~\cite{McFadyen99}.  Evidences of
highly relativistic jets, with bulk Lorentz factor $\Gamma_b \sim
10^2$ -- $10^3$, have been found in recent GRB data~\cite{Fermi_GRB}.

A much larger number of core-collapses with masses of projenitors
$M_\star \lesssim 28 M_{\odot}$ is believed to produce mildly
relativistic, $\Gamma_b \sim 10^{0.5}$ -- $10^1$, {\em slow} jets
which do not break through the stellar envelope unlike the GRB
jets~\cite{McFadyen01}.  The general picture is that materials from
the central engine (a black hole or a highly magnetized neutron star
created from the core-collapse) are emitted in lumps or shells with
mildly relativistic speeds along the rotation axis of the star, thus
forming a slow jet. Some initial shell or shells push out stellar
material (with sub-relativistic or mildly-relativistic speed), thus
make a cavity.  Subsequent shells, with relativistic speed in the
cavity, collide with each other due to variable outflow of accreting
materials similar to the GRB internal shocks which take place well
outside the stellar envelope~(for reviews see e.g. \cite{grb_review}).  Each of these binary
collisions produce shock waves (forward and reverse) in the colliding
shells. The shell(s), initially ejected, can also produce a shock in
the envelope (forward shock), and a reflected shock (reverse shock) in
the shell(s).  The shock in the envelope dissipates very quickly
because of a higher density, but could also accelerate particles
there.

The subsequent binary collisions between the shells take place mainly
at the edge of the jet or at the inner border of the envelope.  The
shock waves that are generated in the leading and trailing shells, in
a binary collision, are mildly relativistic, with the Lorentz factor
$\sim 1$ in the frame of the shells which, in turn, are moving at the
Lorentz factor $\sim 3$ in the observer frame~\cite{Razzaque04}.  As
these shock waves traverse the shells, magnetic field is generated
from turbulence in the upstream and downstream regions of the shock
front.  The magnetic fields in these two regions are similar.

The shocks in the hidden or burried jets are optically thick to
$\gamma$-rays as they are produced under the stellar envelope.  High
energy neutrino fluxes, however, are formed inside the slow jet due to
interactions of shock-accelerated protons with matter, ($pp-$
collisions) and EM radiation ($p\gamma-$ collisions) of
jets~\cite{Razzaque04,Ando05}.

Following Refs.~\cite{Razzaque04,Razzaque05} we adopt a slow jet model
with the following characteristics:

\begin{itemize}

\item
the total kinetic energy of the jet released over its duration is $E_j
\sim 10^{51.5}E_{51.5}$~ergs which is much smaller than the typical
GRB jet energy;

\item 
the jet duration is $t_j \sim 10t_{j,1}$~s, which is the typical
duration of the central engine's activity observed in GRBs;

\item 
the half angle of the jet, $\theta_j \sim 1/\Gamma_b$, is rather large
and implies an isotropic-equivalent total jet energy $E_{j, \rm iso}
\approx 2\Gamma_b^2 E_j$;

\item 
with a variability time scale $t_v \sim 0.1 t_{v,-1}$~s and $\Gamma_b
\sim 10^{0.5}\Gamma_{b,0.5}$, the internal shocks (collision between
two shells) take place at a radius $r_j \approx 2 \Gamma_b^2 c t_{v}
\sim 6.3\cdot 10^{10} \Gamma_{b,0.5}^2 t_{v,-1}$~cm;

\item 
the width of the shocked shells is $\Delta r_j \approx r_j/\Gamma_b^2
\sim 6.3\cdot 10^{9} t_{v,-1}$~cm and there can be $\approx t_j/2t_v
\sim 50$ consecutive collisions between shells which form shocks in
the shells during the jet lifetime.

\end{itemize}

This picture is somewhat idealized since GRB data show more complex
time structure than emission from identical shocked shells.
Nevertheless, it captures the basic scenario of internal shocks.

The pre-shock number density of particles in the jet calculated in the jet
comoving frame (we denote the corresponding characteristics by
``$\prime$'') is 
\be
n'_e \simeq n'_p \approx \frac{E_{j,\rm iso}}
{4\pi r_j^2 \Gamma_b^2 t_j m_p} \sim 3.2\cdot 10^{20} 
~{\rm cm}^{-3}~
\frac{E_{51.5}}{\Gamma_{b,0.5}^4 t_{j,1} t_{v,-1}^2}.
\label{jet_part_density}
\ee
(Here and below $\hbar = c = 1$.)  The post-shock number density is
$\approx 4n'_p$ and the strength of magnetic field that forms due to
turbulence in the shock region is
\be
B' \approx \sqrt{32\pi\epsilon_B n'_p m_p}
\sim 6.3\cdot 10^{8} ~{\rm G}~ 
\sqrt{\frac{\epsilon_{B,-2}E_{51.5}}
{\Gamma_{b,0.5}^4 t_{j,1} t_{v,-1}^2}},
\label{shock_B_field} 
\ee
where $\epsilon_B \sim 10^{-2}\epsilon_{B,-2}$ is the fraction of
shock energy that goes into creating magnetic field.  The magnetic
field drops down to the surrounding value of few Gauss (which is the
field from the ``central engine'', a magnetar or black hole) between
successive shocks.  The temperature of the thermal photons, created by
shocked electrons which carry a fraction $\epsilon_e \sim 10^{-1}
\epsilon_{e,-1}$ of the total energy, in the jet equals
\be
kT' \approx \left( \frac{15}{\pi^2} 
\frac{\epsilon_e E_{j,\rm iso}} 
{4\pi r_j^2 \Gamma_b^2 t_j} \right)^{1/4} \sim 4.3~{\rm keV}
\left( \frac{\epsilon_{e,-1}E_{51.5}}
{\Gamma_{b,0.5}^4 t_{j,1} t_{v,-1}^2} \right)^{1/4}.
\label{jet_temp}
\ee

\subsection{Proton acceleration}

We consider a Fermi acceleration mechanism for protons in shocks.  An
acceleration time scale is proportional to the Larmor time scale and
equals
$$
t'_{p,\rm acc} \approx \frac{\varphi}{m_p^2}\frac{B_{\rm cr}}{B'}
E'_p \sim 10^{-9}~{\rm s}~ \varphi_1 \left( \frac{E'_p}{\rm TeV} \right) 
\left( \frac{B'}{10^9~\rm G} \right)^{-1},
$$
where $B_{\rm cr} =m_p^2/q \approx 1.488\times 10^{20}$~G is the
critical magnetic field and $\varphi \sim 10\varphi_1$ is the number
of gyro-radii required to increase the particle energy by $e$-fold.
We assume an acceleration spectrum $N(E_p) \propto E^{-2}_p$. Then the
differential flux of protons, if they could escape freely from the
jet, at a luminosity distance $d_L$ would be
\be
\Phi_p (E_p) \approx \frac{E_{j,\rm iso}}
{4\pi d_L^2 E_p^2 t_j ~{\rm ln}(E_{p,\rm max}/\Gamma_b m_p)}. 
\label{p_flux}
\ee
The maximum proton energy is determined by the shortest time scale for
energy losses which is the synchrotron cooling time scale
$$
t'_{p,\rm syn} \approx 
\frac{9}{4} \frac{1}{r_e m_e}
\left( \frac{B_{\rm cr}}{B'} \right)^2 \frac{1}{E'_{p}} 
\sim 2\cdot 10^{-2} ~{\rm s}~ 
\left( \frac{E'_p}{\rm TeV} \right)^{-1} 
\left( \frac{B'}{10^9~\rm G} \right)^{-2}. 
$$
In the jet frame this gives 
\be
E^\p_{p,\rm max} \approx \sqrt{ 
\frac{9}{4} \frac{m_p^2} {\varphi r_e m_e}
\frac{B_{\rm cr}}{B^\p} } \sim 2\cdot 10^{3} ~{\rm TeV}~ 
\varphi_1^{-1/2} \left( \frac{B'}{10^9~\rm G} \right)^{-1/2}.
\label{maximum_E} 
\ee
and consequently, in the observer's frame $E_{p,\rm max} = \Gamma_b
E^\p_{p,\rm max} \sim 6.3\cdot 10^{3}$~TeV.  In turn, this energy
determines the maximal energy of the neutrino spectrum.

\subsection{Neutrino fluxes from meson decays}

The rate of $pp-$ interaction by shock-accelerated protons
in the jet is given by
\begin{eqnarray}
K_{pp} (E'_p) &\approx & n'_p \sigma_{pp} (E'_p) \nonumber \\
&\sim & 3\cdot 10^5  ~{\rm s}^{-1}~ 
\left[ 1 + 0.0548\, {\rm ln}\left( \frac{E^\p_p}{\rm TeV} \right)
+ 0.0073\, {\rm ln}^2 \left( \frac{E^\p_p}{\rm TeV} \right)
\right],
\nonumber
\end{eqnarray}
where $\sigma_{pp}$ is the total inelastic cross-section and the
parameterization is valid for $E^\p_p \gtrsim 10~{\rm GeV}$. The
scattering rate of $p\g-$ interactions with thermal photons in the jet
above the threshold of $\pi-$production is given by
\begin{eqnarray}
K_{p\gamma} (E'_p) &\approx & \frac{m_p^2}{2 E_p^{'2}} 
\int_{0}^{\infty} d\varepsilon' 
\frac{n'_\gamma(\varepsilon)}{\varepsilon^{2}} 
\int_{\varepsilon_{\rm th}}^{2E'_p\varepsilon'/m_p} 
d\varepsilon_r ~\varepsilon_r \sigma_{p\gamma} (\varepsilon_r)
\nonumber \\
&\sim & \frac{\sigma_0}{\pi^2} 
\int_{\varepsilon^\p_{th} m_p/2E^\p_p}^{\infty} d\veps^\p
\frac{\veps^{\p 2}}{\exp[\veps^\p /kT^\p] - 1} \sim  10^7 
~{\rm s}^{-1}. 
\nonumber
\end{eqnarray}
Here $\veps^\p_r = \veps^\p (1-\beta_p \cos\theta)E^\p_p/m_p$ is the
photon energy evaluated in the proton's rest frame for the angle
$\theta$ between the directions of the proton and target photon.  We
used for simiplicity a constant, $p\g$ cross-section $\sigma_{p\g}
(\veps^\p_r)\sim \sigma_0 \sim 200\mu b$, above a threshold photon
energy $\veps^\p_r \approx \veps^\p_{\rm th} \approx 0.2$~GeV for pion
production in the rest frame of the proton.  The threshold energy of
protons which produce pions in the $p\g-$ interactions equals
according to (\ref{jet_temp})
$$
E^\p_{p,th} \gtrsim \frac{\veps^\p_{th}m_p}{kT^\p} \approx 42~{\rm
TeV}.
$$
Thus, the $pp-$ scattering rate dominates below $E_{p,th} \lesssim
E^\p_{p,th}\Gamma_b \sim 133$~TeV, that is the whole energy range of
interest.

The fluxes of $\pi-$ and $K-$ mesons from the $pp-$ or $p\gamma-$ 
interaction at production can be calculated as
\be
\Phi_{\pi (K)} (E_{\pi (K)}) = \int dE_p \Phi_p(E_p) 
K_{pp/p\gamma} (E^\p_p) t^\p_{\rm dyn} \frac{Y_{\pi (K)}}{E_p},
\label{pi_K_formula}
\ee
where $Y_{\pi (K)} \equiv E_p (dn_{\pi (K)}/dE_{\pi (K)})$ is the pion
(kaon) yield function from the $pp-$ or $p\gamma-$ interactions, and
$t^\p_{\rm dyn} = r_j/\Gamma_b$ is the dynamic or light crossing time.
The quantity $K_{pp/p\gamma} (E^\p_p) t^\p_{dyn}$ is equivalent to the
optical depth of the respective interactions and is very large for
both the $pp-$ and $p\g-$ processes.  For roughly constant values of
$K_{pp/p\gamma} (E^\p_p) t^\p_{\rm dyn}$ the integral in
(\ref{pi_K_formula}) corresponds to the fraction of the proton beam
energy carried by $\pi^+$ ($f_{\pi^+} \sim 17\%$) and $K^+$ ($f_{K^+}
\sim 2\%$) mesons and follow the primary proton
spectrum~\cite{Gaisser}. Thus the corresponding spectra at production
can be estimated as
\be
\Phi_{\pi^+ (K^+)}^0 (E_{\pi (K)}) \sim \frac{f_{\pi^+ (K^+)} 
E_{j,\rm iso}} {4\pi d_L^2 E_{\pi (K)}^2 t_j
~{\rm ln}(E_{p,\rm max}/\Gamma_b m_p)},
\label{p_K_flux}  
\ee
following Eq.~(\ref{p_flux}). For the $\pi^-$ and $K^-$ fluxes the
fraction of the proton beam energy carried by the mesons are
$f_{\pi^-} \sim 13\%$ and $f_{K^-} \sim 1\%$, respectively.  Charm
production and semi-leptonic decay contribute to neutrino flux at very
high energies~\cite{Enberg09}, which we ignore for the present
discussion.

We use the decay constants listed in Ref.~\cite{Lipari93} to calculate
the neutrino fluxes from the direct decay channels $\pi/K\to \nu_\mu$,
the chain decay channels $\pi/K\to \mu \to \nu_\mu \,\nu_e$ and the
$K^+ \to \pi^0 e^+ \nu_e$ decay channels by taking into account their
energy losses (see Appendix A).  The fluxes are plotted in
Fig.~\ref{fluxes}.  Note that very high-energy neutrinos interact
inelastically with nucleons in the stellar envelope and are subject to
absorption.  We discuss this issue in Sec.\ 2.7.

The ratios of the $\nu_e$ to $\nu_\mu$ and ${\bar \nu}_e$ to ${\bar
\nu}_\mu$ fluxes can be fitted reasonably well as
\begin{eqnarray}
\eps (E) \equiv \frac{\Phi^0_{\nu_e}}{\Phi^0_{\nu_\mu}} &\approx &  
\frac{13.9}{(E_\nu/\rm GeV)^{1.65}} [\Theta (E_\nu - 10~{\rm GeV}) 
+ \Theta (45~{\rm GeV} - E_\nu)] 
\nonumber \\ && \;\;\; + 0.0265 ~[\Theta (E_\nu - 45~{\rm GeV}) 
+ \Theta (10^4~{\rm GeV} - E_\nu)]~;  
\nonumber \\
{\bar \eps} (E) \equiv \frac{\Phi^0_{{\bar \nu}_e}}
{\Phi^0_{{\bar \nu}_\mu}} &\approx &  
\frac{7.85}{(E_\nu/\rm GeV)^{1.5}} [\Theta (E_\nu - 10~{\rm GeV}) 
+ \Theta (45~{\rm GeV} - E_\nu)] 
\nonumber \\ && \;\;\; + 0.0265 ~[\Theta (E_\nu - 45~{\rm GeV})
+ \Theta (10^4~{\rm GeV} - E_\nu)]~,  
\label{nu_e_to_nu_mu}
\end{eqnarray}
where $\Theta(x \geq 0) = 1$ and $\Theta(x < 0) = 0$.  The ratios
decrease from the initial value because of a decreasing contribution
to the $\nu_e-$flux from the chain $\pi - \mu$ decays, and becomes
constant at higher energy because of an approximately constant ratio
between the $\nu_e-$flux from $K^+ \to \pi^0 e^+ \nu_e$ and
$\nu_\mu-$flux from $K^+ \to \mu^+ \nu_\mu$ channels.

The flux of muon neutrinos in the production region (the $r_{j} \sim
6.3\cdot 10^{10}$ cm) and for a jetted source at 10~Mpc can be
parameterized, with a piece-wise power-law function, as
$$
\frac{E^2 \Phi_{\nu_\mu}}{\rm GeV~cm^{-2}~s^{-1}} =  
\cases{
1.64 ~;~ 1 < E/{\rm GeV} < 6, \cr
1.64~(E/{6~\rm GeV})^{-0.55}
~;~ 6 \le E/{\rm GeV} \le 600, \cr
0.13~(E/{600~\rm GeV})^{-1} 
~;~ 600 < E/{\rm GeV} \le 3\times 10^4 ~.
} 
$$ 
This fit reproduces exactly the $\nu_\mu$ flux plotted in
Fig.~\ref{fluxes} below and above 6~GeV and 600~GeV, respectively, and
deviates by at most $30\%$ from the numerical flux within 6-600~GeV.
We use the numerical flux values for all our calculaions.

The ratio of neutrino and antineutrino fluxes can be as large as
$\Phi_\nu/\Phi_{\bar{\nu}} \sim 1.5$ because of the preferential
production of $\pi^+$ and $K^+$ over $\pi^-$ and $K^-$. Indeed, in the
$pp$-collisions the leading pion (which gives main contribution to
neutrino flux) is positive.  In $p\gamma-$ collisions there is a
leading $\pi^+$, although the secondary neutron with roughly $85\%$ of
the initial proton's energy will interact further and produce a
$\pi^-$.

This effect is present in the atmospheric neutrino fluxes, however the
ratio of neutrino and antineutrino fluxes and details of the energy
dependence is different in our situation.  The original flavor content
is thus,
\be
\epsilon(E) : 1 :0 ~~~ {\rm and} ~~~\bar{\epsilon}(E) : 1 : 0 ~,    
\label{flcontent}
\ee
where $\epsilon(E)$ and $\bar{\epsilon}(E)$ decrease from 0.5 at low
energies down to $10^{-2}$ at high energies.

The charged $\pi$, $K$ mesons and muons can also be accelerated in the
shocks~\cite{Koers07} similarly to the protons, with a maximum energy
$E_{\pi;K,\rm max} = (m_{\pi;K}/m_p)^2 E_{p,\rm max}$ following
Eq.~(\ref{maximum_E}).  Although a detail study is lacking, it might
be possible that $\pi$, $K$ and $\mu$ decay without suffering severe
energy losses.  In such a scenario, the flux ratio 1 : 2 : 0 will be
maintained to high energies.  In reality one can expect some
intermediate situation between this flavor content and the one in
Eq.~(\ref{flcontent}) without acceleration of mesons. In what follows
we will present computations for both extreme flavor contents.

The parent mesons and muons are isotropically disributed in the
shocked region because of their small Larmor radius in the large
magnetic field. Therefore the neutrinos are also isotropically
produced in the comoving frame. However in the observer frame they are
emitted mostly along the jet direction similar to the relativistic
beaming effect for photons.  Then neutrinos, produced in the
low-density jet, propagate through the stellar envelope.

\FIGURE{\epsfig{file=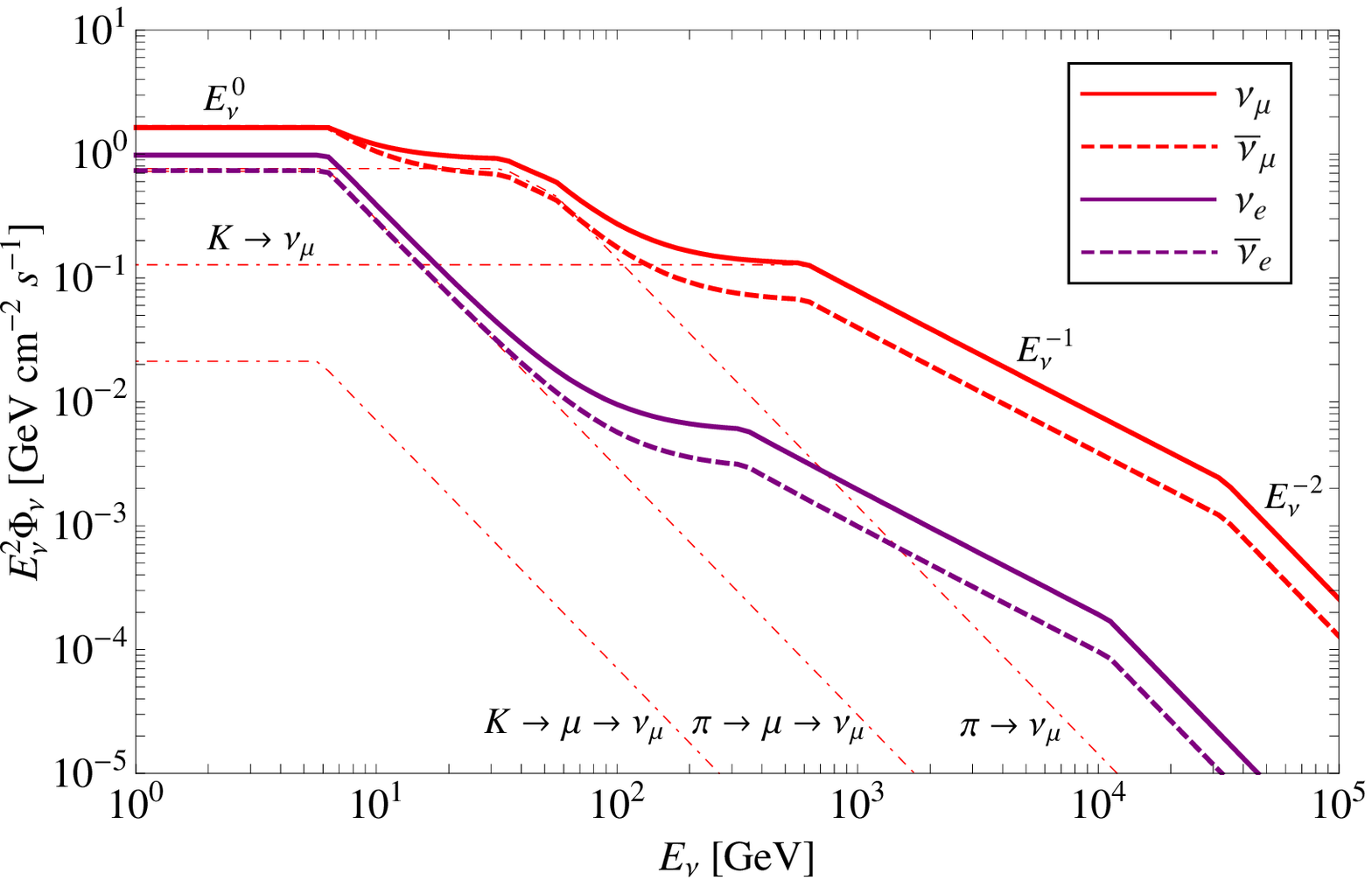,width=10cm}
\caption{Neutrino and antineutrino fluxes (thick curves) at 
production from a hidden source at 10~Mpc. The red (purple) thick 
solid curve corresponds to $\nu_\mu$ ($\nu_e$) flux while the 
red (purple) thick dashed curve corresponds to ${\bar \nu}_\mu$ 
(${\bar \nu}_e$) flux.  The labeled thin dot-dashed curves 
correspond to different components of the $\nu_\mu$ flux. }
\label{fluxes}}

\subsection{Stellar envelope: density profile}

The pre-explosion stars of the GRBs and the associated type Ic
supernovae are widely believed to be He stars with radius $R_\star
\approx 10^{11}$~cm.  Supernovae of type II and Ib are thought to be
explosions of larger stars with radius $R_\star \approx 3\times
10^{12}$~cm, and have a H envelope on the He core.  The envelopes
usually have density profile $\rho (r) \propto r^{-k}$, with $k=1.5$
-- 3~\cite{Matzner99}.  Following Ref.~\cite{Mena:2006eq} we adopt two
stellar profiles, model [A] and [B] parameterized as
\begin{eqnarray}
{\rm [A]} \,\, \rho (r) &=& 3.3\cdot 10^{-6}~{\rm g~cm}^{-3} 
\left( \frac{R_\star}{r} -1\right)^3 ~,
\nonumber \\
{\rm [B]} \,\, \rho (r) &=& 2.8\cdot 10^{-5} ~{\rm g~cm}^{-3}
\cases{ (R_\star/r)^{17/7} \,;\, R_0 < r < r_b \cr
(R_\star/r_b)^{17/7} (r-R_\star)^5/(r_b - R_\star)^5 \,;\, 
r \ge r_b~. }
\label{profiles}  
\end{eqnarray}
The radius of inner border of the envelope equals $R_0 \approx r_j
\approx 6.3\cdot 10^{10}$ cm.  According to Eq.~(\ref{profiles}), the
density at $R_0$ is $\rho_0 \approx 0.33$~g~cm$^{-3}$ but can vary
between (0.1--1)~g~cm$^{-3}$.  The corresponding density of electrons
is $n_0 = \rho_0 N_A Y_e \approx 10^{23}$~cm$^{-3}$ for $Y_e =
0.5$. Thus, neutrinos are produced in a low density region of jet and
propagate through a higher density envelope. We assume that the
boundary between the jet front and the envelope is sharp (much smaller
than any oscillation scale).  Neutrinos are produced in a low density
region of jet. Typical size of this region is about $r_\nu \sim 6.3
\cdot 10^{9}~ {\rm cm}$, that is, about 1/10 of a jet size.

Notice that in general, all characteristics of a star change in time
over jet duration.  In particular, the jet parameters such as
$\Gamma_b$ and $\Delta r_j$ are subject to variation over time. The
jet radius $r_j$ is expected to increase slowly with time as the jet
burrows through the envelope and eventually chokes or successfully
breaks out, the parameters of the envelope $R_0 \approx r_j$ and $n_0$
change with time.

\subsection{Neutrino fluxes and densities}

According to Fig.~\ref{fluxes} the flux integrated over the energy
above 1 GeV, which gives approximately the total flux of neutrinos at
$r_j$, is $F (> 1 {\rm GeV}) \approx 10^{30}~{\rm cm}^{-2}~{\rm
s}^{-1}$.  Then the number density of neutrinos in the source equals
$n^\p_\nu \sim 10^{19} ~{\rm cm}^{-3}$.  Notice that this is a
densitity in the reference frame of the jet where neutrino emission is
approximately isotropic.  Therefore the potential due to $\nu - \nu$
scattering equals $\mu = \sqrt{2} G_F n^\p_\nu$.  It should be
compared with the vacuum frequency $\omega = \Delta
m_{31}^2/2E^{\prime}$. We find that $\mu \sim \omega$ at the energy
$E^{\prime} \sim 10^6$ GeV, which corresponds to $E \sim 3 \cdot 10^6$
GeV in the observer frame.  Consequently, for energies of interest the
neutrino-neutrino effective potential is too small to induce the
collective neutrino effects.  Some collective effect may show up at $E
> 1$ PeV.

\subsection{Coherence and averaging} 

Let us esimate the size of the wave packets of the produced neutrinos
\cite{Farzan}.  Pions mainly decay near shock fronts in the regions
with strong magnetic field and increased density.  Pions undergo
collisions with surrounding photons and gas.  The latter (gas)
dominates at low energies $(E < 4 \cdot 10^{2})$ GeV.  The mean free
path with respect to the collisions can be estimated as $
\lambda_{col} = (10^5,~ 3 \cdot 10^4, ~ 5 \cdot 10^3)$ cm for energies
$E = (0.1, ~1, ~10)$ TeV, correspondingly (in the rest frame of jet
bulk).  For neutrino emission in the forward direction the size of the
wave packet is then \cite{Farzan} $ \sigma_x =
\lambda_{col}/\gamma_\pi, $ where $\gamma_\pi$ is the Lorentz factor
of pion.  This gives $\sigma_x = (10^2,~ 3,~ 5 \cdot 10^{-2})$ cm for
energies $E = (0.1, ~1, ~10)$ TeV, correspondingly.  However, very
strong kinematical shortening of the wave packets occur in presence of
the magnetic field due to bending of pion trajectory. According to
\cite{Farzan}, the size of the wave packet is given by
\be
\sigma_x = 3.5 \cdot 10^{-14} ~{\rm cm}
\left( \frac{\Gamma_{b}}{3} \frac{10^8~ {\rm G}}{B} \right)^{1/2}
\left( \frac{1 ~{\rm TeV}}{E}\right)^{3/2}. 
\label{sigmax-B}
\ee
For $\Gamma_{b} = 3$, $B = 10^8$ ~G and neutrino energies $E = (10^2,
~10^3,~ 10^4)$ GeV we obtain from this formula $\sigma_x = (10^{-12},~
3.5 \cdot 10^{-14}, ~ 10^{-15})$~cm correspondingly.

Separation of the wave packets of different mass states on the way $L$
is given by $d_s = L \Delta m^2/(2 E^2)$.  For typical distance 10 Mpc
and $\Delta m^2_{31}$ we find
$$
d_s = 0.045~{\rm cm} \left(\frac{1~ {\rm TeV}}{E}\right)^2.
$$
Thus, $d_s \gg \sigma_x$, and therefore the wave packets are well
separated at the detection site.  This means that the coherence is
lost in configuration space in the course of neutrino propagation.

The coherence will not be restored at the detection.  Indeed, the
separated packets interact coherenly, if the detector has long memory
(time interval of coherent detection). In this case the packets
interactions would interfere and produce oscillatory pattern in the
energy scale. The period of this oscillatory pattern, that is, the
energy interval over which the phase changes by $2\pi$ is given by
$\Delta E_T = E l_\nu/L$, where $l_\nu$ is the oscillation length.
For $L = 10$ Mpc and $\Delta m^2_{31}$ we obtain
\be
\Delta E_T = 3 \cdot 10^{-3} ~{\rm eV} 
\left( \frac{E}{1 {\rm TeV}} \right)^2. 
\label{enres}
\ee
In practice it is not possible to determine the neutrino energy with
such an accuracy.  According to Eq.~(\ref{enres}) at 1 TeV one needs
to have the energy resolution $\Delta E / E < 10^{-15}$ to see
non-averaged oscillation effect, whereas one may achieve $\Delta E / E
\sim 0.1$.  Thus, incoherent fluxes of mass eigenstates arrive at the
surface of the Earth and interact in a detector.

Let us estimate the coherence length (the distance over which the wave
packets are completely separated): $L_{coh} = \sigma_x 2E^2/\Delta
m^2$.  Using $\sigma_x$ from (\ref{sigmax-B}) we obtain for $\Delta
m^2_{31} = 2.4\cdot 10^{-3}$~eV$^2$
$$
L_{coh} = 1.4 \cdot 10^{13} ~{\rm cm}
\left( \frac{\Gamma_{b}}{3} \frac{10^8~{\rm G}}{B}  \right)^{1/2}
\left( \frac{E}{1~ {\rm TeV}}\right)^{1/2}.
$$
Notice that $L_{coh} \propto \sqrt{E}$ and for $E > 200 $ GeV,
$L_{coh}$ becomes larger than the size of star.

Some small part of pions decays in jet between shocks where the
magnetic field is much smaller. For $B = 1 $~G and $E = 1$ TeV we
obtain $\sigma_x = 3.5 \cdot 10^{-10}$~cm and $L_{coh} = 1.4 \cdot
10^{17}$~cm.  Similar situation is for the 2-body decay of K-mesons.

For muon decays (and also for 3-body $K-$ decays) in the magnetic
field, size of the neutrino wave packets is much larger \cite{Farzan}:
$$
\sigma_x = 1.7 \cdot 10^{-11} ~{\rm cm}
\left(\frac{\Gamma_{b}}{3}\right) 
\left(\frac{10^8 ~{\rm G}}{B} \right)
\left(\frac{1 ~{\rm TeV}}{E}\right)^{2}. 
$$
For $\Delta m^2_{31}$ the coherence length equals 
$$
L_{coh} = 7 \cdot 10^{15} ~{\rm cm}
\left( \frac{\Gamma_{b}}{3}\right) 
\left(\frac{10^8 ~{\rm G}}{B} \right), 
$$
and it does not depend on $E$. Although this length is about 2 - 3
orders larger than for pions, conclusions about loss of coherence are
the same.

\subsection{Effect of inelastic interactions} 

Evolution of the flavor neutrino states $\nu_f^T = (\nu_e, \nu_\mu,
\nu_\tau)$ is described by the equation:
\be
i \frac{d \nu_f}{dt} = (H + H_{int}) \nu_f, 
\label{eveq}
\ee
where $H$ is the standard Hermitian part which includes the vacuum and
refraction terms (the real part of scattering amplitudes, $A^{Re}$)
and $H_{int}$ describes inelastic interactions (the imaginary parts of
scattering amplitudes, $A^{Im}$).  Since the interactions are flavor
diagonal in the lowest order of perturbation theory, we have
$$
H_{int} \equiv - \frac{i}{2} diag(\Gamma_e, \Gamma_\mu, \Gamma_\tau).
$$
For $E > 3 \cdot 10^{3}$~GeV (which corresponds to $ 0.5 \Gamma =
A^{Im} \sim A^{Re}$ for scattering on nucleons) the inelastic
interaction can not be neglected. However, for such high energies in
the first approximation the inelastic amplitudes for all neutrino
species are the same: $\Gamma_e\approx \Gamma_\mu \approx \Gamma_\tau
= \Gamma$.  Indeed, the difference of masses of the charged leptons as
well as the inelastic scattering on electrons can be neglected.
Consequently, the inelastic part of the Hamiltonian becomes
proportional to the unit matrix: $ H_{int} = - \frac{i}{2} \Gamma I.
$ In this case the inelastic interactions and oscillations factor out.
The former does not influence the oscillation pattern.  Indeed, we can
define new flavor wave functions as $ \tilde{\nu}_f = e^{\Gamma t/2}
\nu_f $.  Inserting this relation into (\ref{eveq}) we obtain
$$
i \frac{d \tilde{\nu}_f}{dt} =  H \tilde{\nu}_f 
$$
without the inelastic part. The factor due to inelastic interactions
then appears in the probabilities: $P = e^{- \Gamma t}
\tilde{P}$. This factor describes absorption of neutrinos or
scattering which lead to departure of neutrinos from the coherent
state.  The effect of neutrino absorption in the charged current
processes can be included in the energy spectrum of neutrinos.  Decays
of $\mu$ and $\tau$ produced in the CC interactions of neutrinos will
generate secondary neutrinos.  Due to falling down spectrum with
energy, the contribution of these neutrinos to the total flux is
small.  In the case of neutral current neutrino interactions one
should consider oscillations of the scattered neutrinos. In what
follows we will neglect this scattering since the event rate of
neutrinos with $E>3\cdot 10^3$~GeV is small even for the nearest
plausible source.

\section{Conversion inside the star: general consideration}

\subsection{Propagation basis}

For the mixing matrix in vacuum, $\nu_f = U_{\rm PMNS} \nu$, we use the
standard parameterization
\be
U_{PMNS} = U_{23}(\theta_{23}) \Gamma_\delta U_{13}(\theta_{13}) 
U_{12}(\theta_{12}), 
\label{pmns}
\ee
where $U_{ij} = U(\theta_{ij})$ is the matrix of rotation in the
$ij-$plane over the angle $\theta_{ij}$ and $\Gamma_\delta \equiv
diag(0, 0, e^{i\delta})$.  It is convenient to consider dynamics of
the conversion in the so-called propagation basis, $\nu_f^{\prime}$,
defined as
\be
\nu_f \equiv U_{23} \Gamma_\delta \nu_f^{\prime} ~, ~~~~ 
\nu_f^{\prime}\equiv (\nu_e, \nu_\mu^{\prime}, \nu_\tau^{\prime})~.
\label{propbasis}
\ee
In this basis the CP-phase is eliminated, so that the dynamics of
conversion is not affected by $\delta$. The CP-violation appears only
in projection of the initial and final states onto the propagation
basis.  We have explicitly
\be
U_{23} \Gamma_\delta = \pmatrix{
1   &  0 &  0 \cr
0   &  c_{23}  &  s_{23} e^{i \delta } \cr
0   &   - s_{23}   & c_{23} e^{i \delta}
},
\label{toprop}
\ee
where $s_{23} \equiv \sin \theta_{23}$, $c_{23} \equiv \cos
\theta_{23}$.  According to (\ref{pmns}) and (\ref{propbasis}) the
vacuum mixing matrix in the propagation basis is given by
$\nu_f^{\prime} = U^\prime \nu$, where
\be
U^\prime  = U_{13} U_{12} = \pmatrix{
c_{13} c_{12}   &  c_{13} s_{12} &  s_{13} \cr
- s_{12}   &  c_{12}  &  0 \cr
- s_{13} c_{12}   &   - s_{13} s_{12}   & c_{13}
}.
\label{uprime}
\ee
Here $c_{12} \equiv \cos \theta_{12}$, $s_{12} \equiv \sin
\theta_{12}$, etc.  It is straightforward to show that the mixing
matrix in matter in the propagation basis, $U^\prime_m$, up to
additional small 2-3 rotation has the same form as in (\ref{uprime})
with mixing angles in matter: $\theta_{12} \rightarrow \theta_{12}^m$,
$\theta_{13} \rightarrow \theta_{13}^m$:
\be
U^\prime_m  \approx \pmatrix{
c_{13}^m c_{12}^m   &  c_{13}^m s_{12}^m &  s_{13}^m \cr
- s_{12}^m   &  c_{12}^m  &  0 \cr
- s_{13}^m c_{12}^m   &   - s_{13}^m s_{12}^m   & c_{13}^m
}.
\label{uprimem}
\nonumber
\ee
The level crossing scheme (see Fig.~\ref{level_crossing}) is similar
to the one for low energy supernova neutrinos (see, e.g.,~\cite{sn}).

\FIGURE{\epsfig{file=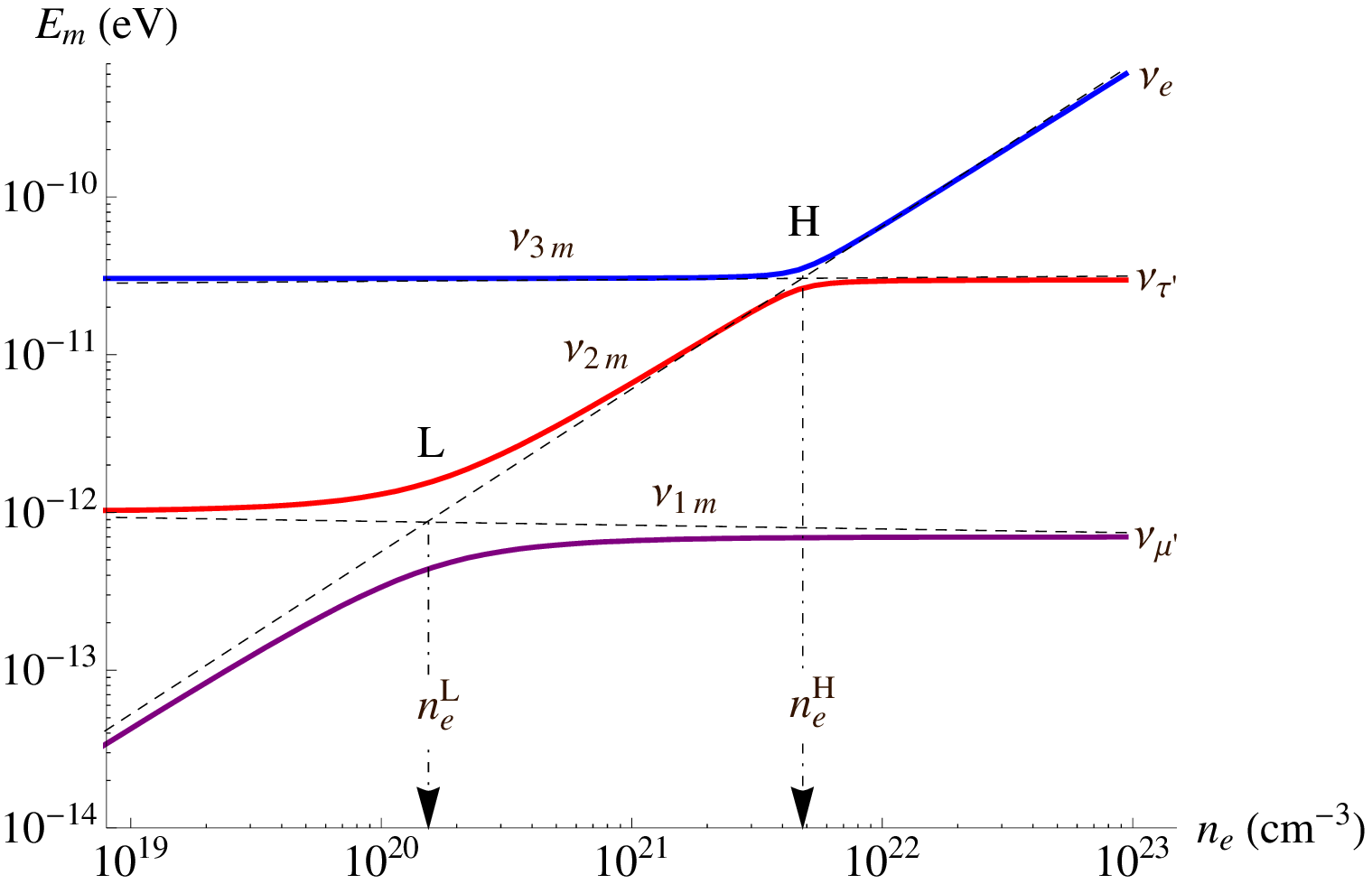,width=8.5cm} [t]
\caption{The level crossing scheme for neutrinos in stellar envelope
in the case of normal mass hierarchy. The solid lines correspond to
the eigenvalues of the effective Hamiltonian as functions of the
electron number density.  The dashed lines correspond to the energy
levels of the flavor states $\nu_e$, $\nu_{\mu^\p}$ and
$\nu_{\tau^\p}$ in the propagation basis.  The vertical dot-dashed
lines indicate the L- and H- resonance densities.  Note that the
positions of the resonances are inversely proportional to the neutrino
energy $E$ (plotted here for $E=2$~TeV).  We used $\sin^2 2\theta_{13}
= 0.08$.}
\label{level_crossing}}

\subsection{General expressions for probabilities}
 
Decays of $\pi-$, $\mu-$ and $K-$ mesons produce incoherent neutrino
and antineutrino fluxes of definite flavors: $\nu_e$, $\nu_{\mu}$,
etc..  Since the mass states are the eigenstates of propagation in
vacuum and coherence between them is lost due to separation of the
corresponding wave packets (see Sec.\ 2.6), we can write for the
$\nu_\alpha \rightarrow \nu_\beta$ conversion probability at the Earth
\be
P(\nu_\alpha \rightarrow \nu_\beta) = 
\sum_i P_* (\nu_\alpha \rightarrow \nu_i) |U_{\beta i}|^2,  
\label{ab-general}
\ee
where $P_* (\nu_\alpha \rightarrow \nu_i)$ is the flavor-to-mass
conversion probability inside the star.  This expression differs from
the one used in Ref.~\cite{Mena:2006eq}, and detailed explanation of
the difference is given in the Appendix B.  If neutrinos cross the
Earth the formula (\ref{ab-general}) is modified as 
\be
P(\nu_\alpha \rightarrow \nu_\beta) =
\sum_i P_* (\nu_\alpha \rightarrow \nu_i) 
P_E(\nu_i \rightarrow \nu_\beta),
\label{ab-generalE}
\ee
where $P_E(\nu_i \rightarrow \nu_\beta)$ is the probability of $(\nu_i
\rightarrow \nu_\beta)$ oscillations in the matter of the Earth.  The
probability $P_E$ can substantially deviate from $|U_{\beta i}|^2$ at
energies $E < 10$ GeV.  For higher energies matter suppresses
oscillations inside the Earth, so that the probability is reduced to
the one in Eq.~(\ref{ab-general}).  We give explicit expressions for
the probabilities in the Earth for constant density in Appendix C.

Inside a star the flavor conversion occurs first in jet and then in
envelope.  Therefore one should take into account oscillations in the
production region of jet and perform integration over this region.
Then the total conversion probability inside the star equals
$$
P_*(\nu_\alpha \rightarrow \nu_i) = \frac{1}{r_\nu} \int_0^{r_\nu} dx 
\left| \sum_\xi A_{jet} (\nu_\alpha \rightarrow \nu_\xi)(x)
\cdot A_{env} (\nu_\xi \rightarrow \nu_i) \right|^2,
$$
where $A_{jet} (\nu_\alpha \rightarrow \nu_\xi)(x)$ is the amplitude
of probability of $\nu_\alpha$ transition to the flavor state
$\nu_\xi$ inside jet: between the production point $x$ and the inner
border of the envelope.  The integration is performed over the
production region and in what follows we will use the one-dimensional
(1D) integration for simplicity.  Note that 1D consideration is
valid if we consider neutrinos from a given source, although in this
case the density profile can differ from the one in radial direction
(recall that jet has rather large cone angle). If $\theta_{tr} \gtrsim
\theta_{jet}$ is the angle between the radial (jet) direction and
direction to observer, the radial distance equals $r = x \cos
\theta_{tr}$, where $x$ is the distance along the trajectory. Then the
density profile which neutrino experiences is $n(x \cos \theta_{tr}) =
n_0 (x \cos \theta_{tr}/r_0)^{-k}$.  In this case the density profile
is flatter and the evolution will be more adiabatic.  For neutrinos
produced by all hidden sources (the diffuse flux) we need to perform
also integration over $\cos \theta_{tr}$.

In the propagation basis the probability can be written as 
\be
P_*(\nu_\alpha \rightarrow \nu_i) = \frac{1}{r_\nu}
\int_0^{r_\nu} dx \left| \sum_\xi \sum_\beta 
(U_{23} \Gamma_\delta)_{\alpha \beta} 
\cdot 
A_{jet} (\nu_\beta^\prime \rightarrow  \nu_\xi^\prime)(x) 
\cdot 
A_{env} (\nu_\xi^\prime \rightarrow \nu_i) \right|^2. 
\label{star-totcp}
\ee
Here $A_{jet} (\nu_\beta^\prime \rightarrow \nu_\xi^\prime)$ is the
amplitude of transition inside jet in the propagation basis.  The
probability averaged over the production region equals
$$
P_*(\nu_\alpha \rightarrow \nu_i) = \frac{1}{r_\nu}
\int_0^{r_\nu} dx \left|S_{i \alpha} \right|^2, 
$$
where 
$S_{i \alpha } \equiv  A_*(\nu_\alpha \rightarrow \nu_i)$. 
The matrix of total flavor-to-flavor probabilities equals 
\be
\hat{P}_{tot} = \hat{P}_* \hat{P}_E,   
\nonumber
\label{tot-m}
\ee
or if there is no Earth matter effect: 
\be
\hat{P}_{tot} = \hat{P}_* \cdot P_{PMNS}^T, 
\label{ptot}
\ee
where $P_{PMNS} \equiv ||~|(U_{PMNS})_{ ij}|^2||$ is the matrix of
moduli square of the mixing matrix elements in vacuum.

According to (\ref{star-totcp}) the amplitudes of transitions inside
the star is given by
$$
A_*(\nu_\alpha \rightarrow \nu_i) = 
(U_{23} \Gamma_\delta)_{\alpha \beta}
A_{jet} (\nu_\beta^\prime \rightarrow  \nu_\xi^\prime)
A_{env} (\nu_\xi^\prime \rightarrow \nu_i). 
$$
In matrix form we have  
\be
S_* = (U_{23} \Gamma_\delta) S_{jet} S_{env}.  
\label{total-s}
\ee
Here 
$$
(S_{jet})_{\xi \beta} \equiv A_{jet} (\nu_\beta^\prime 
\rightarrow  \nu_\xi^\prime) ~~{\rm and} ~~
(S_{env})_{i \xi} \equiv  A_{env} (\nu_\xi^\prime \rightarrow \nu_i)  
$$
are the evolution matrices inside jet and inside an envelope in the
propagation basis correspondingly.

For $E > 2\cdot 10^3$ GeV or/and smaller production region the phase
of oscillations inside jet is small and these oscillations can be
neglected.  Furthermore, as we will show in Sec.\ 4.5, the effect of
oscillations inside jet is zero or small and so in the first
approximation can be neglected.  Therefore we will first consider
oscillations inside the star ignoring oscillations inside jet. In this
case $S_{jet} = I$ and $ S_* \approx (U_{23} \Gamma_\delta) S_{env}$.

\subsection{Matter affected range }

In Fig.~\ref{prob-l} we show the conversion probabilities in different
channels for two different initial densities in the envelope.  There
are two key energies in the problem: the two resonance energies which
correspond to maximal electron density in the envelope $n_0$ (density
at the border between jet and envelope) \footnote{This is $2\nu-$
definition, in fact there is some small shift of the 1-2 resonance due
to 1-3 mixing.}:
$$ 
E_R^L \approx \frac{\Delta m^2_{21}}{ 2 V_0} \cos 2 \theta_{12}, ~~~~ 
E_R^H \approx \frac{\Delta m^2_{31}}{ 2 V_0} \cos 2 \theta_{13},
$$ 
where 
$
V_0 =  \sqrt{2} G_F n_0. 
$
For $n_0 = 10^{23}$ cm$^{-3}$ we find 
$$
E_R^L = 1.3~ {\rm GeV}, ~~~ E_R^H = 75~ {\rm GeV}. 
$$
These energies determine the borders of energy regions with 
different dynamics of flavor conversion (see Fig.~\ref{prob-l}) 
inside the star: 
\begin{enumerate}
\item 
$E < E_R^L$ - the vacuum oscillations (VO) region: VO dominate for
both 1-2 and 1-3 modes.
\item 
$E_R^L \lesssim E \lesssim E_R^H$ - the intermediate energy range, $(7
- 70)$ GeV.  The corresponding neutrinos are produced between the two
resonances.  The 1-3 mixing and split lead to averaged vacuum
oscillations (the mass eigenstate $\nu_3$ decouples from the rest of
system), whereas for 1-2 mixing and mass split the matter effects
dominate.
\item  
$E_R^H \lesssim E \lesssim E_{na}$, where $E_{na}$ is the energy of
strong adiabaticity breaking.  Neutrinos are produced above the 1-3
resonance in the density scale; here matter effects are important for
both mixings.  For the power dependence of the spectrum we have $
E_{na} \sim R_* \Delta m_{31}^2 /4\pi \sim 10^2 E_R^H$.
\item 
$E > E_{na}$ - matter suppresses oscillations inside the star, here
the flavor conversion is due to oscillations in vacuum from the
surface of a star to the earth.  Loss of coherence leads to the
averaged oscillation result.
\end{enumerate}

\FIGURE{\epsfig{file=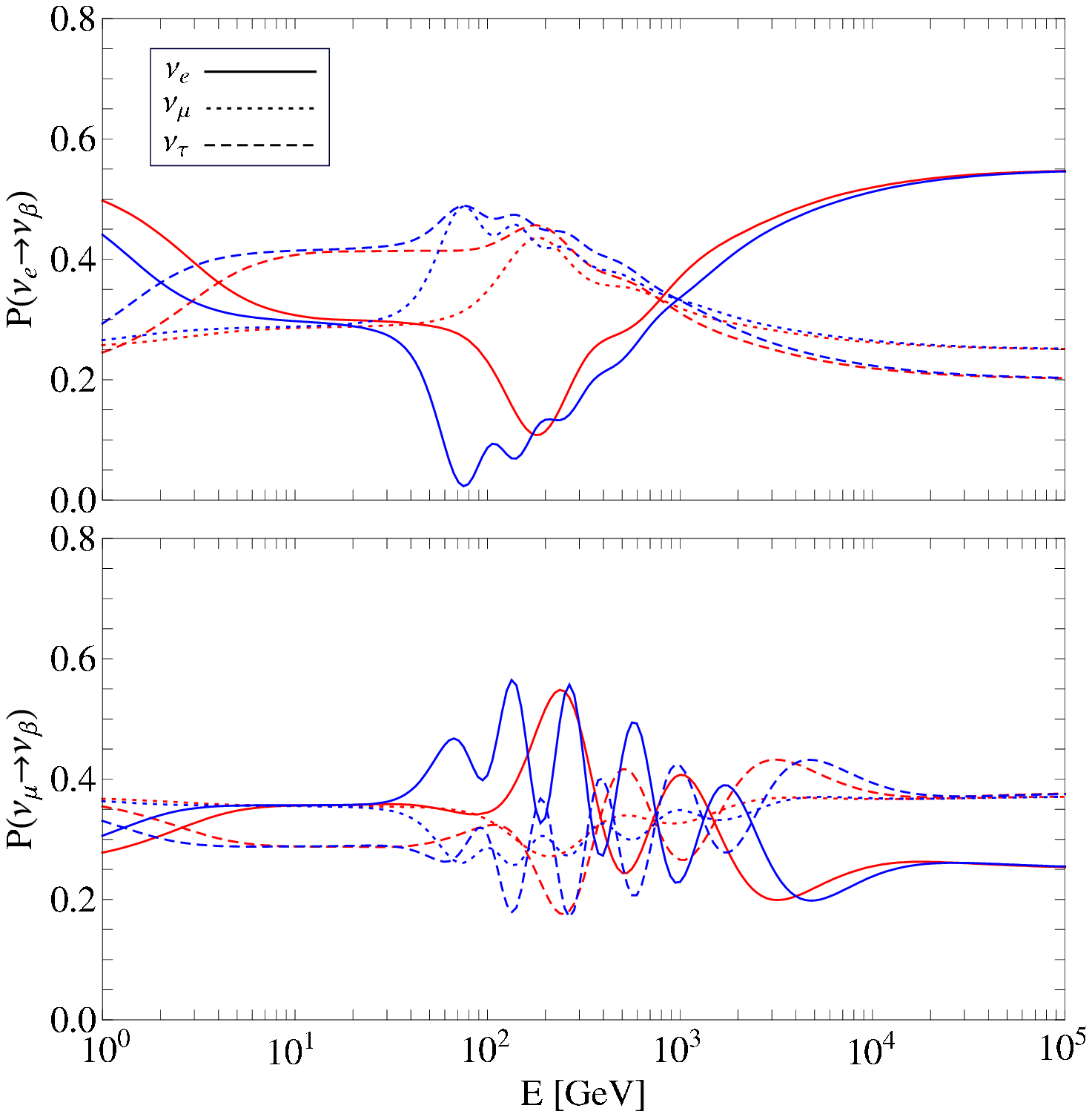,width=10cm}
\caption{Probabilities of the $\nu_e \to \nu_\beta$ (top panel) and
$\nu_\mu \to \nu_\beta$ (bottom panel) transitions for the density
profile A with two different inner densities $n_0 = 10^{23}$ cm$^{-3}$
(red lines) and $n_0 = 2\cdot 10^{23}$ cm$^{-3}$ (blue lines).  We
used $\sin^2 2\theta_{13} = 0.08$, $\sin^2 \theta_{23} = 0.5$
$\delta_{cp} =0$ and normal mass hierarchy. }
\label{prob-l}}

The matter affected range spans over 5 orders of magnitude from about
0.3 GeV to $3 \cdot 10^{4}$ GeV. With increase of $n_0$ the range
expands in both directions.~\footnote{Here we discuss whole energy
affected range and discuss relevant physics in spite of the fact that
the present detectors have rather high energy threshold $E \gtrsim
10^2$ GeV. Future experimental developments may, in principle,
substantially reduce this threshold.}  According to Fig.~\ref{prob-l}
generic features of the energy dependences of the probabilities can be
summarized as follows.
\begin{itemize}
\item 
Plateau in the intermediate range above the vacuum-to-matter
transition region at (0.3 - 3)~ GeV.
\item 
Peak or dip (depending on channel) at $E \gtrsim E_R^H$ due to the 1-3
resonance.
\item 
Wiggles at $E \gtrsim E_R^H$ in the nonadiabatic edge, above the peak
(dip).  As we will show the wiggles are due to interference induced by
adiabaticity breaking.
\item 
The wiggles are different in channels with initial $\nu_e$ and
$\nu_\mu$.  According to Fig.~\ref{prob-l} the wiggles are small in
 $\nu_e \rightarrow \nu_\beta$ channels (upper
panel), and large in $\nu_\mu \rightarrow \nu_\beta$ channels. 
\item  
The wiggles modulate the non-adiabatic edge of the energy profile of
the effect.
\end{itemize}

Note that the $\nu_e \to \nu_e$ probability as function of energy can
be viewed as two dips in the energy profile, one embedded into
another, which corresponds to crossing of two resonances: due to the
1-2 and 1-3 mixings.  If the 1-3 mixing is small or zero, the 1-3 dip
is absent and the plateau extends to higher energies.  With further
increase of energies the probabilities will converge without wiggles
to the asymptotic values as in the case of solar neutrinos.  In what
follows we will explain these features.

\subsection{Asymptotic values of probabilities}

Outside the matter affected region the probabilities converge to the
asymptotic values which are given by the averaged vacuum oscillation
probabilities.  Indeed, for $E \ll E_R^L$ the matter effects can be
neglected and we deal with vacuum oscillations inside the star,
(essentially the flavor state is projected onto mass eigenstates in
the production region and then these eigenstates propagate inside the
star without changes), so
\be 
P_* (\nu_\alpha \rightarrow \nu_i) = |U_{\alpha i}|^2 , 
\label{alphaprob}
\ee
and therefore Eq.~(\ref{ab-general}) becomes
\be
P(\nu_\alpha \rightarrow \nu_\beta) =
\sum_i |U_{\alpha i}|^2 |U_{\beta i}|^2. 
\label{abprob}
\ee
The same result (\ref{alphaprob}), (\ref{abprob}) are valid for very
high energies: $E \gg E_R^H$. In this region matter suppresses
oscillations inside the star.  In other terms, for very high energies
the adiabaticity is strongly broken and no flavor conversion occurs,
$P_* = I$.  Here we project $\nu_\alpha$ onto mass eigenstates at the
surface of the star, which is equivalent to the coherence lost.  Again
$P_* (\nu_\alpha \rightarrow \nu_i)$ equals (\ref{alphaprob}) and the
total probability is given by (\ref{abprob}).  The probabilities reach
the asymptotic at $E \sim 10^5$ GeV and converge to the same
asymptotic values below 1 GeV (in this sence the picture is symmetric)
as in Fig.~\ref{prob-l}.

Since for antineutrinos in vacuum $U_{\alpha i} \rightarrow U_{\alpha
i}^*$, the result (\ref{abprob}) is also valid for antineutrinos: $
P(\bar{\nu}_\alpha \rightarrow \bar{\nu}_\beta) = P(\nu_\alpha
\rightarrow \nu_\beta)$. There is no CP-asymmetry, although the
probabilities depend on the phase $\delta$.  Since the asymptotics
does not depend on $\Delta m^2$, for the inverted mass hierarchy we
have again the same result (\ref{abprob}).

\subsection{Adiabatic conversion}

At low energies ($E \lesssim E_R^H \sim 100$~GeV) the adiabaticity
condition is satisfied and we can use results in the adiabatic
approximation.  Let $\nu_m^0$ be the eigenstates of the Hamiltonian in
the initial moment (at the inner border of the envelope). Then the
flavor states can be represented in terms of $\nu_m^0$ as $ \nu_f =
U^{0}_m \nu_m^0$.  The adiabatic evolution means that the transitions
between the eigenstates can be neglected and therefore $\nu_m^0
\rightarrow \nu$, or explicitly
\be
\pmatrix{
\nu_{1m}^0 \cr
\nu_{2m}^0 \cr
\nu_{3m}^0  
}  
\rightarrow 
\pmatrix{
\nu_{1} \cr
\nu_{2} \cr
\nu_{3}
}.
\label{adtra}
\ee
Consequently, inside the star the flavor state $\nu_\alpha$ evolves as
$\nu_\alpha \rightarrow U^{0}_{\alpha i} \nu_i$, and therefore the
adiabatic probability is given by
$$
P_* (\nu_\alpha \rightarrow \nu_i) = |U^0_{\alpha i}|^2.  
$$
Then the flavor transition probability at the Earth equals 
$$ 
P(\nu_\alpha \rightarrow \nu_\beta) = |U^0_{\alpha i}|^2 
|U_{\beta i}|^2. 
$$

Let us describe the adiabatic transitions in the propagation basis.
Now $ \nu^{\prime}_f = U^{\prime 0}_m \nu_m^0, $ where $U^{\prime
0}_m$ is the mixing matrix in matter in the intial moment (at the
density $n_0$).  According to Eq.~(\ref{propbasis}), the flavor states
can be written as $ \nu_f = (U_{23} \Gamma_\delta) S_{jet} U^{\prime
0}_m \nu_m^0.  $ Therefore according to (\ref{adtra}) $ S \nu_f =
(U_{23} \Gamma_\delta) S_{jet} U^{\prime 0}_m \nu, $ and
$$
S_{*} =  \langle \nu| S | \nu_f \rangle  =  (U_{23} \Gamma_\delta) 
S_{jet} U^{\prime 0}_m. 
$$
Neglecting oscillations inside  jet we have $S_{jet} = I$ and 
\be
S_{*} =  \langle \nu | S |\nu_f \rangle  =  
(U_{23} \Gamma_\delta) U^{\prime 0}_m.
\label{withoutj}
\ee
The expression in (\ref{withoutj}) coincides with the mixing matrix in
matter in the initial moment: $U^0_{\alpha i} \approx (U_{23}
\Gamma_\delta) U^{\prime 0}_m$.

The adiabaticity condition can be satisfied at low energies, in
particular, in the intermediate region.  This region corresponds to
neutrino production above the 1-2 resonance density.  Far above the
1-2 resonance
$\theta_{12}^m \approx \pi/2$.  Therefore using Eq.~(\ref{uprimem}) we
obtain the mixing matrix in the initial state
\be
U^{\prime 0}_m  \approx \pmatrix{
0    &  c_{13}^0      &  s_{13}^0 \cr
- 1  &  0             &  0 \cr
0    &   - s_{13}^0   & c_{13}^0
},
\label{uprimeint}
\ee
where $c_{13}^0$ and $s_{13}^0$ are the mixing parameters in the
initial moment.  Inserting this matrix into Eq.~(\ref{withoutj}) we
obtain the matrix of amplitudes
\be
S_*  \approx U_{\alpha i}^0 = \pmatrix{
0    &  c_{13}^0      &  s_{13}^0 \cr
- c_{23}  &  -s_{13}^0 s_{23} e^{i\delta}  &  c_{13}^0 s_{23} e^{i\delta} \cr
+ s_{23}  &  -s_{13}^0 c_{23} e^{i\delta}  & - c_{13}^0 c_{23} e^{i\delta}
}.
\label{adampl}
\ee
Consequently,  the matrix of probabilities inside the star equals 
\be
\hat{P}_*  \approx \pmatrix{
0         &  c_{13}^{0~2}           &  s_{13}^{0~2} \cr
c_{23}^2  &  s_{13}^{0~2} s_{23}^2  &  c_{13}^{0~2} s_{23}^2 \cr
s_{23}^2  &  s_{13}^{0~2} c_{23}^2  &  c_{13}^{0~2} c_{23}^2
},
\label{adprob}
\ee
where the rows correspond to the initial flavor states $\nu_e,
\nu_\mu, \nu_\tau$ and the columns correspond to the final mass
states. According to (\ref{adprob}), the probabilities do not depend
on $\delta$ and this is the consequence of the adiabaic propagation at
the energies above the 1-2 resonance. So, any dependence on $\delta$
is manifestation of the adiabaticity violation or/and closeness to the
1-2 resonance.  The flavor transition probabilities at the detector
are then given by Eq.~(\ref{ptot}).  Some dependence on $\delta$
follows from dependence of the projections of the mass states to
flavor states back at the detection, i.e., from factors $|U_{PMNS}|^2$
as we will discuss in Sec.\ 4.6.

Let us consider some particular cases of Eq.~(\ref{adprob}).

\noindent
1. In the intermediate range (plateau) we have $\theta_{13}^0 \approx
\theta_{13}$ and the probability matrix becomes
\be
\hat{P}_*  \approx \pmatrix{
0         &  c_{13}^{2}           &  s_{13}^{2} \cr
c_{23}^2  &  s_{13}^{2} s_{23}^2  &  c_{13}^{2} s_{23}^2 \cr
s_{23}^2  &  s_{13}^{2} c_{23}^2  &  c_{13}^{2} c_{23}^2
}.
\label{adprobb}
\ee

\noindent
2. Above the H-resonance, $E \gtrsim E_R^H$, we have $\theta_{13}^0
\approx \pi/2$, so that $c_{13}^{0} \approx 0$ and $s_{13}^{0} \approx
1$, and consequently,
$$
U^{\prime 0}_m  \approx \pmatrix{
0    &  0      &  1 \cr
- 1  &  0             &  0 \cr
0    &   - 1    &   0
}.
$$
For large enough $\theta_{13}$ the adiabaticity is fulfilled, and
using the adiabatic result (\ref{withoutj}) we obtain
$$
S_*  \approx \pmatrix{
0         &  0                      &  1 \cr
- c_{23}  &  -  s_{23} e^{i\delta}  &  0 \cr
  s_{23}  &  - c_{23} e^{i\delta}   &  0 
}
$$
which leads to 
\be
\hat{P}_*  \approx \pmatrix{
0         &  0         &  1  \cr
c_{23}^2  &  s_{23}^2  &  0  \cr
s_{23}^2  &  c_{23}^2  &  0
}.
\label{adprobb3}
\ee
Then, according to general formula (\ref{ptot}) the flavor probability
matrix equals
\be
\hat{P}  \approx \pmatrix{
|U_{e3}|^2  & |U_{\mu 3}|^2   &   |U_{\tau 3}|^2 \cr
c_{23}^2|U_{e1}|^2 + s_{23}^2|U_{e2}|^2  & 
c_{23}^2|U_{\mu 1}|^2 +  s_{23}^2|U_{\mu 2}|^2 &  
c_{23}^2|U_{\tau 1}|^2 + s_{23}^2|U_{\tau 2}|^2 \cr
s_{23}^2|U_{e1}|^2 + c_{23}^2|U_{e2}|^2  &  
s_{23}^2|U_{\mu 1}|^2 + c_{23}^2|U_{\mu 2}|^2   &  
s_{23}^2|U_{\tau 1}|^2 + c_{23}^2|U_{\tau 2}|^2
}.
\label{adprobfl}
\ee

\noindent
3. For antineutrinos, the 1-2 mixing is suppressed at $E \gtrsim
E_R^L$, so that $\theta_{12}^0 \approx 0$.  Therefore, according to
(\ref{uprimem}) we obtain
$$
\bar{U}^{\prime 0}_m  \approx \pmatrix{
c_{13}^0  &  0      &  s_{13}^0 \cr
0         &  1      &  0 \cr
- s_{13}^0    &  0   & c_{13}^0
}.
$$
The matrix of amplitudes in the flavor basis equals  
$
\bar{S}_{*} = 
(U_{23} \Gamma_{-\delta}) \bar{U}^{\prime 0}_m , 
$
or explicitly 
$$
\bar{S}_*  \approx \pmatrix{
 c_{13}^0 & 0      &  s_{13}^0 \cr
-s_{13}^0 s_{23} e^{- i\delta} & c_{23}  & c_{13}^0 s_{23} e^{-i\delta} \cr
-s_{13}^0 c_{23} e^{-i\delta}  & s_{23}  & c_{13}^0 c_{23} e^{-i\delta}
}.
$$
Consequently,  the matrix of probabilities inside the star is  
\be
\hat{P}_*  \approx \pmatrix{
c_{13}^{0~2} & 0           &  s_{13}^{0~2} \cr
s_{13}^{0~2} c_{23}^2  &  c_{23}^2  &  c_{13}^{0~2} s_{23}^2 \cr
s_{13}^{0~2} c_{23}^2 &   s_{23}^2  &  c_{13}^{0~2} c_{23}^2
}.
\label{adprobun}
\ee

\noindent
4. For the inverted mass hierarchy (IH) above the 1-2 resonance we
have the same initial mixing matrix as for the normal mass hierarchy
(NH) given in Eq.  (\ref{uprimeint}), the same S-matrix (\ref{adampl})
and the same matrix of probabilities (\ref{adprob}).

In the intermediate energy range (below 1-3 resonance) the 1-3 mixing
equals approximately the vacuum mixing as in NH case and therefore the
probability matrix is as in Eq.~(\ref{adprobb}). The difference
appears at higher energies, since there the matter effect on 1-3
mixing is different for the normal and inverted hierarchies.  In
particular, above the 1-3 resonance we have in the case of IH:
$\theta_{13}^0 \approx 0$, $c_{13}^0 \approx 1$, $s_{13}^0 \approx 0$
and
\be
\hat{P}_*  \approx \pmatrix{
0         &  1  &  0 \cr
c_{23}^2  &  0  &  s_{23}^2 \cr
s_{23}^2  &  0  &  c_{23}^2
}.
\label{adprobih}
\noindent
\ee
Then the matrix of flavor transition probabilities in case of IH
becomes
$$
\hat{P}  \approx \pmatrix{
|U_{e2}|^2  & |U_{\mu 2}|^2   &   |U_{\tau 2}|^2 \cr
c_{23}^2|U_{e1}|^2 + s_{23}^2|U_{e3}|^2  &
c_{23}^2|U_{\mu 1}|^2 +  s_{23}^2|U_{\mu 3}|^2 &
c_{23}^2|U_{\tau 1}|^2 + s_{23}^2|U_{\tau 3}|^2 \cr
s_{23}^2|U_{e1}|^2 + c_{23}^2|U_{e 3}|^2  &
s_{23}^2|U_{\mu 1}|^2 + c_{23}^2|U_{\mu 3}|^2     &
s_{23}^2|U_{\tau 1}|^2 + c_{23}^2|U_{\tau 3}|^2
},
$$
as compared to the NH case (\ref{adprobfl}). 

For antineutrinos, the situation is similar: in the intermediate
region the results coincide with those for NH, whereas above the
H-resonance energy the results change.  Now this is the resonance
channel and $c_{13}^0 \approx 0$, $s_{13}^0 \approx 1$.  Consequently,
the matrix of probabilities of the flavor to mass transitions becomes:
\be
\hat{P}_*  \approx \pmatrix{
0         &  0         &  1  \cr
s_{23}^2  &  c_{23}^2  &  0  \cr
c_{23}^2  &  s_{23}^2  &  0
}.
\label{adprobbih}
\ee

\subsection{Adiabaticity breaking} 

At high energies, not far above the H-resonance for the profiles A and
B, the adiabaticity is broken and in contrast to Eq.~(\ref{adtra}) the
transitions between the eigenstates of propagation in matter occur.
The adiabaticity is broken mainly in resonance regions, so that in
resonances jumps from one eigenstate to another occur.

We will consider evolution in the so-called factorization
approximation when two level crossings (1-3 and 1-2) occur
independently one after another and do not influence each other (see
Fig.~\ref{level_crossing}).  Probability of crossing of two resonances
equals the product probabilities of crossings of individual
resonances.

The ``jump'' transition probability in each crossing is well described
by the double exponent formula~\cite{petcov}
\be
P_{jump} = \frac{e^{2\pi \lambda_n 
\omega (1 - \sin^2 \theta) } - 1}{e^{2\pi \lambda_n \omega} - 1}~,  
\label{petcov}
\ee 
where $\omega \equiv \Delta m^2 /2E$ and $\lambda_n \equiv n/ (dn/dr)$
is the scale of the electron density change.  For 1-3 level crossing
the jump probability $P_H = P_{jump}(\Delta m^2_{31}, \theta_{13})$
and for the 1-2 level crossing the jump probability $P_L(\Delta
m^2_{21}, \theta_{12})$ (see Fig.~\ref{level_crossing}).  In
Fig.~\ref{jump} we show the jump probabilities computed for different
neutrino parameters.

Let us consider $\nu_e$ production with energies substantially larger
than the resonance energy, so that in initial state $\nu_e \approx
\nu_{3m}$.  In factorization approximation we have
\be
P_* (\nu_e \rightarrow \nu_1) = P_{H} P_{L}, ~~~
P_* (\nu_e \rightarrow \nu_2) = P_{H}(1 - P_{L}) , ~~~
P_* (\nu_e \rightarrow \nu_3) = 1 - P_{H}. 
\label{stare1}
\ee
Then the ($\nu_e \rightarrow \nu_e$) probability equals 
$$
P(\nu_e \rightarrow \nu_e)=  
(1- P_{H})|U_{e3}|^2 +  
P_{H}(1 - P_{L}) |U_{e2}|^2 + 
P_{H} P_{L} |U_{e1}|^2. 
$$
It can be rewritten as 
$$
P(\nu_e \rightarrow \nu_e)= |U_{e3}|^2 +
P_{H}(|U_{e2}|^2 - |U_{e3}|^2)  +
P_{H} P_{L} (|U_{e1}|^2 - |U_{e2}|^2).
$$
For other channels we have 
\be
P(\nu_e \rightarrow \nu_\mu) = 
(1- P_{H})|U_{\mu 3}|^2 +
P_{H}(1 - P_{L}) |U_{ \mu 2}|^2 +
P_{H} P_{L} |U_{\mu 1}|^2, 
\label{prob-nonadmu}
\ee
and similarly for transition to $\nu_{\tau}$. 

The jump probabilities have the following asymptotics for $E
\rightarrow \infty$:
\be
P_{H} \rightarrow 1 - |U_{e3}|^2, ~~~ 
P_{L} \rightarrow  \frac{|U_{e1}|^2}{1 - |U_{e3}|^2}~. 
\label{asympt}
\ee
Indeed, the limit $E \rightarrow \infty$ corresponds to absense of
flavor transformation, i.e., $\nu_e \rightarrow \nu_e$. If the initial
state is $\nu_e = \nu_{3m}$, the result for $P_{H}$ is obtained in the
$2\nu-$ approximation neglecting matter effect on the 1-3 mixing in
the intermediate point between the two resonances.  Then the
expression for $P_{L}$ can be found from the condition that the
$3\nu-$ probability $P_* (\nu_e \rightarrow \nu_1) = |U_{e1}|^2$
(since $\nu_e \rightarrow \nu_e$), and therefore according to
(\ref{stare1}): $P_L = |U_{e1}|^2/P_H$. Finally, using asymptotic
expression for $P_H$ we arrive at the result shown in
Eq.~(\ref{asympt}).

It is interesting to note that the adiabaticity is broken quite
similarly in the same energy range for both resonances. This feature
is the consequence of the fact that for $\sin^2 2\theta_{13} \sim
0.08$: $\sin^2 2\theta_{13} \Delta m^2_{31} \sim \sin^2 2\theta_{12}
\Delta m^2_{21}$.

Due to breaking of adiabaticity new interference effects emerge which
lead to oscillatory behavior (``wiggles'') of the conversion
probabilities (see Fig.~\ref{prob-l}). There are two types of the
interference effects which are related to the adiabaticity violation
in the H-resonance and L-resonance. We will call the results of these
interferences the H-wiggles and the L-wiggles correspondingly.  In
general both L- and H- wiggles are present in all the channels
simultaneously, however, as we will show in a given channel one type
of wiggles dominates.

\FIGURE{\epsfig{file=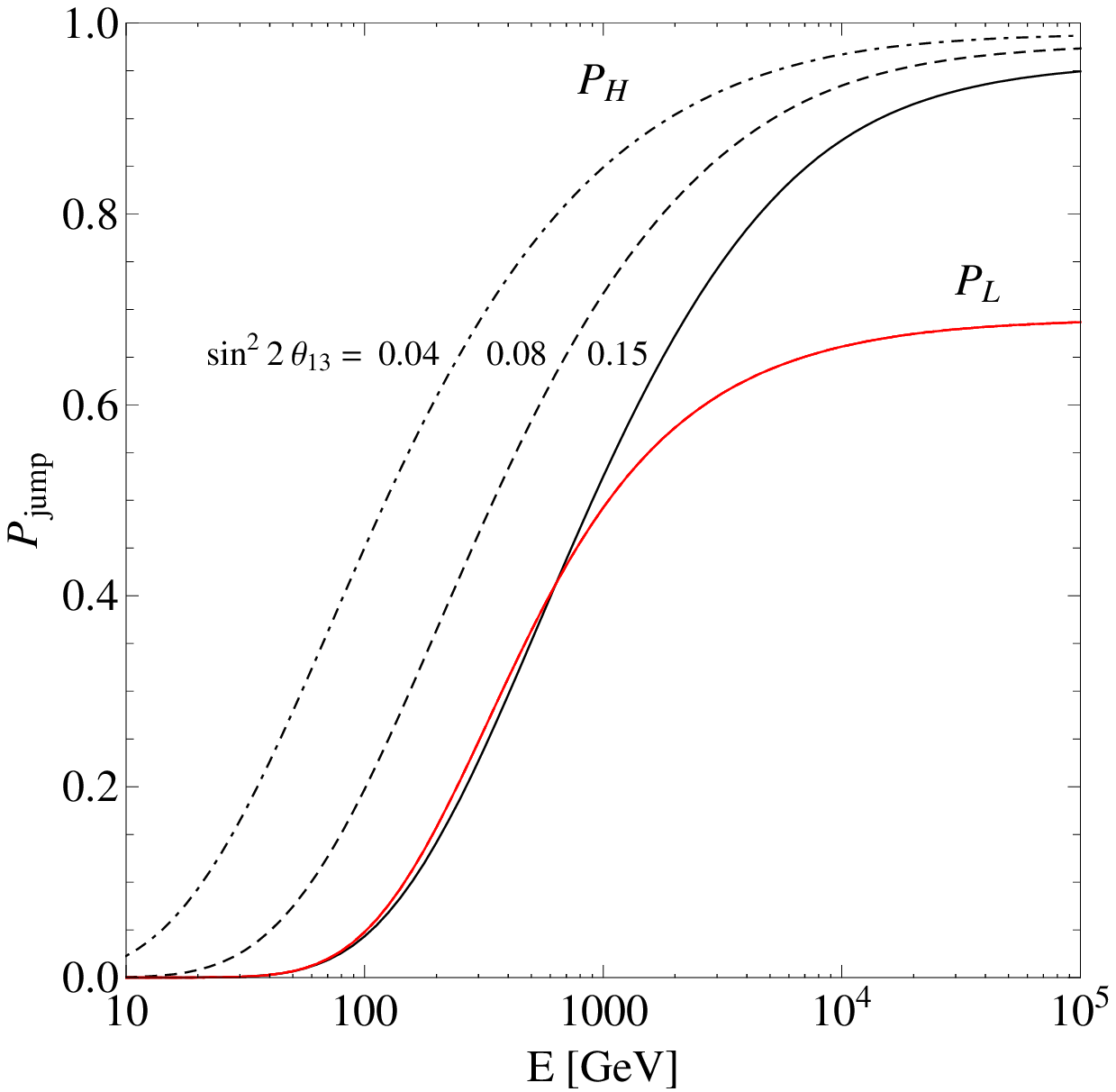,width=9.cm}
\caption{The jump probabilities in the H-resonance $P_H$ (black lines)
and in the L-resonance $P_L$ (red line) as functions of neutrino
energy.  $P_H$ is shown for three different values of $\sin^2
\theta_{13}$.  We use the profile A and neutrino parameters: $\Delta
m^2_{21} = 8\cdot 10^{-5}$ eV$^2$, $\Delta m^2_{31} = 2.4 \cdot
10^{-3}$ eV$^2$, $\sin^2 2 \theta_{12} = 0.86$.  }
\label{jump}}

\subsection{H-wiggles}

The H-wiggles are related to mixing of the eigenstates $\nu_{2m}$ and
$\nu_{3m}$ in a given flavor state and to the adiabaticity violation
in the H-resonance.  Above the H-resonance substantial admixture of
$\nu_{2m}$ and $\nu_{3m}$ appear in the $\nu_e-$state, and therefore
the H-wiggles dominate in $\nu_e$ channels.

Let us neglect 1-2 mixing and consider transition $\nu_e \rightarrow
\nu_{2m}$ in the region of H-resonance, i.e. compute the probability
to find $\nu_{2m}$ at densities below the H-resonance but above the
L-resonance.  The problem is reduced to $2\nu$-problem according to
the factorization approximation.  In the H-resonance there are two
channes of transition $\nu_e \rightarrow \nu_{2m}$:
$$
\nu_e  \rightarrow  \nu_{2m} \rightarrow \nu_{2m},~~~~~
\nu_e  \rightarrow  \nu_{3m} \rightarrow \nu_{2m}, 
$$
where the first arrow denotes projection of $\nu_e$ state on the
corresponding eigenstate.  It is the interference of the corresponding
amplitudes that leads to the H-wiggles.

Let us quantify the interference picture.  At the production point
\be
\nu_e \approx \cos \theta_{13}^0 \, \nu_{2m} + \sin \theta_{13}^0 \, 
\nu_{3m}, 
\label{eq:nueh}
\ee
where $\theta_{13}^0$ is the mixing angle in matter at the production
point:
$$
\sin^2 2\theta_{13}^0 \approx 
\frac{\sin^2 2\theta_{13}}{\cos^2 2\theta_{13}
(1 - E/E_R^H)^2 + \sin^2 2\theta_{13}}. 
$$
Above the resonance, $E > E_R^H$, the angle $\theta_{13}^0 \rightarrow
\pi/2$ and the admixture of $\nu_{2m}$, given by $\cos \theta_{13}^0$,
decreases.
According to Eq.~(\ref{eq:nueh}), below the H-resonance the total
amplitude of probability to find $\nu_{2m}$ is given by
$$
A(\nu_e \rightarrow \nu_{2m}) = 
\cos \theta_{13}^0 A_{22} + \sin \theta_{13}^0 A_{32},
$$
where $A_{32}$ is the amplitude of $\nu_{3m} \rightarrow \nu_{2m}$
transition in the H-resonance, $|A_{32}| = \sqrt{P_H}$, and $|A_{22}|
= \sqrt{1 - P_H}$ is the amplitude of survival probability.  Then the
probability to find $\nu_{2m}$ equals
\be
P(\nu_e \rightarrow \nu_{2m})  =
\cos^2 \theta_{13}^0 (1 - P_H)
+ \sin^2 \theta_{13}^0 P_H
- \sin 2\theta_{13}^0 \sqrt{P_H (1 - P_H)} \cos \phi_H,
\label{eq:prob}
\ee
where
$$
\phi_H \equiv Arg(A_{22} A_{32}^*).
$$
The last term in Eq. (\ref{eq:prob}) is the interference term of two
amplitudes which is responsible for the wiggles.  The amplitude of
wiggles is given by
$$
W = \sin 2 \theta_{13}^0 \sqrt{P_H (1 - P_H)}.
$$
Using the formula (\ref{petcov}) for $P_H$ and for $\sin^2 \theta_{13}
= 0.02$ we find the following values: $W = (0.05,~ 0.035,~ 0.025,~
0.013)$, for $E = (200,~ 300,~ 500,~ 1000)$~GeV respectively; in
agreement with results of the numerical computations in
Fig.~\ref{prob-l}.

Let us estimate the phase $\phi_H$. 
Note that transition between the eigenstates $\nu_{3m} \rightarrow
\nu_{2m}$ is driven by the derivative $\dot{\theta}_{13}^m$ and occurs
mainly in the resonance region $r_R \pm \Delta r_R$. Here $r_R = r_R
(E)$ is determined from the resonance condition, $\sqrt{2} G_F
n_e(r_R) = \Delta m^2_{31}/(2E)$, and the half-size of the resonance
region is given by
$$
\Delta r_R =  \lambda_n \tan 2 \theta_{13} = 
\frac{n}{d n/dr} \tan 2 \theta_{13}.  
$$ 
For the power dependence of the density on distance,   
$$
n = n_0 \left(\frac{r}{R_0}\right)^{-k}, 
$$
the scale factor equals $\lambda_n = r/k$. Therefore $ \Delta r_R =
r_R \tan 2 \theta_{13}/k $.

The whole evolution of the phase $\phi$ can be devided into three
parts: (i) from $R_0$ to $r_R - \Delta r_a$, where the system evolves
adiabatically and transitions $\nu_{3m} \leftrightarrow \nu_{2m}$ can
be neglected; (ii) from $r_R - \Delta r_a$ to $r_R + \Delta r_a$,
where the transitions $\nu_{3m} \leftrightarrow \nu_{2m}$ occur; and
(iii) from $r_R + \Delta r_a$ to $R_*$, where the adiabaticity is
restored and states $\nu_{3m}$, $\nu_{2m}$ evolve independently again.
For moderate adiabaticity breaking $\Delta r_a \lesssim \Delta r_R$;
for high energies the adiabaticity is broken even beyond the resonance
layer.
 
The total amplitudes $A_{22}$ and $A_{32}$ introduced above can be
written as $A_{22} = A_{22}^{iii} A_{22}^{ii} A_{22}^{i}$ and $A_{32}
= A_{22}^{iii} A_{32}^{ii} A_{33}^{i}$ since $A_{32} \neq 0$ only in
the region (ii).  Note that the amplitudes in the region (ii) and
consequently phases coincide for both $A_{22}$ and $A_{32}$.
Correspondingly, the total phases $\phi_2 \equiv arg(A_{22})$ and
$\phi_3 \equiv arg(A_{32})$ equal
\be
\phi_2  =  \phi_2^{(i)} + \phi_2^{(ii)} + \phi_2^{(iii)},~~~~  
\phi_3 =  \phi_3^{(i)} + \phi_3^{(ii)} + \phi_2^{(iii)},
\label{phi23}
\ee
where $\phi_j^{(i)}$ and $\phi_j^{(iii)}$ ($j = 2,3$) are the
adiabatic phases in the regions (i) and (iii) correspondingly, whereas
$\phi_2^{(ii)} \equiv arg (A_{22}^{(ii)})$ and $\phi_3^{(ii)} \equiv
arg (A_{32}^{(ii)})$ are non-adiabatic phases.  Note that the last
terms in Eqs.~(\ref{phi23}) (contributions from the region (iii)) are
identical.
Thus, the phase difference we are looking for equals: 
$$
\phi_H = \phi_3 - \phi_2 = 
\Delta \phi^{(i)} - Arg (A_{22}^{(ii)} A_{23}^{(ii)*}), 
$$
where 
$
\Delta \phi^{(i)} \equiv \phi_3^{(i)} - \phi_2^{(i)} 
$
is the adiabatic phase difference  acquired in the region (i): 
\be
\Delta \phi^{(i)} \approx \int_{R_0}^{r_R(E) - \Delta r_a(E)} dx 
\sqrt{\left[V(x) - \cos 2 \theta_{13} 
\frac{\Delta m^2 }{2E} \right]^2 + 
\sin^2 2\theta_{13} \left(\frac{\Delta m^2 }{2E}\right)^2}~. 
\label{ad1}
\ee 
For power-law density profile the upper limit of integration equals
$r_R(E)(1 - \tan 2 \theta_{13}/k)$.

Few comments are in order. The phase $\phi_H$ does not depend on
evolution above (outside) the resonance (region (iii)) and therefore
it does not depend on $R_\star$.  The phase depends on $r_{j}$ via the
adiabatic phase contribution. The non-adiabatic contribution $Arg
[A_{22}^R A_{23}^{R*}] < 2 \pi$ (or much smaller) since the
adiabaticity violation corresponds to the width of the resonance layer
to be smaller than the oscillation length.  To a good approximation
the phase in region (ii) can be taken into account by extending the
integral in Eq.~(\ref{ad1}) to this region, that is, taking $r_R(E) +
\Delta r_a(E)$ as the upper limit of integration.

If $V \gg \Delta m^2/(2E)$, the adiabatic phase can be estimated in
the following way:
\be
\Delta \phi^{(i)} \approx \int_{R_0}^{r_R - \Delta r_R} V(x) dx - 
\frac{\Delta m^2}{2E}\cos 2 \theta_{13} 
\left[r_R - \Delta r_R - R_0\right],
\label{adphase}
\ee
where we have taken $\Delta r_a = \Delta r_R$.  This allows us also to
estimate the period of wiggles in the energy scale.  The first term in
(\ref{adphase}) only weakly depends on energy.  This is because $V(x)$
is a steeply decreasing function of $x$ (the power $k \approx 3$) and
the integral is given mainly by the lower limit $V(R_0)$, which is
independent of $E$.  The upper limit depends on $E$ but its
contribution to $\Delta\phi^{(i)}$ is smaller than the second term in
(\ref{adphase}) and therefore will not change the period of wiggles
substantially.  Hence it is enough to study energy dependence of the
second term which can be rewritten as
\be 
\phi_{ad}(E) \approx \frac{2\pi}{l_\nu} 
\left[r_R (E) - \Delta r_R (E) - R_0 \right].
\label{adphaseH}
\ee 
This dependence on energy agrees well with periods of wiggles obtained
from exact numerical computations (Fig.~\ref{prob-l}).  The key point
that the phase should be computed not down to $R_\star$ (which would
introduce much faster oscillations) but down to $\sim r_R + \Delta
r_a$.

The H-wiggles have not been observed in the probabilities for other
objects.  In the case of SMA solution of the solar neutrino problem,
due to smallness of the vacuum mixing angle the wiggles are so small
that they are simply unobservable.  For the LOW solution the mixing at
the production point is very strongly suppresed.  For supernova
neutrinos, again the amplitude of the wiggles is very strongly
suppresses and, moreover, the period is so small that the wiggles are
averaged out.

\subsection{L-wiggles}

The L-wiggles are the consequence of mixing of $\nu_{1m}$ and
$\nu_{2m}$ in a given flavor state $\nu_\alpha$ and the interference
of the transition amplitudes
\be
\nu_\alpha  \rightarrow  \nu_{1m} \rightarrow \nu_i ~~{\rm and} ~~
\nu_\alpha  \rightarrow  \nu_{2m} \rightarrow \nu_i, 
\label{int-L}
\ee
where $i = 1, 2$. For energies above the H-resonance the states
$\nu_{1m}$ and $\nu_{2m}$ are mixed mainly in $\nu_\mu$ and
$\nu_\tau$, and their admixture in $\nu_e$ is negligible. Therefore
the L-wiggles are realized in channels with initial states of
$\nu_\mu$ and $\nu_\tau$.  (see Fig.~\ref{prob-l}, upper panel).  The
transitions (\ref{int-L}) appear due to adiabaticity violation in the
L-resonance.

For definiteness we will describe the wiggles in the $\nu_\mu$
channels in the case of normal mass hierarchy.  According to
(\ref{adampl}) in the production point $\nu_\mu$ has the following
matter eigenstate content
\be
\nu_\mu = - c_{23} \nu_{1m}  - s_{13}^0 s_{23} e^{i\delta} \nu_{2m}  
+ c_{13}^0 s_{23} e^{i\delta}\nu_{3m}.  
\label{content}
\ee

Note that below the H-resonance energy the admixture of $\nu_{2m}$ is
suppressed by smallness of $s_{13}^0 \approx s_{13}$. In contrast, for
$E \gtrsim E_R^H$ we have $s_{13}^0 \approx 1$ and according to
(\ref{content})
$$
\nu_\mu \approx - c_{23} \nu_{1m} - s_{23} e^{i\delta} \nu_{2m}. 
$$
Consider evolution of this state.  Its $\nu_{1m}$ and $\nu_{2m}$
components undergo various transformations with the following
amplitudes in factorization approximation:
\bea 
|A(\nu_{1m} \rightarrow \nu_2)| & = & \sqrt{P_L}~, ~~~~~~~~
|A(\nu_{2m} \rightarrow  \nu_{2m} \rightarrow \nu_2)| 
=  \sqrt{(1 - P_H)(1 - P_L)}~, 
\nonumber\\
|A(\nu_{1m}  \rightarrow \nu_1)| & = & \sqrt{1 - P_L}~, ~~~
|A(\nu_{2m}  \rightarrow  \nu_{2m} \rightarrow \nu_1)|   
= \sqrt{(1 - P_H) P_L}~. 
\nonumber
\eea
Furthermore, at the resonance crossing $A(\nu_{1m} \rightarrow
\nu_{2m}) = - A(\nu_{2m} \rightarrow \nu_{1m})$.  (For completeness:
$|A(\nu_{2m} \rightarrow \nu_{3})| = \sqrt{P_H}$, which is irrelevant
for this interference.)  Note that $\nu_{1m}$ crosses only
L-resonance, whereas $\nu_{2m}$ crosses both resonances which is
indicated by two arrows.  Summing up the amplitudes from different
channels of transitions we obtain
\bea
P_* (\nu_\mu \rightarrow \nu_1) & = &   
\left|c_{23} \sqrt{1 - P_L} 
- e^{-i\phi_L}  s_{23} e^{i\delta} \sqrt{(1 - 
P_H)P_L}\right|^2,  
\nonumber\\
P_* (\nu_\mu \rightarrow \nu_2) & = & 
\left|c_{23} \sqrt{P_L}
+ e^{-i\phi_\mu}  s_{23} e^{i\delta} \sqrt{(1 - P_H)(1 - 
P_L)}\right|^2,  
\nonumber \\
P_* (\nu_\mu \rightarrow \nu_3) & = & \sin^2 \theta_{23} P_H ~.
\label{starproav}
\eea
Here $\phi_L$ is the phase difference of the amplitudes collected from
the production point to the end of L-resonance region where the
adiabaticity is restored.  To a good approximation one can use for
$\phi_L$ the adiabatic phase difference:
\be 
\phi_L \approx \int_{R_0}^{r_R^L + \Delta r_R^L} dx~ (H_{2m} - H_{1m}). 
\label{phase-mu}
\ee

The probabilities (\ref{starproav}) can be rewritten as 
\bea
P_* (\nu_\mu \rightarrow \nu_1) & = &
c^2_{23} (1 - P_L)
+ s^2_{23} (1 - P_H) P_L - I_\mu,  
\nonumber\\
P_* (\nu_\mu \rightarrow \nu_2) & = &
c^2_{23} P_L + 
s^2_{23} (1 - P_H)(1 - P_L) + I_\mu,
\nonumber\\
P_* (\nu_\mu \rightarrow \nu_3) & = & s^2_{23} P_H,
\label{starproav1}
\eea
where the interference term equals 
\be
I_\mu \equiv \sin 2\theta_{23} \cos (\phi_L  + \delta) 
\sqrt{(1 - P_H) P_L (1 - P_L)}~. 
\label{int-term}
\ee
Note that the CP-violation phase enters together with the oscillation
(``strong'' phase) and to disentangle the former one needs to know
$\phi_L$.  Let us first summarize properties of the interference term.

\noindent
1. The interference term is not suppressed by small mixing in contrast
to the H-wiggles considered in the previous section.  Here the
admixture is determined by the vacuum mixing agle $\theta_{23}$ which
is close to maximal.  In fact, in the energy range where $P_L \sim
1/2$, this term can have the amplitude of the order 1.

\noindent
2. The interference term appears in the transitions to $\nu_1$ and
$\nu_2$ and not to $\nu_3$. It has an opposite sign in $\nu_\mu
\rightarrow \nu_1$ and $\nu_\mu \rightarrow \nu_2$ probabilities (as
it should be according to unitarity).

\noindent
3. The interference term vanishes if $P_H = 1$ which, in turn, is
realized for $\theta_{13} \rightarrow 0$. Non-zero 1-3 mixing is a
necessary condition for its appearance.  The amplitude is maximal if
the transition in H-resonance is adiabatic. With increase of energy
the adiabaticity in H-resonance is broken, $P_H \rightarrow 1$, and
this suppresses the H-wiggles.

\noindent
4. A necessary condition for appearance of wiggles are $P_L \neq 0$
and $P_L \neq 1$, that is, the adiabaticity should be broken in the
L-resonance but it should not be broken very strongly.

\noindent
5. The energy region where the L-wiggles are realized (more precisely,
the lower border of this region) is determined by the following two
conditions:
\begin{itemize}
\item 
the mixing of the eigenstates $\nu_{1m}$ and $\nu_{2m}$ should be
large enough which happens in the region of the H-resonance and above
it;
\item 
the adiabaticity in the L-resonance should be broken.  For the
density profiles we are discussing the adiabaticity starts to be
broken in the energy range of H-resonance and above it (which is to
some extent accidental).  As a result, the L-wiggles appear in the
same energy region as the H-wiggles.
\end{itemize}

The flavor probabilities can be obtained plugging expressions
(\ref{starproav1}) in (\ref{ab-general}). In particular,
\bea
P({\nu}_\mu \rightarrow {\nu}_e) & = & 
\cos^2 \theta_{23}\left[ |U_{e1}|^2 - 
P_L (|U_{e1}|^2 -  |U_{e2}|^2) \right] + 
\nonumber\\
& \,  &  \sin^2 \theta_{23} (1 - P_H) 
\left[ |U_{e2}|^2 + P_L (|U_{e1}|^2 - |U_{e2}|^2) \right]  
+  (|U_{e2}|^2 -  |U_{e1}|^2) I_\mu ~. 
\nonumber
\eea

Let us consider separately the interference terms in the flavor
probabilities. According to (\ref{starproav1})
\bea
P(\nu_\mu \rightarrow \nu_e)_{int} & = & 
(|U_{e2}|^2 -  |U_{e1}|^2) I_\mu ,  
\nonumber\\
P(\nu_\mu \rightarrow \nu_\mu)_{int} & = &  
(|U_{\mu 2}|^2 - |U_{\mu 1}|^2) I_\mu , 
\nonumber\\
P(\nu_\mu \rightarrow \nu_\tau)_{int} & = 
& (|U_{\tau 2}|^2 - |U_{\tau 1}|^2) I_\mu.
\nonumber
\eea
For the selected values of the mixing angles 
$s_{23}^2 = 0.5$ and  $\sin^2 2 \theta_{13} = 0.08$ 
we have  
$$
P(\nu_\mu \rightarrow \nu_e)_{int} = -0.372 I_\mu ~, ~~ 
P(\nu_\mu \rightarrow \nu_\tau)_{int} = 0.318 I_\mu ~, ~~ 
P(\nu_\mu \rightarrow \nu_\mu)_{int} = 0.054 ~, 
$$
in agreement with the results of numerical computations (see
Fig.~\ref{prob-l}, lower panel).  The wiggles in the $\nu_\mu
\rightarrow \nu_e$ and $\nu_\mu \rightarrow \nu_\tau$ channels are
large and in opposite phase and their amplidude decreases with
increase of energy because $P_L \rightarrow 1$ and $P_H \rightarrow
1$.  The L-wiggles are suppressed in the $\nu_\mu \rightarrow \nu_\mu$
channel.

Let us now consider the phase of oscillatory behavior
(\ref{phase-mu}). The phase $\phi_L$ is collected (i) from the
production point $r_{j}$ to the H-resonance, $r_R^H$.  Here the
difference of eigenvalues is essentially given by the vacuum
frequency: $\Delta m^2_{13}/2E$; (ii) in the H-resonance region and
below: here the difference of eigenvalues start to decrease down to
$\sim \Delta m^2_{12}/2E$; (iii) in the L-resonance, where the
frequency is $\sim \Delta m^2_{12}/(2E \sin 2 \theta_{12})$.  The
phases collected in regions (ii) and (iii) are small. In the
L-resonance the oscillation length becomes larger than whole baseline.
For instance for $E = 200$ GeV, we obtain $ l_\nu \sim 6\cdot 10^{11}$
cm $\gg r_R^L$.  Although now the phase is collected from much larger
region than in the case of H-wiggles: from $r_{j}$ to $r_R^L + \Delta
r_R^L$ (e.g. $2 \cdot 10^{11}$ cm, for $E \sim 200$ GeV), the phase
$\phi_L$ is comparable to $\phi_H$ (collected by $0.7 \cdot 10^{11}$
cm).  The reason is that the phase between $r_R^H + \Delta r_R^H$ and
$r_R^L + \Delta r_R^L$ is relatively small. Consequently, the period
of L-wiggles is comparable to the period of H-wiggles but certain
phase shift is present.

\subsection{Dependence on the density profile}

The dependence of the probabilities on characteristics of density
profile of the star (the initial density $n_0$ (see Fig.~\ref{prob-l})
and gradient of density change in the envelope) is shown in
Fig.~\ref{prob-profiles} for profiles A and B.

Let us first consider dependence of probabilities on $k$.  With
decrease of $k$ the adiabaticity determined by $\lambda_n = r_R/k$
becomes stronger.  Indeed, for fixed energy the resonance layer has
larger radius $r_R$ as $k$ decreases. Consequently $P_H$ becomes
smaller and
\begin{itemize}
\item 
the 1-3 dip (in $\nu_e \to \nu_e$ channel) becomes deeper and reaches
the adiabatic minimum, $P_{min} = s_{13}^2$, even for small values of
1-3 mixing (see Fig.~\ref{prob-profiles}, upper panel);
\item 
adiabaticity is broken at higher energies ($E > 170$ GeV for $\sin^2 2
\theta_{13} = 0.08$), and correspondingly, the wiggles (which are
manifestations of the adiabaticity breaking) shift to higher
energies;
\item 
the amplitude of L-wiggles does not change significantly; 
\item 
period of wiggles decreases. This follows immediately from
Eq.~(\ref{adphaseH}): $r_R$ is larger and therefore the phase for a
given $E$ increases.
\end{itemize}

Similar features are present for $\nu_\mu \rightarrow \nu_\beta$ modes
(see Fig.~\ref{prob-profiles}, lower panel).  With decrease of $k$ the
adiabaticity both in H- and L- resonances improves.  This means that
in the range $E \gtrsim E_R^H$ the probabilities reach their adiabatic
values.  The amplitude of wiggles becomes smaller and period is
substantially smaller.  The latter is due to increase of $r_R^L$ as in
the case of H-wiggles.

For antineutrinos, decreases of gradient leads to 
stronger matter effect in the 
range $(10^2 - 10^4)$  GeV,  essentially extending region of 
the adiabatic conversion in the L-resonance to higher energies.

Effect of the inner density increase is illustrated in
Fig.~\ref{prob-l}.  There are two consequences of this increase: (i)
shift of the resonance energies and therefore whole energy region of
matter effects to smaller energies; (ii) improvement of adiabaticity
in both resonances due to the fact that for a given energy the radii
of resonance layers increase. This, in turn, leads to the following
observational consequences:
\begin{itemize}
\item 
the dip due to 1-3 resonance shifts to smaller energies and starts at
$E = 60 - 80$ GeV for the factor of 2 density increase;
\item 
the dip (peak) reaches the adiabatic minimum (maximum) for the
survival (transition) channels; e.g. for $P(\nu_e \rightarrow \nu_e)
\approx \sin^2 \theta_{13}$ and $P(\nu_\mu \rightarrow \nu_e) \approx
0.5(1 - \sin^2 \theta_{13})$;
\item 
the H-wiggles have smaller period and so there are more wiggles on the
nonadiabatic edge;
\item 
the L-wiggles have smaller amplitude at low energy which then
increases with energy. This is related to good adiabaticity in the
L-channel and small $P_L$ which increase with energy.  Period of
wiggles becomes smaller.
\item 
Due to better adiabaticity the asymptotics are achieved at higher
energies.
\end{itemize}

The antineutrino probabilities are affected very weakly and mainly in
the range $(10^2 - 10^3)$ GeV.

On the contrary, with decrease of the initial density in the envelope
and increase of the gradient, the matter-affected region shrinks: the
region shifts to higher energies but its non-adiabatic edge - to lower
energies.

\FIGURE{\epsfig{file=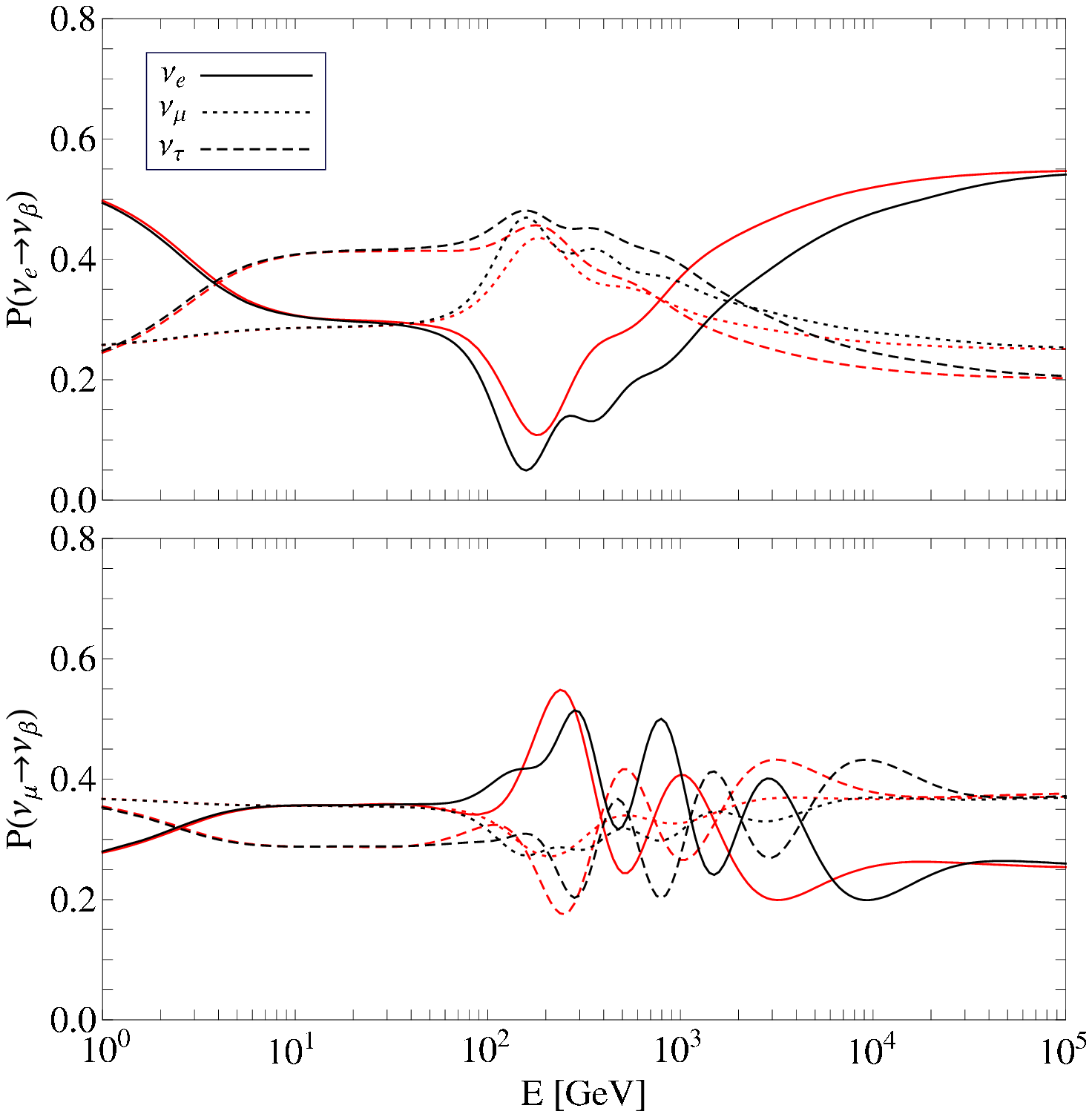,width=10cm}
\caption{The $\nu_e \to \nu_\beta$ (top panel) and $\nu_\mu \to
\nu_\beta$ (bottom panel) transition probabilities as functions of the
neutrino energy for two different stellar density profiles, $\rho
\propto r^{-3}$ (red curves) and $\rho \propto r^{-17/7}$ (black
curves), but for the same initial density at $R_0$.  We used $\sin^2
2\theta_{13} = 0.08$, $\delta_{cp} =0$ and normal mass hierarchy. }
\label{prob-profiles}}

\section{Properties of conversion probabilities}

Let us consider dependence of conversion probabilities in specific
channels on neutrino parameters $\theta_{13}$, $\theta_{23}$ as well
as on the type of mass hierarchy.

\subsection{Probabilities of $\nu_e \rightarrow \nu_\beta$
transitions}

In Fig.~\ref{prob-13mix} we show the $\nu_e \rightarrow \nu_\beta$,
$\beta = e, \mu, \tau$ probabilities as functions of neutrino energy
for different values of the 1-3 mixing angle.  Properties of these
probabilities can be well understood using the considerations in the
previous sections.

\noindent
1. In asymptotics according to (\ref{abprob}) we have the averaged
$3\nu-$ oscillation probability in vacuum:
\be
P(\nu_e \rightarrow \nu_e) =
\sum_i |U_{e i}|^4 =  \cos^4 \theta_{13} \left( 1 - \frac{1}{2} 
\sin^2 2\theta_{12} \right) + \sin^4 \theta_{13}. 
\label{ee-prob1}
\ee
For $\nu_e \rightarrow \nu_\mu$ channel the probability equals
$P(\nu_e \rightarrow \nu_\mu) = \sum_i |U_{e i}|^2 |U_{\mu i}|^2$,
etc.

\noindent
2. In the range $E_R^L \ll E \ll E_R^H$, inserting the matrix of
probabilities (\ref{adprobb}) into Eq.~(\ref{ptot}) we obtain
\be
P(\nu_e \rightarrow \nu_e) = 
|U_{e2}|^2 (1 - |U_{e3}|^2)  + |U_{e3}|^4 =   
|U_{e2}|^2  - |U_{e3}|^2(|U_{e2}|^2  - |U_{e3}|^2), 
\label{ee-norm}
\ee
or in terms of the angles: 
$P(\nu_e \rightarrow \nu_e)=  \cos^4 \theta_{13} 
\sin^2 \theta_{12} + \sin^4 \theta_{13}$.  
For the $\nu_e \rightarrow \nu_\mu$  channel we have    
\be
P(\nu_e \rightarrow \nu_\mu) =  |U_{\mu 2}|^2 (1 - |U_{e3}|^2)
+  |U_{\mu 3}|^2  |U_{e3}|^2  = 
|U_{\mu 2}|^2 + |U_{e3}|^2 (|U_{\mu 3}|^2 - |U_{\mu 2}|^2) ,
\label{emu-norm}
\ee
and similar expression for $P(\nu_e \rightarrow \nu_\tau)$ with
substitition $|U_{\mu i}|^2 \rightarrow |U_{\tau i}|^2$.

\noindent
3. For $E > E_R^H$ and large enough 1-3 mixing the adiabatic
evolution gives according to (\ref{adprobb3}) and (\ref{ptot})
\be
P(\nu_e \rightarrow \nu_e)= |U_{e3}|^2, ~~~ 
P(\nu_e \rightarrow \nu_\mu)= |U_{\mu 3}|^2, ~~~
P(\nu_e \rightarrow \nu_\tau)= |U_{\tau 3}|^2.
\label{emt-norm}
\ee
These results reproduce the probabilities at $E \sim 200$ GeV for the
1-3 mixing $\sin^2 2\theta_{13} > 0.1$. For smaller 1-3 mixings the
adiabaticity is broken already at $E \sim E_R^H$ and therefore
$P(\nu_e \rightarrow \nu_e) > |U_{e3}|^2$.

\FIGURE{\epsfig{file=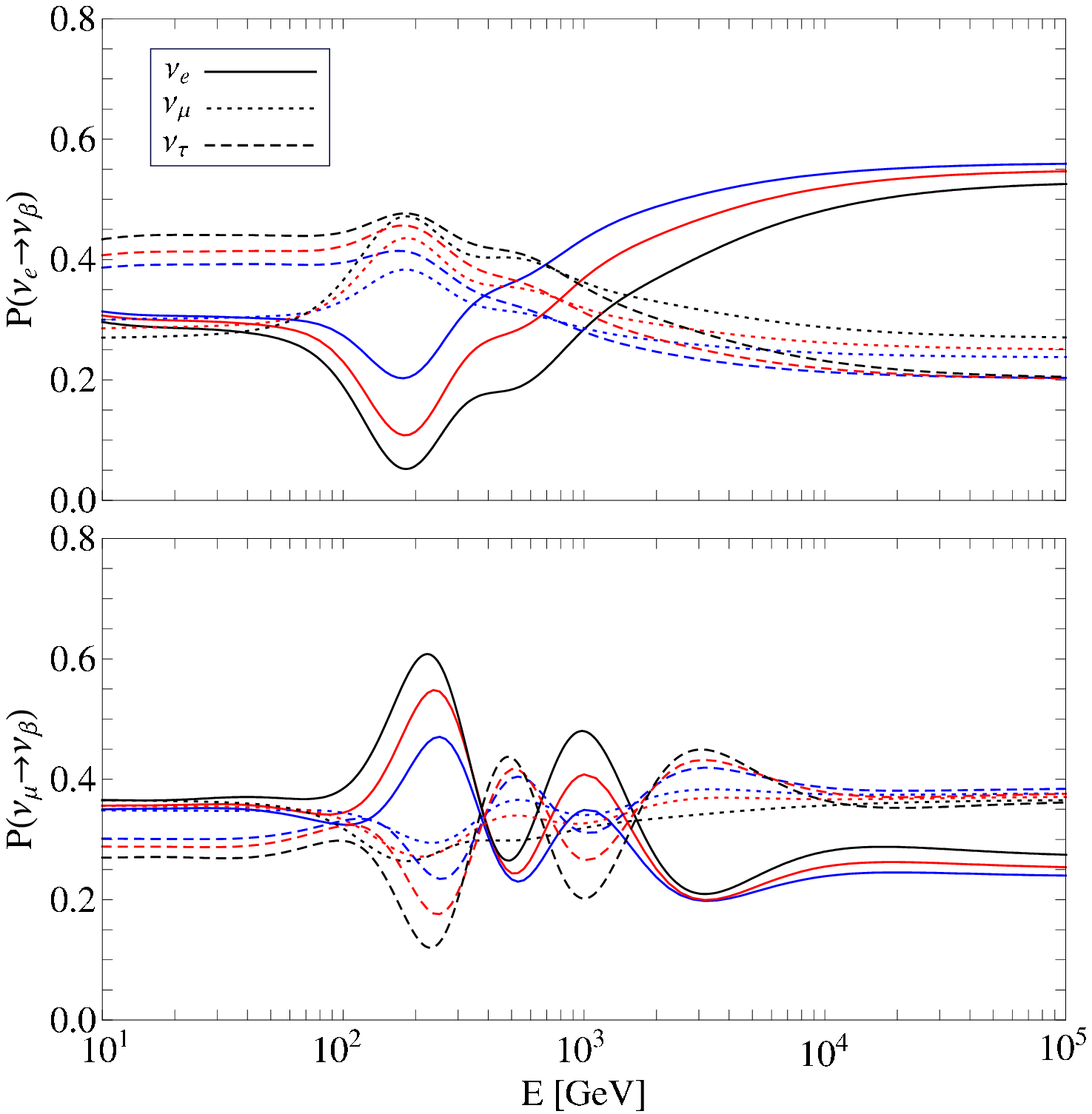,width=10cm}
\caption{Probabilities for $\nu_e \to \nu_\beta$ (top panel) and
$\nu_\mu \to \nu_\beta$ (bottom panel) transitions for different
values of 1-3 mixing: $\sin^2 2\theta_{13} = 0.04$ (black lines), 0.08
(red) and 0.15 (blue lines). We take profile A, $\delta_{cp} =0$,
$\sin^2 \theta_{23} = 0.5$ and normal mass hierarchy.}}
\label{prob-13mix}

If neutrinos are produced not too far (in energy scale) from the
H-resonance, the $\nu_e-$state contains non-negligible admixture of
the $\nu_{2m}$ eigenstate:
\be
\nu_e = \cos \theta_{13}^0 \, 
\nu_{2m} + \sin \theta_{13}^0 \, \nu_{3m} ~.
\label{eq:nuclose}
\ee
Adiabatic evolution of this combination will give then 
$$
P_* (\nu_e \rightarrow \nu_1) = 0, ~~~
P_* (\nu_e \rightarrow \nu_2) = \cos^2 \theta_{13}^0, ~~~
P_* (\nu_e \rightarrow \nu_3) = \sin^2 \theta_{13}^0 , 
$$
and instead of (\ref{emt-norm}) for $\nu_e \rightarrow \nu_e$  
probability we obtain 
$$
P(\nu_e \rightarrow \nu_e) =  
\cos^2 \theta_{13}^0 \cos^2 \theta_{13} \sin^2 \theta_{12} + 
\sin^2 \theta_{13}^0 \sin^2 \theta_{13} ~.
$$

For $ E > E_R^H$ with increase of energy the adiabaticity violation
becomes important.

The analytic results presented here allow one to understand the
dependence of the probabilities on $\theta_{13}$ and $\theta_{23}$.
In asymptotics and in the intermediate region (plateau) the
probabilities only weakly depend on $\theta_{13}$ (see
Fig.~\ref{prob-13mix}): the corresponding corrections are proportional
to $|U_{e3}|^2 = s_{13}^2$ (see (\ref{ee-prob1}), (\ref{ee-norm}),
(\ref{emu-norm})).  The strongest effect is in the range of
H-resonance and above it, where the 1-3 mixing is enhanced. For the
dip in the adiabatic case we have $P \sim s_{13}^2$ (\ref{emt-norm}).
With decrease of $s_{13}^2$ the adiabaticity becomes broken and the
survival probability increases. In the limit of very small $s_{13}^2$
the second dip disappears and we will have only one dip due to 1-2
mixing. The probabilities in other channels have an ``inverted''
dependence.

\FIGURE{\epsfig{file=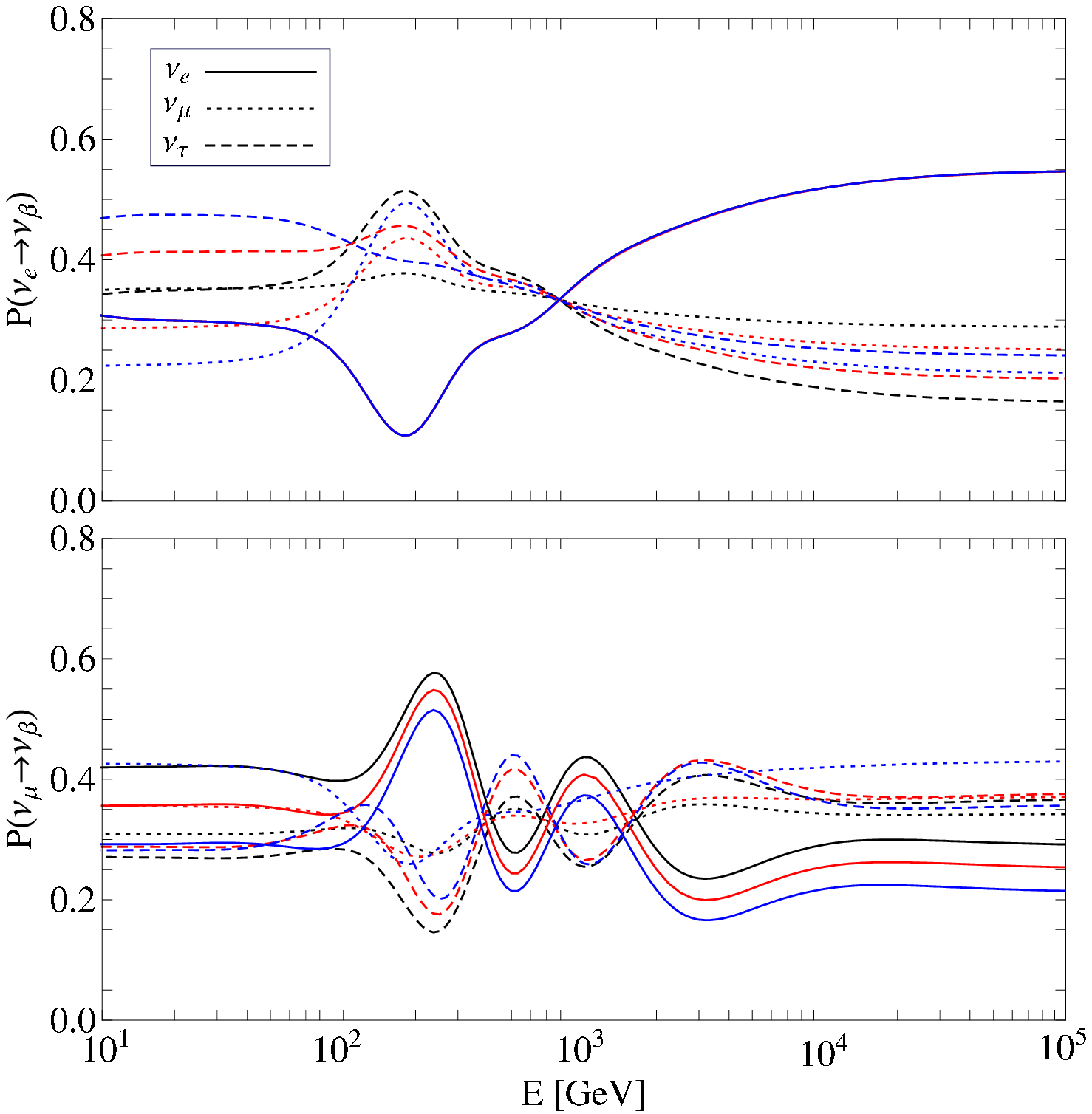,width=10cm}
\caption{The $\nu_e\to\nu_\beta$ (top panel) and $\nu_\mu\to\nu_\beta$
(bottom panel) probabilities as functions of energy for different
values of 2-3 mixing: $\sin^2 \theta_{23} = 0.4$ (blak curves), 0.5
(red curves) and 0.6 (blue curves). We used profile A, $\sin^2
2\theta_{13} = 0.08$, $\delta_{CP}=0$ and normal mass hierarchy.}
\label{prob-23mix}}

The dependence of probabilities on the 2-3 mixing is shown in
Fig.~\ref{prob-23mix}. Note that the $\nu_e \to \nu_e$ probability
does not depend on $\theta_{23}$.  The dependences of other
probabilities have rather interesting feature: there are two energies
$E_1 \approx 100$ GeV and $E_2 \approx 800$ GeV at which the
probabilities do not depend on $\theta_{23}$, and moreover, at $E_2
\approx 800$ GeV the probabilities in all the channels are equal.
With increase of $s_{23}^2$ the $\nu_e \to \nu_\mu$ probability
decreases at $E < E_1$ and $E > E_2$ and it increases in the interval
$E_1 - E_2$. Indeed, according to (\ref{emu-norm}) in the intermediate
range $P(\nu_e \to \nu_\mu) = |U_{\mu 2}|^2 \approx c_{23}^2$, and in
the 1-3 dip (\ref{emt-norm}): $P(\nu_e \to \nu_\mu) = |U_{\mu 3}|^2
\approx s_{23}^2$.  In asymptotics, a weaker change can be immediately
understood from expression $P(\nu_e \rightarrow \nu_\mu) \approx |U_{e
1}|^2 |U_{\mu 1}|^2 + |U_{e 2}|^2 |U_{\mu 2}|^2$.  The strongest
dependence is in the plateau.  The dependence for $\nu_e \rightarrow
\nu_\tau$ channel is just inverted.

\subsection{Probabilities of $\nu_\mu \rightarrow \nu_\beta$ and
$\nu_\tau \rightarrow \nu_\beta$ transitions}

The probabilities $P(\nu_\mu \rightarrow \nu_\beta)$ as functions of
neutrino energy for different channels and two different values of 1-3
mixing are shown in Fig.~\ref{prob-13mix}.  The main difference from
the $\nu_e-$ channels is that $\nu_\mu$ is not the eigenstate of the
propagation basis (in contrast to $\nu_e$) but combination of these
states. This leads to more complicated expressions for the
probabilities and additional interference effects.  The probabilities
equal
$$
P(\nu_\mu \rightarrow \nu_\beta) =
\sum_i P_* (\nu_\mu \rightarrow \nu_i) |U_{\beta i}|^2,
$$
and explicit expressions for probabilities in different energy regions
are given below.

The asymptotic values of probabilities for $E \ll E_L$ and $E \gg E_L$
equal
\be
P(\nu_\mu \rightarrow \nu_\beta) =
\sum_i |U_{\mu i}|^2 |U_{\beta i}|^2.
\label{va-general}
\ee 
In the plateau we obtain from (\ref{ptot}) and (\ref{adprobb})
$$
P(\nu_\mu \rightarrow \nu_\beta) =
c_{23}^2 |U_{\beta 1}|^2 + 
s_{23}^2 |U_{e3}|^2 |U_{\beta 2}|^2 + 
s_{23}^2 c_{13}^3 |U_{\beta 3}|^2, 
$$
or 
\be
P(\nu_\mu \rightarrow \nu_\beta) =
\frac{|U_{\tau 3}|^2 |U_{\beta 1}|^2}{1 - |U_{e3}|^2} + 
\frac{|U_{\mu 3}|^2 |U_{e3}|^2 |U_{\beta 2}|^2}{1 - |U_{e3}|^2} + 
|U_{\beta 3}|^2 |U_{\mu 3}|^2,  
\label{generalint}
\ee
where we used  that 
\be
\sin^2 \theta_{23} \equiv \frac{|U_{\mu 3}|^2}{1 - |U_{e3}|^2}, ~~~
\cos^2 \theta_{23} \equiv \frac{|U_{\tau 3}|^2}{1 - |U_{e3}|^2}. 
\label{angle-elem}
\ee
Above the H-resonance for large 1-3 mixing, which satisfies the
adiabaticity condition, we find from (\ref{ptot}) and (\ref{adprobb3})
\bea
P(\nu_\mu \rightarrow \nu_\beta) & = & 
|U_{\beta 1}|^2 c^2_{23} + 
|U_{\beta 2}|^2 s^2_{23} 
 \nonumber\\ 
 & = & \frac{1}{1 - |U_{e3}|^2} 
\left( |U_{\beta 2}|^2|U_{\mu 3}|^2  + 
|U_{\beta 1}|^2 |U_{\tau 3}|^2 \right). 
\label{generalmin}
\eea
It reproduces in the first approximation correct values of the
probabilities in the dip at $E \sim (150 - 170)$ GeV for large 1-3
mixing and for small gradients.  Certain deviation of numerical
results from analytic ones is a manifestation of the adiabaticity
violation.  For maximal 2-3 mixing we obtain from (\ref{generalmin}):
$P(\nu_\mu \rightarrow \nu_\beta) = 0.5 (1 - |U_{\beta 3}|^2)$.

As in the case of $\nu_e \rightarrow \nu_\beta$ the change of 1-3
mixing mainly affects the probabilities in the energy range above the
H-resonance. With decrease of $\theta_{13}$ the jump probability
increases and according to (\ref{int-term}) the amplitude of wiggles,
$I_\mu \propto \sqrt{1 - P_H}$, decreases.

The $\nu_{\tau}-$ amplitudes and probabilities (they appear at the
border of envelope due to oscillations inside jets) can be found from
the $\nu_\mu-$ amplitudes obtained in the previous subsection by
substitutions: $U_{\mu 3} \rightarrow U_{\tau 3}$, $\sin \theta_{23}
\rightarrow \cos \theta_{23}$, $\cos \theta_{23} \rightarrow - \sin
\theta_{23}$.

\subsection{Probabilities in antineutrino channels }

There is no resonances in antineutrino channel for normal mass
hierarchy and the dependence of the probabilities on energy is simpler
(see Fig.~\ref{prob-anti13}). The 1-3 mixing is not enhanced and the
conversion effects are mainly due to large 1-2 mixing at low energies
(where this mixing is not suppressed).

The asymptotic values of the probabilities are the same as in the
neutrino channels (\ref{va-general}). There is no CP-violation
asymmetries.  In the intermediate region ($ E^L_R < E < E_R^H$) we
obtain from (\ref{ptot}) and (\ref{adprobun})
\bea
P(\bar{\nu}_e \rightarrow \bar{\nu}_\beta) &  = &  
|U_{\beta 1}|^2 (1 - |U_{e3}|^2) + |U_{\beta 3}|^2  |U_{e3}|^2 
= |U_{\beta 1}|^2  - |U_{e3}|^2 (|U_{\beta 1}|^2 - |U_{\beta 3}|^2),
\nonumber\\
P(\bar{\nu}_\mu \rightarrow \bar{\nu}_\beta) & = & 
\frac{1}{1 - |U_{e3}|^2} \left(|U_{\tau 3}|^2 |U_{\beta 2}|^2 +
|U_{\mu 3}|^2 |U_{e3}|^2 |U_{\beta 1}|^2 \right)
+ |U_{\mu 3}|^2 |U_{\beta 3}|^2, 
\label{bareb-inv}
\eea
($\beta = e, \mu, \tau$). They reproduce well the result of numerical
computations in the range $(10-50)$ GeV (see
Fig.~\ref{prob-anti13}).

In the range $E \gtrsim E_R^H$, (150-180 )~GeV, the 1-3 mixing is
suppressed by matter, so that $c_{13}^0 \approx 1$, $s_{13}^0 \approx
0$ in (\ref{adprobun}), and the flavor transition probabilities in the
adiabatic approximation are
\bea
P(\bar{\nu}_e \rightarrow \bar{\nu}_\beta) & =& |U_{\beta 1}|^2, 
\nonumber\\
P(\bar{\nu}_\mu \rightarrow \bar{\nu}_\beta) & = & 
\frac{1}{1 - |U_{e3}|^2} 
\left(|U_{\tau 3}|^2 |U_{\beta 2}|^2 + 
|U_{\mu 3}|^2 |U_{\beta 3}|^2 \right). 
\label{bargeneralmin}
\eea
Comparing the first probability with the one in (\ref{bareb-inv}) we
conclude that $P(\bar{\nu}_e \rightarrow \bar{\nu}_e)$ only slightly
increases in comparison with the intermediate region.  With further
increase of energy the probability decreases due to the adiabaticity
violation, approaching the vacuum oscillation result. Correspondingly,
$P(\bar{\nu}_e \rightarrow \bar{\nu}_\tau)$ is ``inverted''.

The probabilities in antineutrino channels (in contrast to the
neutrino resonance case) rather weakly depend on energy.  Their
dependence on the 1-3 mixing is weak (Fig.~\ref{prob-anti13}):
according to (\ref{bareb-inv}), the corrections to $P(\bar{\nu}_e
\rightarrow \bar{\nu}_\beta)$ are of the order $|U_{e3}|^2$, and
corrections to $P(\bar{\nu}_\mu \rightarrow \bar{\nu}_\beta)$ are even
smaller. The probabilities differ from their asymptotic values mainly
in the intermediate range.  Interestingly, the probabilities with
$\bar{\nu}_\mu$ in final state (at least for zero $\delta$ and maximal
2-3 mixing) are not affected by matter.

\FIGURE{\epsfig{file=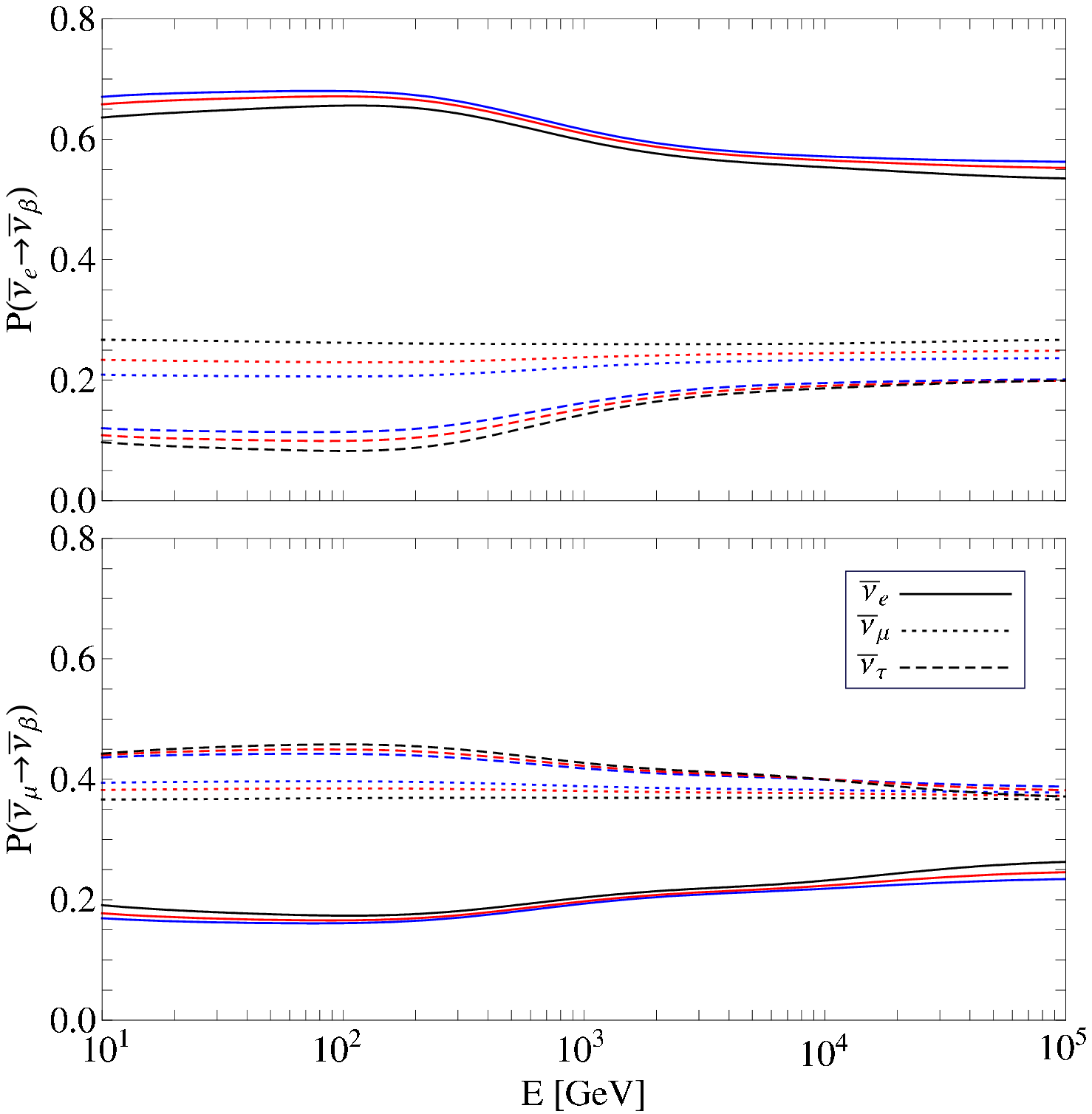,width=10cm}
\caption{The same as in Fig.~\ref{prob-13mix} for the antineutrino
channels.}
\label{prob-anti13}}

According to Fig.~\ref{prob-23anti} the $\nu_e \rightarrow \nu_e$
probability does not depend on $\theta_{23}$. Dependences of
probabilities on 2-3 mixing are not very strong and well described by
Eqs.~(\ref{bareb-inv}) and (\ref{bargeneralmin}).

\FIGURE{\epsfig{file=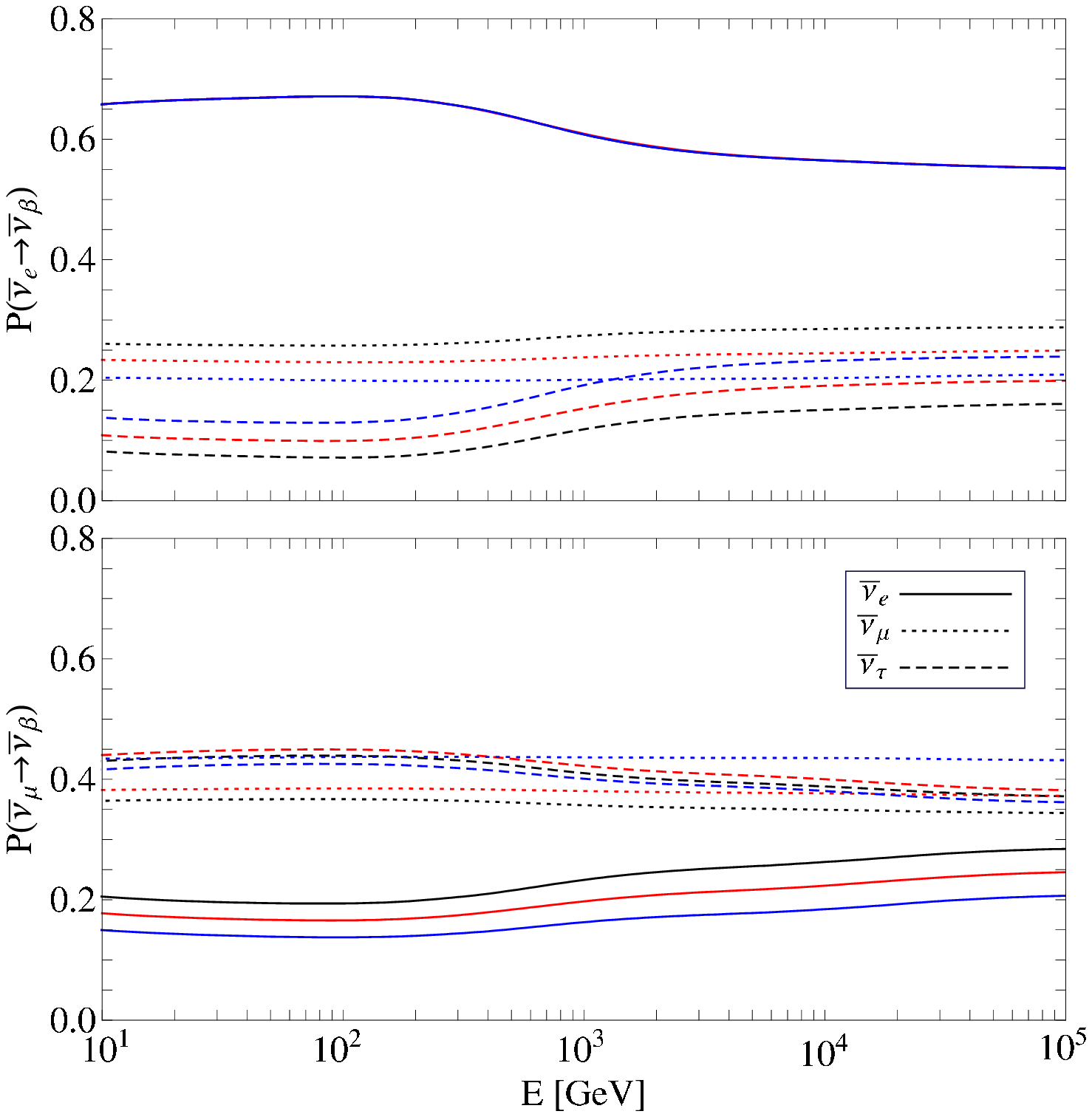,width=10cm}
\caption{The same as in Fig.~\ref{prob-23mix} for the antineutrino
channels.  }
\label{prob-23anti}}

\subsection{Probablities for the inverted mass hierarchy}

Now the L-resonance is in the neutrino channels, whereas the
H-resonance is in the antineutrino channels. Therefore the
probabilities are given by some combinations of the results obtained
for $\nu$ and $\bar{\nu}$ for the normal mass hierarchy (see
Fig.~\ref{CPinv}).  The differences appear when interplay of the two
resonances (in the case of NH) becomes important.

Let us consider first the neutrino probabilities. (i) The asymptotic
values are the same as in the NH case.  (ii) In the intermediate
region we can use the same result as for NH (\ref{emu-norm}) and
(\ref{ee-norm}).  (iii) Above the H-resonance according to
(\ref{adprobb3}) and (\ref{ptot}) we have
\be
P(\nu_e \rightarrow \nu_\beta) = |U_{\beta2}|^2, 
~~~(\beta = e, ~\mu, ~\tau),
\label{ih-eb}
\ee
which should be compared with the results in Eq.~(\ref{emt-norm}) for
the normal mass hierarchy.  In particular, $P(\nu_e \rightarrow \nu_e)
= |U_{e2}|^2$.  Comparing this result with (\ref{emt-norm}) we see
that the probability slightly increases in the H-resonance in contrast
to the decrease in the NH case.  According to Fig.~\ref{CPinv} for
$\sin^2 2\theta_{13} = 0.08$ the adiabaticity is broken already at $E
= E_R^H$ and therefore values of probabilities deviate from those in
Eq.~(\ref{ih-eb}).

In the intermediate energy region for the $(\nu_\mu \to \nu_\beta)$
channels we have the same results as in the case of normal mass
hierarchy (\ref{generalint}). Above the H-resonance, again the
difference from the NH case appears since there is no level crossing.
Using (\ref{adprobih}) we obtain flavor probabilities
\bea
 P(\nu_\mu \rightarrow \nu_\beta) & = &
 |U_{\beta 3}|^2 \sin^2 \theta_{23} + 
|U_{\beta 1}|^2 \cos^2 \theta_{23}
 \nonumber\\
 & = & \frac{1}{1 - |U_{e3}|^2}
\left( |U_{\beta 3}|^2|U_{\mu 3}|^2  +
|U_{\beta 1}|^2 |U_{\tau 3}|^2 \right),
\nonumber
\eea
which should be compared with Eq.~(\ref{generalmin}).

Let us consider the antineutrino channels.  In the intermediate energy
range for the initial $\bar{\nu}_e$ we have the same results as in the
NH case (\ref{bareb-inv}).  Above the H-resonance (which is now in the
antineutrino channel) using Eq.~(\ref{adprobbih}) we obtain
\bea 
P (\bar{\nu}_e \rightarrow \bar{\nu}_\beta)& = &|U_{\beta 3}|^2,\\ 
P(\bar{\nu}_\mu \rightarrow \bar{\nu}_\beta) & = & 
\frac{1}{1 - |U_{e3}|^2} \left(|U_{\tau 3}|^2 |U_{\beta 2}|^2 +
|U_{\mu 3}|^2 U_{\beta 1}|^2 \right).
\nonumber
\eea
They reproduce correct values of the probabilities in the interval $E
\sim (150 - 170)$ GeV.

\subsection{Effects of oscillations inside jets}

In Fig.~\ref{probjetnu} we compare the probabilities with and without
oscillations in jet. As follows from these plots, the jet effect is
very small and appears in the region above the H-resonance.  This can
be explained in the following way.  Oscillations between neutrino
production point inside jet and inner part of the envelope are
described by $A_{jet} (\nu_\alpha \rightarrow \nu_\xi)$ introduced in
Eq.~(\ref{star-totcp}).  We assume that in the neutrino production
region a jet has an average density $n_j \sim 1.5 \cdot 10^{20}$
cm$^{-3}$.  The corresponding resonance energies equal: $E_{R j}^L =
10^{3}$ GeV and $E_{R j}^H = 6 \cdot 10^{4}$ GeV for the 1-2 and 1-3
mass splits correspondingly.  Total matter width (column density)
equals $n_j r_{\nu} \sim 9 \cdot 10^{29}$ cm$^{-2}$ which is much
smaller than the refraction length \cite{Luna} and therefore matter
effect can be neglected. Indeed, for $E < E_R$ the potential is
smaller than the kinetic (vacuum) term, whereas for $E > E_R$ the
oscillation length is much smaller than the baseline, so that vacuum
mimicking situation is realized \cite{vacmim}.  In general, amplitude
of vacuum oscillations in the flavor basis can be written as
\be
A_{jet} (\nu_\alpha^\prime \rightarrow \nu_\xi^\prime) 
= \delta_{\alpha \xi} + 
U_{\alpha 3}^\prime U_{\xi 3}^{\prime *} \left(e^{i2\phi_{31}}  
- 1 \right) + 
U_{\alpha 2}^\prime  U_{\xi 2}^{\prime *} \left(e^{i2\phi_{21}} -1 
\right), 
\label{vo-ampl}
\ee 
where $\phi_{31} \equiv \pi x/l_{31}$ and $\phi_{21} \equiv \pi
x/l_{21}$ are half oscillation phases inside jet.  Vacuum oscillation
lengths for the two modes equal $l_{13} = 10^{10}~{\rm cm}~ (E/
100~{\rm GeV})$, and $l_{12} = 3\cdot 10^{11}~{\rm cm}~ (E/100~{\rm
GeV})$.  The typical length of the neutrino trajectory inside jet
$r_\nu \sim 6 \cdot 10^{9}$~cm, and therefore for $E \gtrsim 60$ GeV
the phase $\phi_{21} \approx 0$. So, we can neglect oscillations due
to 1-2 mass split and consider non-averaged vacuum oscillations due to
1-3 mass splitting only.  Then the amplitudes (\ref{vo-ampl}) become
\be
A_{jet} (\nu_\alpha^\prime \rightarrow \nu_\xi^\prime) 
= \delta_{\alpha \xi} + 
U_{\alpha 3}^\prime U_{\xi 3}^{\prime *}~ \eta(x) ~,  
\label{vo-ampl3}
\ee 
where 
\be
\eta(x) \equiv \left[e^{i2\phi_{31}(x)}  -1 \right].
\label{eta-def}
\ee

\FIGURE{\epsfig{file=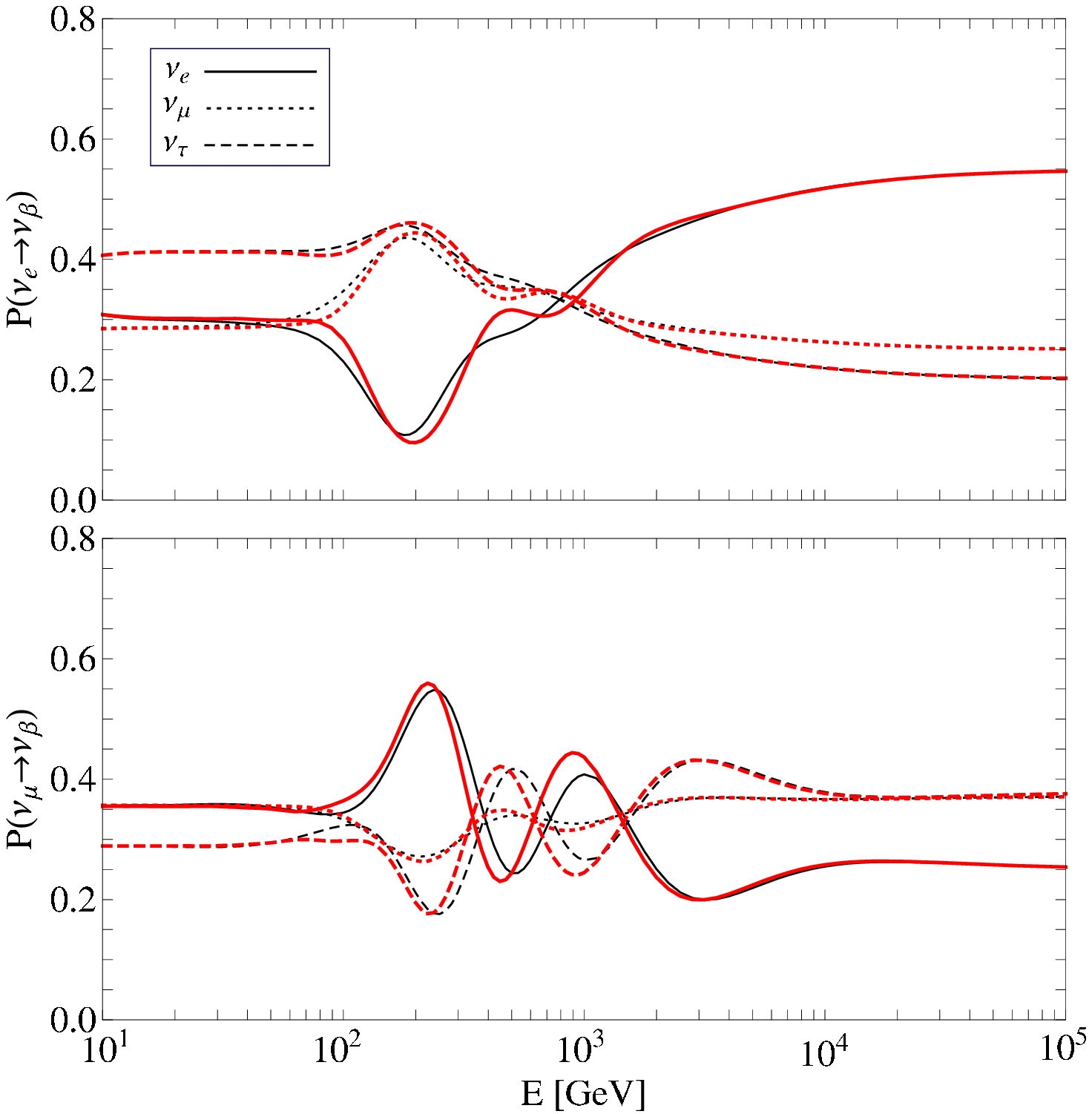,width=10cm}
\caption{ The $\nu_e \to \nu_\beta$ (top panel) and $\nu_\mu \to
\nu_\beta$ (bottom panel) transition probabilities with oscillations
in the jet (thick red curves) and without oscillations in jet (thin
black curves).  We used $\sin^2 2\theta_{13} = 0.08$, $\sin^2
\theta_{23} = 0.5$ and $\delta_{CP} = 0$ in normal mass hierarchy. }
\label{probjetnu}}

According to (\ref{uprime}) the mixing matrix elements in the
propagation basis are given by
$$
U_{\alpha 3}^{\prime} = \vec{U}_3 \equiv 
(s_{13}, 0 ,  c_{13}), ~~~~ \alpha = 
(e, \mu, \tau). 
$$
Thus, the S-matrix of transitions inside jet (\ref{vo-ampl3}) can be
written as
$$
S_{jet} = I + \eta \vec{U}_3 \vec{U}_3^T = \pmatrix{
1 +  \eta s_{13}^2  &  0         &  \eta s_{13}c_{13} \cr
0                   &  1         &  0  \cr
\eta s_{13}c_{13}   &  0 &  1 +  \eta c_{13}^2 
}.
$$
Inserting this matrix into (\ref{total-s}) we obtain the total
S-matrix inside the star:
\be
S_{*} = \pmatrix{
1 +  \eta s_{23}^2                  & 0      & \eta s_{13}c_{13}  \cr
\eta s_{13}c_{13}s_{23} e^{i\delta} & c_{23} & s_{23} e^{i\delta}(1 + \eta c_{23}^2) \cr
\eta s_{13}c_{13} c_{23} e^{i\delta}  & -s_{23}  & c_{23} e^{i\delta}(1 + \eta c_{23}^2) 
}
\times S_{env}, 
\label{sss-tot}
\ee
where $(S_{env})_{\alpha i} \equiv A_{env} (\nu_\alpha^{\prime}
\rightarrow \nu_i)$.

In the adiabatic approximation the S-matrix in the envelope equals the
mixing matrix at the inner part of the envelope: $S_{env} =
U_{m}^{\prime 0}$ and the latter is given in Eq.~(\ref{uprimeint}).
Inserting this matrix into (\ref{sss-tot}) we obtain explicitly
\be
S_{*} = \pmatrix{
0 &  c_{13}^0 (1 +  \eta s_{23}^2) - \eta s_{13}^0 s_{13}c_{13} & 
s_{13}^0 (1 + \eta s_{23}^2) +  \eta c_{13}^0 s_{13}c_{13} \cr
-c_{23} & 
\eta c_{13}^0 s_{13}c_{13} s_{23} e^{i\delta} - s_{13}^0 s_{23}e^{i\delta}(1 + \eta c_{23}^2) & 
\eta s_{13}^0 s_{13}c_{13} s_{23} e^{i\delta} + c_{13}^0 s_{23}e^{i\delta}(1 + \eta c_{23}^2) \cr
- s_{23} & 
\eta c_{13}^0 s_{13}c_{13} c_{23} e^{i\delta} - s_{13}^0 c_{23}e^{i\delta}(1 + \eta c_{23}^2) &
\eta s_{13}^0 s_{13}c_{13} c_{23} e^{i\delta} + c_{13}^0 c_{23}e^{i\delta}(1 + \eta c_{23}^2) 
}.
\label{sss-totad}
\ee
Consider this matrix in specific energy ranges.  In the intermediate
energy range ($c_{13}^0 = c_{13}$, $s_{13}^0 = s_{13}$) it is
$$
S_{*} = \pmatrix{
 0       &  c_{13}                   &  s_{13}(1 + \eta) \cr
-c_{23}  & -s_{13} s_{23}e^{i\delta} & c_{13} s_{23}e^{i\delta}(1 + \eta) \cr
-s_{23}  & -s_{13} c_{23}e^{i\delta} & c_{13} c_{23}e^{i\delta}(1 + \eta) 
}.
$$
Corrections due to oscillation inside jet are given by the terms
$\propto \eta$.  Since $(1 + \eta) = e^{i 2 \phi_{13}}$ and $|(1 +
\eta)|^2 = 1$, no corrections to the probabilities appear due to
oscillations in jet.  An absence of the corrections is related to the
particular initial state (mixing in matter) in the envelope, the
adiabatic evolution in the envelope and loss of coherence on the way
from the star to the Earth.

Above the H-resonance ($c_{13}^0 \approx 0$, $s_{13}^0 \approx 1$) we
have in the adiabatic approximation
\be
S_{*} = \pmatrix{
 0       &  -\eta s_{13}c_{13}                   &  1 + \eta s_{13}^2 \cr
-c_{23}  & -s_{23}e^{i\delta}(1 + \eta c_{13}^2) &  \eta s_{13} c_{13} s_{23}e^{i\delta} \cr
-s_{23}  & -c_{23}e^{i\delta}(1 + \eta c_{13}^2) &  \eta s_{13} c_{13} c_{23}e^{i\delta}  
}.
\label{sss-totadab}
\ee
The probabilities are given by moduli squared of the $S_{*}$ elements:
$P_*(\nu_\alpha \rightarrow \nu_i) \equiv |(S_*)_{\alpha i}|^2$.
Then, according to (\ref{sss-totadab}) corrections appear. Note that
$|1 + \eta c_{13}^2|^2 = |1 + \eta^* s_{13}^2|^2$. Explicitly we
obtain from (\ref{sss-totadab}) the probabilities for
$\nu_e-$channels:
\be
P_*(\nu_e \rightarrow \nu_1) = 0,~~~
P_*(\nu_e \rightarrow \nu_2) = s^2_{13} c^2_{13}  |\eta|^2, ~~~
P_*(\nu_e \rightarrow \nu_3) = |1 +  s^2_{13} \eta |^2.
\label{nhnueh}
\ee
Now corrections are non-zero but they are suppressed by small factor
$s^2_{13}$.

For the $\nu_\mu-$channels we obtain from (\ref{sss-totadab})  
\be
P_*(\nu_\mu \rightarrow \nu_1) = c^2_{23},~~~
P_*(\nu_\mu \rightarrow \nu_2) = s^2_{23} |1 + c^2_{13} \eta|^2, ~~~
P_*(\nu_\mu \rightarrow \nu_3) = s^2_{23} s^2_{13} c^2_{13} | \eta |^2, 
\label{nhnumuh}
\ee
and again the corrections are small, being suppressed by $s^2_{13}$.

From expression for $S_*$ it follows that in the adiabatic case the
probabilities do not depend on $\delta$: the $S_*-$ matrix elements
either do not depend on $\delta$ or are proportional to the overal
phase factor $e^{i\delta}$.

Similarly, for antineutrinos the probabilities of transitions are not
affected by oscillations inside jet in the intermediate range.  Above
the H-resonance, according to (\ref{sss-totadab}),
$$
P_*(\bar{\nu}_e \rightarrow \bar{\nu}_1) = |1 +  s_{13}^2 \eta |^2, ~~~
P_*(\bar{\nu}_e \rightarrow \bar{\nu}_2) = 0, ~~~
P_*(\bar{\nu}_e \rightarrow \bar{\nu}_3) = s_{13}^2 c_{13}^2  |\eta |^2.
$$
$$
P_*(\bar{\nu}_\mu \rightarrow \bar{\nu}_1) = s_{13}^2 c_{13}^2 s_{23}^2  |\eta |^2,  ~~~
P_*(\bar{\nu}_\mu \rightarrow \bar{\nu}_2) = c_{23}^2, ~~~
P_*(\bar{\nu}_\mu \rightarrow \bar{\nu}_3) = s_{23}^2|1 +  c_{13}^2 \eta |^2, 
$$
(compare with the corresponding results for neutrinos).  As for the
neutrino case, here we have non-zero corrections which are suppressed
by small factor $s_{13}^2$.

In the case of inverted mass hierarchy the amplitudes of oscillation
inside jet are the same as for normal mass hierarchy, at least in our
approximation ($2\nu-$ vacuum oscillations).  So, the amplitudes
$A_{jet}$ equal the amplitudes in Eq.~(\ref{vo-ampl3}).  Consequently,
general formulas for the total amplitudes inside a star are given in
Eq.~(\ref{sss-tot}).

For intermediate energy range we have exactly the same results as in
the case of NH: there is no corrections due to oscillations inside
jet.  The difference appears at energies above H-resonance since now
the level crossing and mixing in initial state are changed (see
Fig.~\ref{level_crossing}).  We obtain the following expressions for
probabilities:
$$
P_*(\nu_e \rightarrow \nu_1) = 0,~~~
P_*(\nu_e \rightarrow \nu_2) = |1 +  s^2_{13} \eta |^2, ~~~
P_*(\nu_e \rightarrow \nu_3) = s^2_{13} c^2_{13}  |\eta|^2,  
$$
$$
P_*(\nu_\mu \rightarrow \nu_1) = c^2_{23},~~~
P_*(\nu_\mu \rightarrow \nu_2) = s^2_{23} s^2_{13} c^2_{13} |\eta|^2, ~~~
P_*(\nu_\mu \rightarrow \nu_3) = s^2_{23} |1 + c^2_{13} \eta|^2. 
$$
Note that probabilities of transitions to $\nu_1$ are the same as in
NH case, whereas the probabilities for transitions to $\nu_2$ and
$\nu_3$ have been interchanged.  The conclusion is the same as for NH:
corrections are non-zero but they are suppressed by small factor
$s^2_{13}$.  In the antineutrino channels in the intermediate energy
range results are the same as in the case of NH.  Above the
H-resonance all the probabilities coincide with those we had obtained
in Eqs.~(\ref{nhnueh}) and (\ref{nhnumuh}) for neutrinos.

An overall conclusion is that there is no corrections due to
oscillations inside jets in the energy range between resonances and
small corrections appear above the H-resonance.  In general, the jet
effect appears if the adiabaticity is broken.

\subsection{CP-violation effects}

Let us consider dependence of the oscillation probabilities on the
CP-violation phase (see Figs.~\ref{CPnorm} and \ref{CPinv}).  Outside
the matter affected energy interval, $E = (1 - 10^5)$ GeV, the
probabilities equal the averaged oscillation probabilities in vacuum.
Important feature is that the coherence of the mass states is lost
after evolution inside star.  Effects of the $(\nu_\alpha \rightarrow
\nu_i)$ transitions sum up incoherently. So, projection back to final
flavor state depends on moduli $|U_{\beta i}|^2$ and no interference
effects appear.  These probabilities depend on moduli squared of the
mixing matrix elements and the interference terms are absent.  In this
case no CP-odd effects appear, the CP- as well as T-asymmetries vanish
and $P (\nu_\alpha \rightarrow \nu_\beta) = P (\bar{\nu}_\alpha
\rightarrow \bar{\nu}_\beta)$.  The dependence on CP-violation phase
originates from $|U_{\alpha i}|^2$ ($\alpha = \mu, \tau$, and $i =
1,2$) and is rather weak since the terms with $\delta$ are
proportional to $\sin \theta_{13}$.

The interference and CP-violation can be related to projection of the
initial state onto the propagation basis states.  To describe this let
us first neglect oscillations inside jet and consider specific
conversion channels.

\begin{enumerate}
\item  
{\it $\nu_e \rightarrow \nu_\beta$ channels:} since $\nu_e$ is the
component of the propagation basis, no CP- violation appears in the
projection and further convertions.  $P(\nu_e \rightarrow \nu_e)$ does
not depend on $\delta$ (see Figs.~\ref{CPnorm} and \ref{CPinv} top
panels).  $P(\nu_e \rightarrow \nu_\beta)$ ($\beta = \mu, \tau$) do
depend on $\delta$ via $|U_{\beta i} |^2$ (i = 1,2) (see
Eq.~(\ref{prob-nonadmu})) via projection of the mass states onto the
flavor state in the detector. Furthermore, for maximal 2-3 mixing with
change of $\delta$ the $(\nu_e \rightarrow \nu_\mu)$ probability is
transformed into $(\nu_e \rightarrow \nu_\tau)$ and {\it vice versa},
when $\delta$ increases from $0$ to $\pi$, and at $\delta = \pi/2$
$$
P(\nu_e \rightarrow \nu_\mu) = P(\nu_e \rightarrow \nu_\tau). 
$$
\item 
{\it $\nu_\mu \rightarrow \nu_\beta$ channels:} now according to
(\ref{toprop})
\be
\nu_\mu = c_{23} \nu_\mu^{\prime} + 
s_{23}e^{i\delta} \nu_\tau^{\prime}, 
\label{pres-mu}
\ee
and the transition $\nu_\mu \rightarrow \nu_i$ has two channels via
$\nu_\mu^{\prime}$ and $\nu_\tau^{\prime}$ which can interfere.  So,
as follows from (\ref{pres-mu}) the probability equals
\be
P_*(\nu_\mu \rightarrow \nu_i) = 
\left|c_{23} A_*(\nu_\mu^{\prime} \rightarrow \nu_i) +     
s_{23}e^{i\delta} A_*(\nu_\tau^{\prime} \rightarrow \nu_i)
\right|^2. 
\label{cp-int}
\ee
The interference term depends on $\delta$.  As we saw before in the
adiabatic case $A_*(\nu_\mu^{\prime} \rightarrow \nu_i) = \delta_{i1}$
and $A_*(\nu_\tau^{\prime} \rightarrow \nu_i) = (0, -s_{13}^0,
c_{13}^0 )$ (see Eq.~(\ref{uprimeint})).
\end{enumerate}

\FIGURE{\epsfig{file=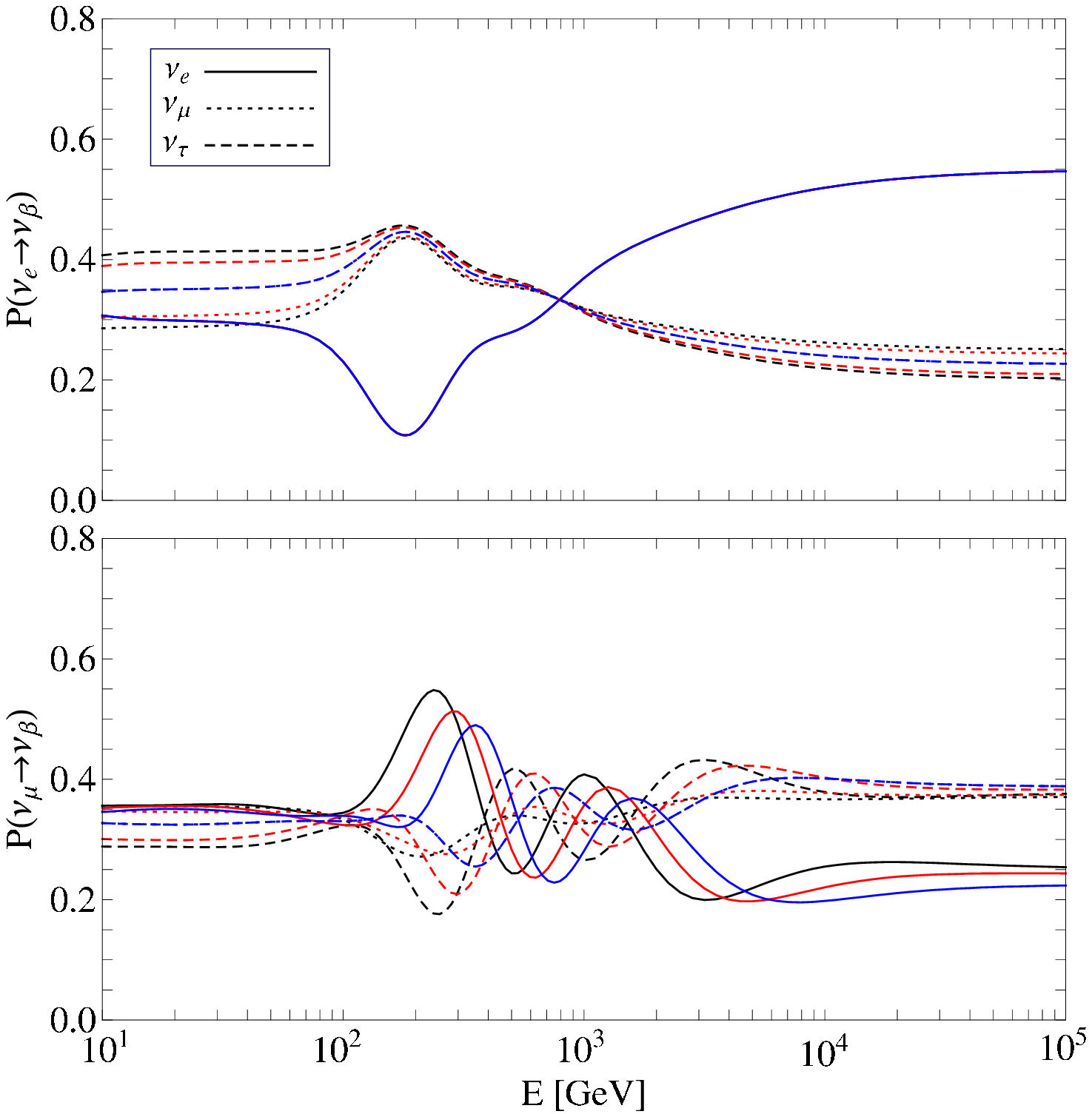,width=10cm}
\caption{The probabilities as functions of energies for different
values of the CP-phase $\delta_{cp} = 0$ (black curves), $\pi/4$ (red
curves), and $\pi/2$ (blue curves).  We take normal mass hierarchy and
$\sin^2 2\theta_{13} = 0.08$.  }
\label{CPnorm}}

Let us consider the interference in various energy ranges.  In the
intermediate energy range, according to (\ref{uprimeint})
$\nu_\mu^{\prime} \approx - \nu_{1m}$ and $\nu_\tau^{\prime} \approx -
s_{13} \nu_{2m} + c_{13} \nu_{3m}$. Since in this range $\nu_{1m}$
evolves adiabatically to $\nu_{1}$ and the latter is orthogonal to the
rest of the state, no interference appear between $\nu_\mu^{\prime}-$
and $\nu_\tau^{\prime}-$ channels in (\ref{cp-int}), and consequently,
$$
P_*(\nu_\mu \rightarrow \nu_i) =
\left|c_{23} A_*(\nu_\mu^{\prime} \rightarrow \nu_i)\right|^2 +
\left|s_{23} A_*(\nu_\tau^{\prime} \rightarrow \nu_i)
\right|^2.
$$
This is in accordance with results (\ref{adprob}).  So, the
interference and dependence on $\delta$ require adiabaticity
violation, which is realized above the H-resonance. For $E \gtrsim
E_R^H$ we have $s_{13}^0 \approx 1$, $\nu_\mu^{\prime} \approx -
\nu_{1m}$, and $\nu_\tau^{\prime} \approx - \nu_{2m}$. Therefore
$$
P_*(\nu_\mu \rightarrow \nu_i) =
\left|c_{23} A_*(\nu_{1m} \rightarrow \nu_i) +
s_{23} e^{i\delta} A_*(\nu_{2m} \rightarrow \nu_i)
\right|^2, 
$$
and due to adiabaticity violation, both amplitudes in this equation
are non zero simultaneously for $i = 1$ and 2. The phase $\delta$
appears in the interference term in combination with the oscillation
phase and therefore with change of $\delta$ the wiggles above the
H-resonance shift (see Fig.~\ref{CPnorm}).  It might be practically
impossible to disentangle the effect of $\delta$ from $\phi_L$.
Similar consideration holds for $\nu_{\tau}$.

\FIGURE{\epsfig{file=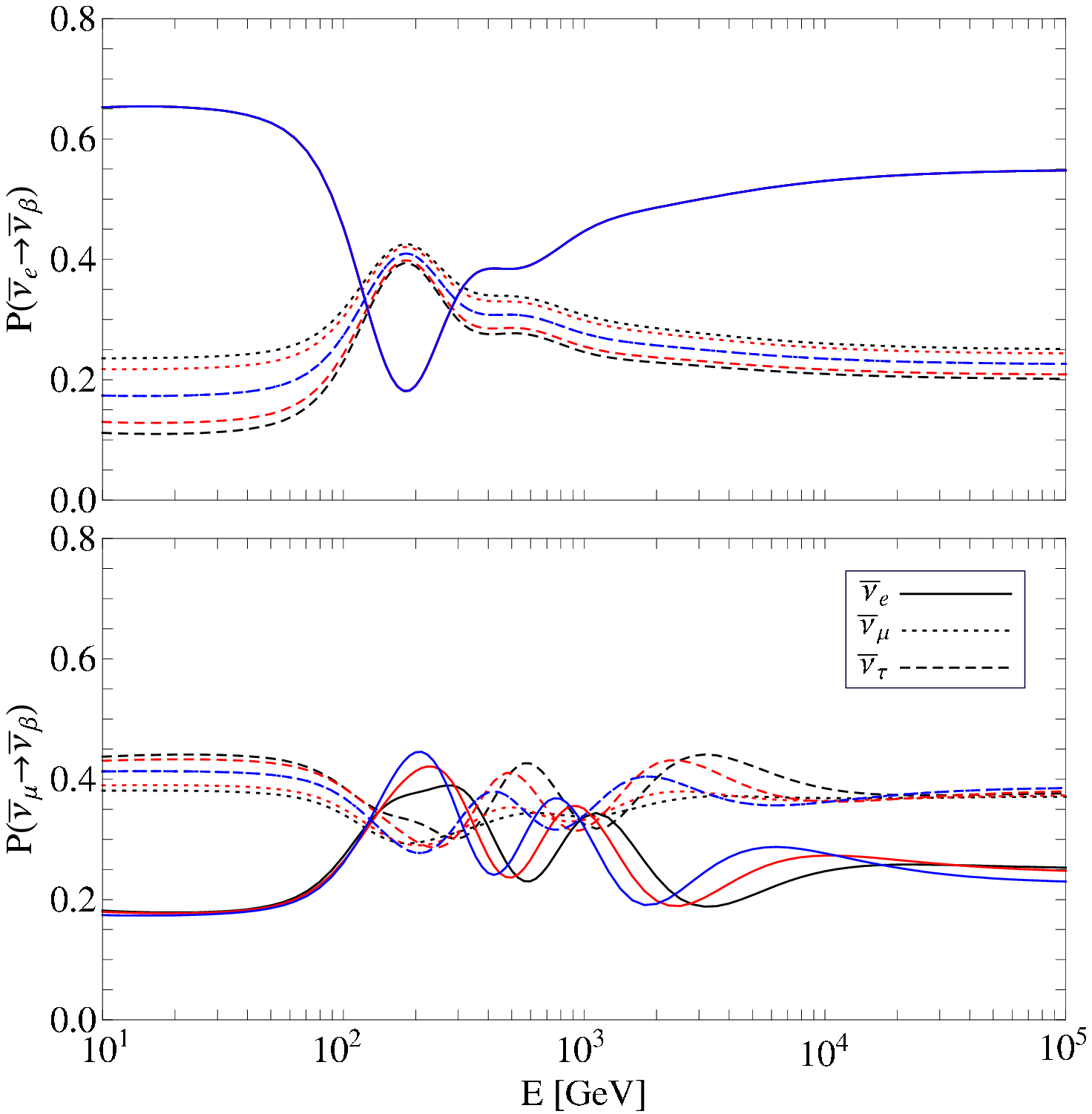,width=10cm}
\caption{The same as in Fig.~\ref{CPnorm} for antineutrinos and the
case of inverted mass hieararchy.  }
\label{CPinv}}

Thus, in adiabatic energy range the dependence of the flavor
probabilities on $\delta$ follows from the projection of mass states
onto the flavor states in the detector only. The projection of the
original flavor state onto the eigenstate of propagation gives an
additional dependence on $\delta$ only if the adiabaticity is broken
inside the star.  In this connection let us take into account
oscillations inside jets.  From Eq.~(\ref{sss-tot}) we have
\be
A_*(\nu_e \rightarrow \nu_i) =
A_{env} (\nu_e \rightarrow \nu_i)  + 
\eta s_{13} \left[
s_{13} A_{env} (\nu_e \rightarrow \nu_i) + 
c_{13} A_{env} (\nu_\tau^\prime \rightarrow \nu_i)  
\right], 
\label{nue-pr2}
\ee
where the second term is the jet effect. No dependence on the CP-phase
appears here.  For the $\nu_\mu \rightarrow \nu_i$ transitions,
according to (\ref{sss-tot}),
\bea
A_*(\nu_\mu \rightarrow \nu_i) & = &
c_{23} A_{env} (\nu_\mu^\prime \rightarrow \nu_i) +
s_{23} e^{i \delta} A_{env} (\nu_\tau^\prime \rightarrow \nu_i) + 
\nonumber\\
 & + & s_{23} c_{13}  e^{i \delta} \eta 
\left[s_{13} A_{env} (\nu_e \rightarrow \nu_i) +
c_{13} A_{env} (\nu_\tau^\prime \rightarrow \nu_i) \right], 
\label{numu-pr3}
\eea

where the same combination of the amplitudes as in (\ref{nue-pr2})
enters the correction term.  For $\eta = 0$ (no oscillations inside
jet) this expression is reduced to the one in Eq.~(\ref{cp-int}). The
dependence on the CP-phase here is explicit. All other factors and
amplitudes do not depend on $\delta$.

Let us consider the antineutrino channels.  General form of the
amplitudes is the same as for neutrinos with the following changes:
$\delta \rightarrow - \delta$, and $\theta_{ij}^m \rightarrow
\bar{\theta}_{ij}^m$ which corresponds to change $V \rightarrow - V$.
The amplitudes of oscillations inside jet are unchanged; they are
essentially the $2\nu-$ oscillations in vacuum in our approach.

\section{Fluxes at the Earth}

\subsection{Flavor ratios}

We compute the fluxes at the Earth separately for neutrinos and
antineutrinos keeping in mind that in future experiments the signs of
charged leptons may, in principle, be determined.  The fluxes at the
Earth equal:
$$
\Phi_{\nu_\alpha} = \Phi_{\nu_\mu}^0 P(\nu_\mu \rightarrow \nu_\alpha)
+ \Phi_{\nu_e}^0 P(\nu_e \rightarrow \nu_\alpha)
= \Phi_{\nu_\mu}^0 [ P(\nu_\mu \rightarrow \nu_\alpha)
+ \epsilon P(\nu_e \rightarrow \nu_\alpha)],  
$$
where $\alpha = e, \mu, \tau$, and $\epsilon$ is given by
Eq.~(\ref{nu_e_to_nu_mu}).  Similar expressions will be used for
antineutrinos with substitution $\epsilon \rightarrow \bar{\epsilon}$.
Let us introduce the flavor ratios as
$$
r_{\alpha/\mu} \equiv \frac{\Phi_{\nu_\alpha}}{\Phi_{\nu_\mu}}
= \frac{P(\nu_\mu \rightarrow \nu_\alpha)
+ \epsilon P(\nu_e \rightarrow \nu_\alpha)}{P(\nu_\mu \rightarrow \nu_\mu)
+ \epsilon P(\nu_e \rightarrow \nu_\mu)},  
$$ 
and for antineutrinos  
$$
r_{\bar{\alpha}/\mu} \equiv \frac{\Phi_{{\bar \nu}_\alpha}}{\Phi_{\nu_\mu}}
= \frac{\Phi_{\bar{\nu}_\mu}^0}{\Phi_{\nu_\mu}^0}
~\frac{P(\bar{\nu}_\mu \rightarrow \bar{\nu}_\alpha)
+ \bar{\epsilon} 
P(\bar{\nu}_e \rightarrow \bar{\nu}_\alpha)}{P(\nu_\mu \rightarrow \nu_\mu) + 
\epsilon P(\nu_e \rightarrow \nu_\mu)},   
$$ 
normalizing all the fluxes at the Earth to the $\nu_\mu-$flux at the
Earth.

According to Fig.~\ref{fluxes} the ratio of original fluxes decrease
with energy as
\be
\frac{\Phi_{\bar{\nu}_\mu}^0}{\Phi_{\nu_\mu}^0} = 
\cases{
1 ~;~~~ E \lesssim 10 ~{\rm GeV} \cr
0.5 ~;~ E \gtrsim 300 ~{\rm GeV}\,.}
\label{anti-fl}
\ee
The ratios contain complete information relevant for observations.
Properties of the flavor ratios can be easily undertstood from
properties of probabilities.  If the charge of lepton produced by a
neutrino is not defined, and thus, the experiment sums up signals of
neutrinos and antineutrinos, observables are determined by the ratio:
$$
r_{(\alpha + \bar{\alpha})/\mu} \equiv 
\frac{\Phi_{\nu_\alpha} + \xi_\alpha \Phi_{\bar{\nu}_\alpha}}{\Phi_{\nu_\mu} 
+ \xi_\mu \Phi_{\bar{\nu}_\mu}} = 
\frac{r_{\alpha/\mu} + \xi_\alpha r_{\bar{\alpha}/\mu}}{1 + \xi_\mu 
r_{\bar{\mu}/\mu}},  
$$ 
where $\xi_\alpha$ describe ratios of antineutrino and neutrino
cross-sections and corresponding efficiencies of detection.  Note
that contribution of antineutrinos is suppressed by about 0.2 - 0.3
due to smaller cross-section ($\xi_\alpha \sim 1/2$) and smaller
original flux (another factor 1/2).

At energies below $10^{5}$ GeV, $\nu_e$ and $\nu_\tau$ can not be
distinguished: both produce the showering events.  The double-bang
events which are signatures of the $\nu_{\tau}-$interaction can be
identified at much higher energies~\cite{double-bang}.  Thus, with
existing detectors one can study the ratio of the showering and
tracking events (with muon track in the final state) $r_{sh/tr}$. In
the first approximation it is determined by the ratio of the sum of
$\nu_e-$ and $\nu_\tau-$ fluxes to the $\nu_\mu-$ fluxes:
\be
r_{sh/tr} =  
\frac{\Phi_{\nu_e} + \xi_e \Phi_{\bar{\nu}_e} + 
\Phi_{\nu_\tau} + \xi_\tau \Phi_{\bar{\nu}_\tau}}{\Phi_{\nu_\mu} 
+ \xi_\mu \Phi_{\bar{\nu}_\mu}} 
= 
\frac{r_{e/\mu} + \xi_e r_{\bar{e}/\mu} + r_{{\tau}/\mu} + 
\xi_\tau r_{\bar{\tau}/\mu} }{1 + \xi_\mu r_{\bar{\mu}/\mu}}.  
\label{sh-trr}
\ee 
This expression should be corrected (even neglecting misidentification
of events): (i) certain part of $\nu_\tau-$flux contributes via
transitions: $\nu_\tau \rightarrow \tau \rightarrow \mu$ to the
traking events~(see e.g. \cite{tau-mu}); (ii) neutral current interactions (not
affected by oscillations) contribute to the showering events.

To a good approximation, for $E > 100$~GeV, the charged current
cross-sections are equal for all three flavors and therefore we can
sum up the $\nu_e$ and $\nu_\tau$ fluxes at the detector.  (Note
that still one should subtract the channel $\nu_{\tau} \rightarrow
\tau \rightarrow \mu$ which leads to the tracking event.)  Therefore
if the initial flux is composed of $\nu_\mu$, the total flux which
produces showering events is determined by the probability
$$
P_{sh} =  \sum_i P_*(\nu_\mu \rightarrow \nu_i) (|U_{ei}|^2 + 
(1 - b_\mu)|U_{\tau i}|^2), 
$$
where $b_\mu \sim 0.2$ is the fraction of $\tau \rightarrow \mu$
decays.  Using unitarity we find
$$
P_{sh} \equiv \sum_i P_*(\nu_\mu \rightarrow \nu_i) 
(1 - |U_{\mu i}|^2) = 1 - \sum_i P_*(\nu_\mu \rightarrow \nu_i)|U_{\mu 
i}|^2  - b_\mu \sum_i P_*(\nu_\mu \rightarrow \nu_i) |U_{\tau i}|^2,  
$$
and the latter equals $P_{sh} = 1 - P_{tot}(\nu_\mu \rightarrow
\nu_\mu) - b_\mu \sum_i P_*(\nu_\mu \rightarrow \nu_i) |U_{\tau
i}|^2$.  If $\nu_{\mu}$ dominates in the initial state,
$P_{tot}(\nu_\mu \rightarrow \nu_\mu)$ determines the probabilities of
all observable events when $\tau \rightarrow \mu$ decays are
neglected.  In the presence of the original $\nu_e-$flux we have
$$
P_{sh} \equiv 1 - P(\nu_\mu \rightarrow \nu_\mu) 
+\epsilon [1 -  P(\nu_e \rightarrow \nu_\mu)]. 
$$

\subsection{Properties of the flavor ratios}

In Figs.~\ref{fr-13} - \ref{fr-n0} we show the flavor ratios
$r_{\alpha/\mu}$ and $r_{\bar{\alpha}/\mu}$ ($\alpha = e, \mu, \tau$)
as functions of neutrino energies, values of neutrino parameters and
density profiles.  For illustration we use two extreme original flavor
contents: (i) $\epsilon : 1 : 0$, where $\epsilon = \epsilon (E)$ and
according to Fig.~\ref{fluxes} it decreases from 0.5 at $E \lesssim
10$ GeV down to $10^{-2}$ at high energies.  This content would
corespond to stong suppression of muon decays in medium of jet without
secondary acceleration.  (ii) $1 : 2 : 0$ which corresponds to
$\epsilon = 0.5$ and does not depend on energy.  This could be
realized when additional acceleration of secondary muons occurs in
jets.

Consider first $\epsilon : 1 : 0$. For very small $\epsilon$ we have
$$
r_{\alpha/\mu}  \approx  \frac{\Phi_{\nu_\alpha}}{\Phi_{\nu_\mu}} 
\approx \frac{P(\nu_\mu \rightarrow \nu_\alpha)}
{P(\nu_\mu \rightarrow \nu_\mu)}~. 
$$
Furthermore, since $P(\nu_\mu \rightarrow \nu_\mu)$ has weak
dependence on energy, the ratio essentially repeats behavior of
$P(\nu_\mu \rightarrow \nu_\alpha)$. According to Fig.~\ref{fr-13}
(upper panel), and in agreement with our analytic considerations, the
ratio $r_{e/\mu}$ has an asymptotic value $\sim 0.7$ and the deviation
of this value from the flavor equilibration is due to the original
flavor content.  In the intermediate region (plateau), due to equality
$P(\nu_\mu \rightarrow \nu_e) \approx P(\nu_\mu \rightarrow \nu_\mu)$
for maximal 2-3 mixing we have $r_{e/\mu} \approx 1$.  The 1-3 peak
(at $E = 200$ GeV) is slightly enhanced in comparison to the peak in
probability.

In the plateau for maximal 2-3 mixing we have $P(\nu_\mu\to\nu_\tau) <
P(\nu_\mu\to\nu_\mu)$, and consequently, $r_{\tau/\mu} \sim 0.8$. The
wiggles of $P(\nu_\mu\to\nu_\tau)$ and $r_{\tau/\mu}$ have an opposit
phase as compared with wiggles of $P(\nu_\mu \to \nu_e)$ and
$r_{e/\mu}$, and the amplitude of wiggles of showers/tracks ratio is
suppressed.  For maximal 2-3 mixing the asymptotic value equals
$r_{\tau/\mu} \approx 1$.  Due to opposit phase the $r_{\tau/\mu}$ and
$r_{e/\mu}$ wiggles compensate each other in the ratio $r_{sh/tr}$,
leading to substantially smaller amplitude of $r_{sh/tr}$ wiggles (see
Fig.~\ref{sh-trk}).

\FIGURE{\epsfig{file=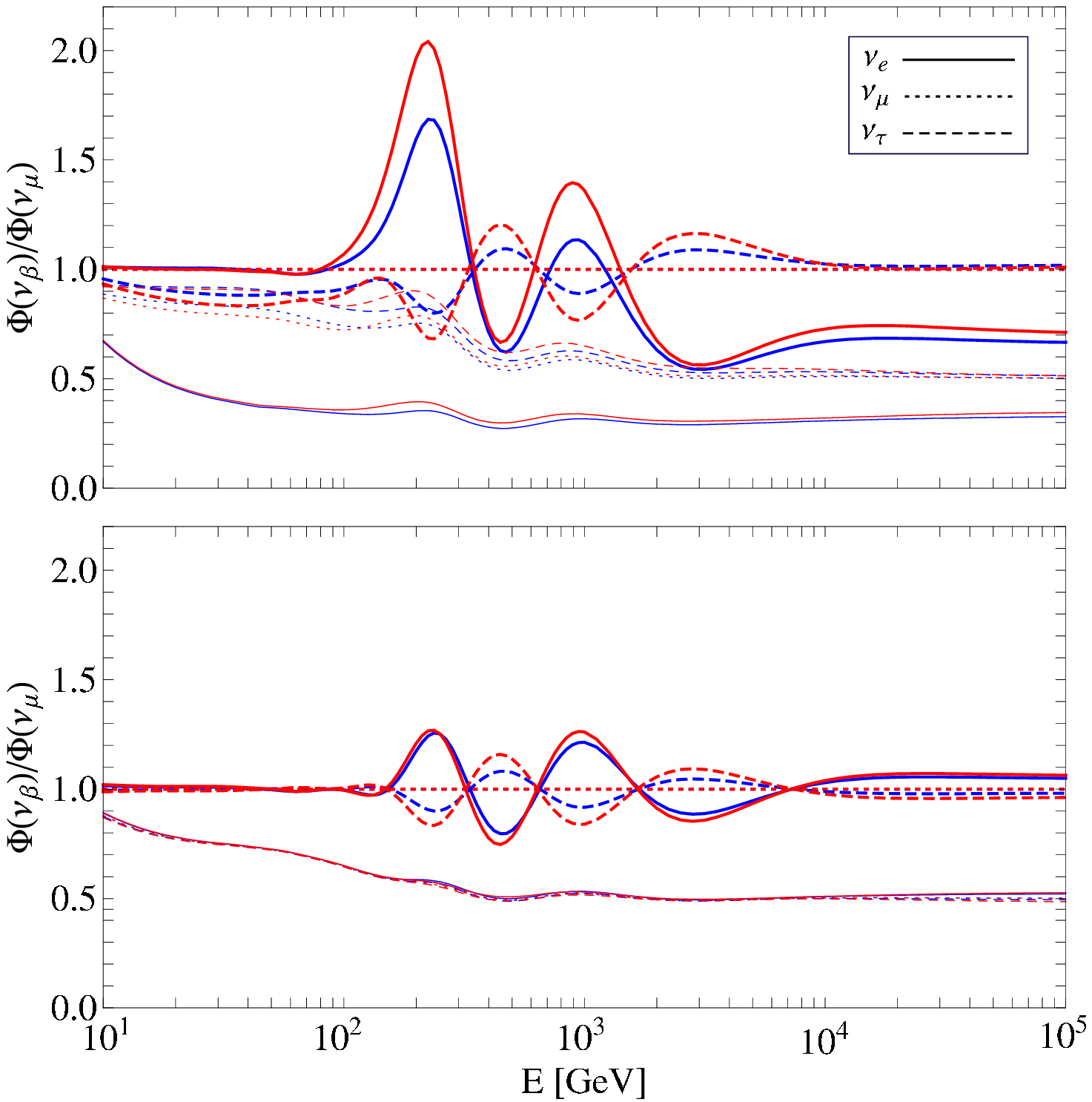,width=10cm}
\caption{ Neutrino and antineutrino flavor ratios (fluxes normalized
to the $\nu_\mu$ flux) as functions of the neutrino energy for two
different values of 1-3 mixing, $\sin^2 2\theta_{13} = 0.04$ (blue
lines) and 0.08 (red lines) and two different original flavor
contents: $\eps : 1 : 0$ (upper panel) and $1 : 2 : 0$ (lower panel).
The thick (thin) lines correspond to the neutrino (antineutrino)
fluxes.  We used profile A, $\sin^2 \theta_{23} = 0.5$, $\delta_{cp} =
0$ and normal mass hierarchy. }
\label{fr-13}}

\FIGURE{\epsfig{file=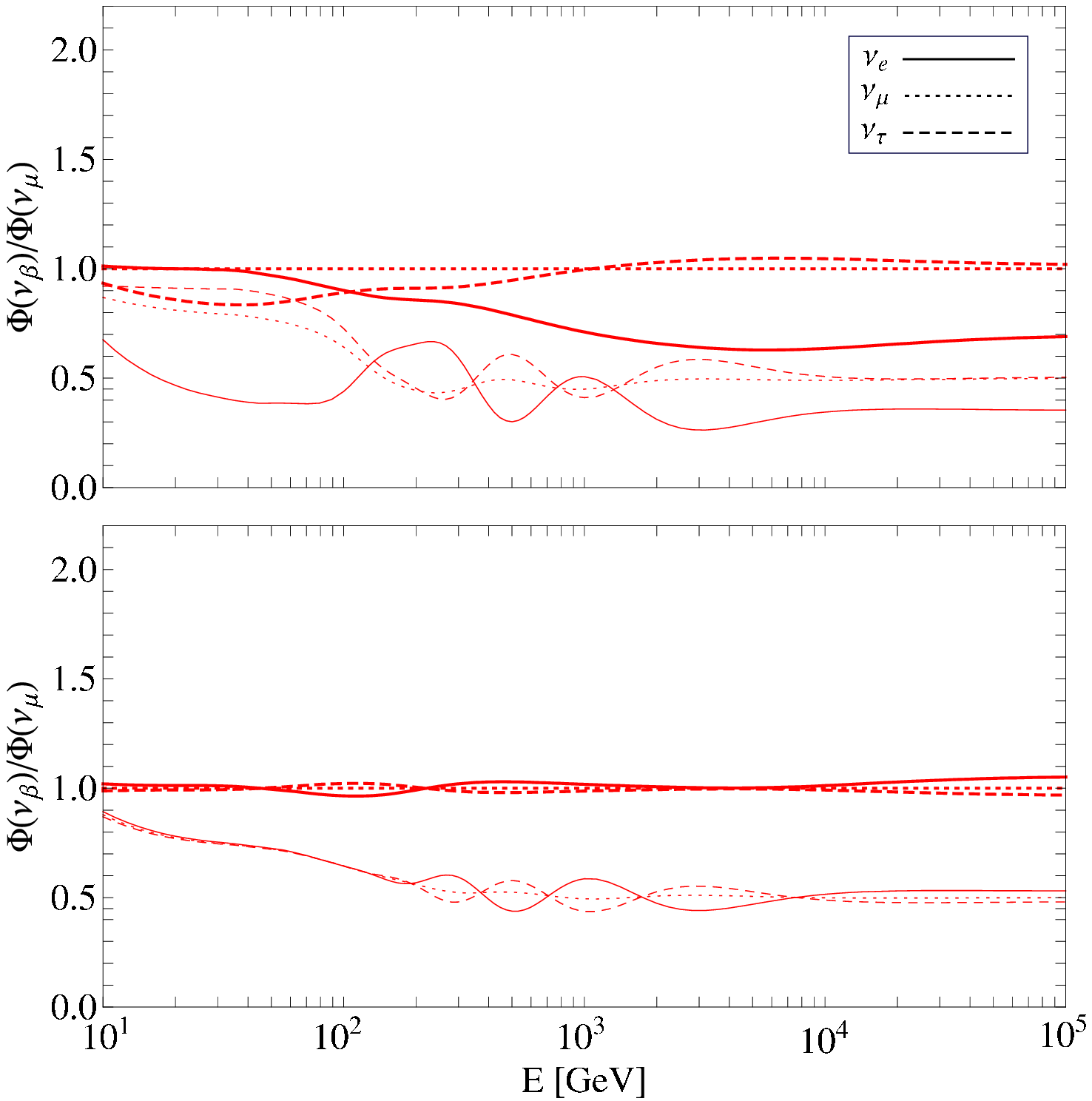,width=10cm}
\caption{The same as in Fig.~\ref{fr-13} for inverted mass
hierarchy and only for $\sin^2 2\theta_{13} = 0.08$. }
\label{fr-ih}}

For anineutrinos (see Fig.~\ref{fr-13}) the ratio 
$$
r_{\bar{\alpha}/\mu}  \approx  
\frac{\Phi^0_{\bar{\nu}_\mu}}{\Phi^0_{\nu_\mu}} \cdot 
\frac{P(\bar{\nu}_\mu \rightarrow \bar{\nu}_\alpha)}
{P(\nu_\mu \rightarrow \nu_\mu)} 
$$
follows to a large extent the ratio of original fluxes in
(\ref{anti-fl}) in the case of NH. An additional distortion due to
oscillations is rather small.  In particular, difference of high and
low energy asymptotics in flavor ratios is related to difference of
ratio $\Phi^0_{\bar{\nu}_\mu}/\Phi^0_{\nu_\mu}$ at high and low
energies.

For the original flavor content $(1:2:0)$ (bottom panels) the
$\nu_e-$flux and its transformations substantially change the flavor
ratios at the Earth.  In the intermediate range, still $r_{e/\mu}
\approx 1$ as a consequence of the equality $P(\nu_e \to \nu_e)
\approx P(\nu_e \to \nu_\mu)$ for maximal 2-3 mixing. At and above the
energy of H-resonance the probability $P(\nu_e \to \nu_e)$ has a dip,
whereas $P(\nu_e \to \nu_\mu)$ has a peak and above the peak/dip they
have wiggles of the opposit sign. Therefore the wiggles of ratio
$r_{e/\mu} $ are strongly attenuated. This ratio deviates from 1
(equilibration) which is the signature of the matter effect only in
the region $(10^2 - 10^4)$ GeV and maximal effect is about $30\%$.
Similarly, for the ratio $r_{\tau/\mu}$ substantial attenuation
occurs.  Moreover, the wiggles of $r_{e/\mu} $ and $r_{\tau/\mu}$ have
opposite phase, and therefore the wiggles in the $r_{sh/tr}$ are
further suppressed (see Fig.~\ref{sh-trk}).

\FIGURE{\epsfig{file=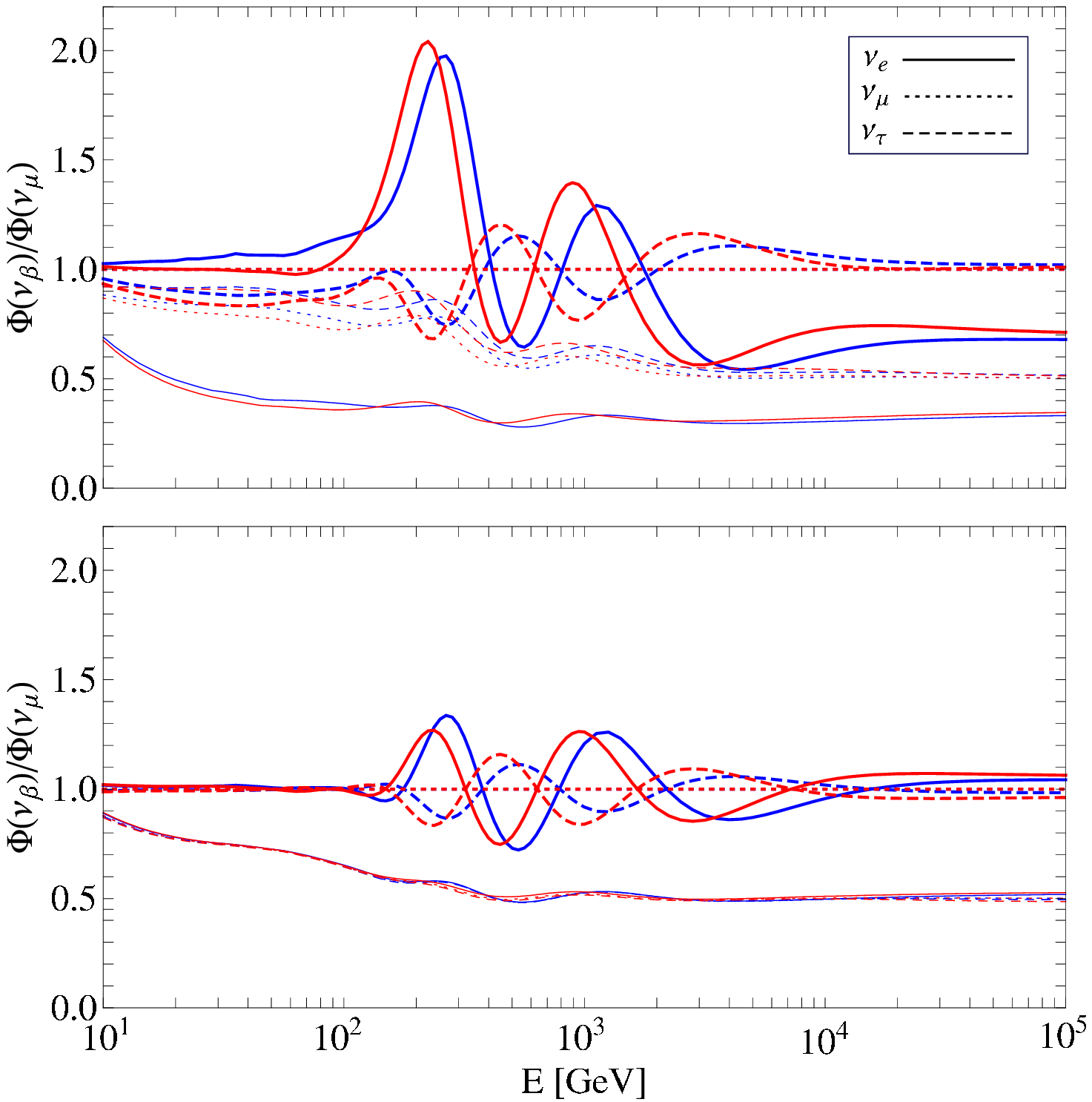,width=10cm}
\caption{Neutrino and antineutrino flavor ratios as functions of the
neutrino energy for two different values of the CP phase $\delta$: 0
(red lines) and $\pi/4$ (blue lines) and two different original flavor
contents: $\eps : 1 : 0$ (upper panel) and $1 : 2 : 0$ (lower panel).
The thick (thin) lines correspond to the neutrino (antineutrino)
fluxes.  We used profile A, $\sin^2 \theta_{23} = 0.5$, $\sin^2
\theta_{13} = 0.08$ and normal mass hierarchy. }
\label{fr-delta}}

As follows from Fig.~\ref{fr-13}, the highest sensitivity to
$\theta_{13}$ is in the range of H-resonance peak and wiggles: $E =
(10^2 - 10^3)$ GeV.  (Recall that in the $\nu_\mu-$channels the peak
is also due to resonance enhancement of the 1-3 mixing.)  With
decrease of $\theta_{13}$ adiabaticity is broken stronger and the
amplitude of wiggles becomes smaller.

Let us consider the case of inverted mass hierarchy (see
Fig.~\ref{fr-ih}).  For the $(\epsilon:1 : 0)$ original content, as in
the case of NH the asymptotics is $r_{e/\mu} \approx 0.7$ and in the
plateau $r_{e/\mu} \approx 1$. In the H-resonance region there is no
peak, the ratio decreases with increase of energy from 1 to the
asymptotic value. This dependence is modulated by small wiggles.  The
peak appears in the H-resonance region in the antineutrino ratio
$r_{\bar{e}/\mu}$.

For the original flavor ratio $(1:2:0)$, due to strong compensations
of contributions from different channels the matter effect on the
flavor ratios is small.
   
Let us consider dependence of the flavor ratios on other neutrino
parameters.

\noindent
1.  The phase of wiggles changes with $\delta$: wiggles shift
(see Fig.~\ref{fr-delta}).  Changes in plateau as well as in the
asymptotics are rather weak.

\noindent
2. Dependence of flavor ratios, as functions of energy, on the 2-3
mixing is strong (see Fig.~\ref{fr-23}).  The strongest dependence is
for $r_{e/\mu} $ (NH) in the plateau region: as we already mentioned,
for maximal 2-3 mixing the probabilities of $(\nu_\mu \to \nu_e)$ and
$(\nu_\mu \to \nu_\mu)$ transitions coincide.  However with change of
$\theta_{23}$ they change in the opposite way: one increases and
another decreases, as can be seen in Fig.~\ref{fr-23}. The $\nu_\mu
\to \nu_\tau$ probability changes much weaker.  Note that asymptotic
values also vary rather substantially, whereas the change in the 1-3
peak is small.  Thus, determination of $r_{e/\mu} $ in the plateau
region would be the most sensitive to search for deviation of the 2-3
mixing from maximal.  For the original flavor content $(1:2:0)$ the
dependence on $\theta_{23}$ is weaker.

\FIGURE{\epsfig{file=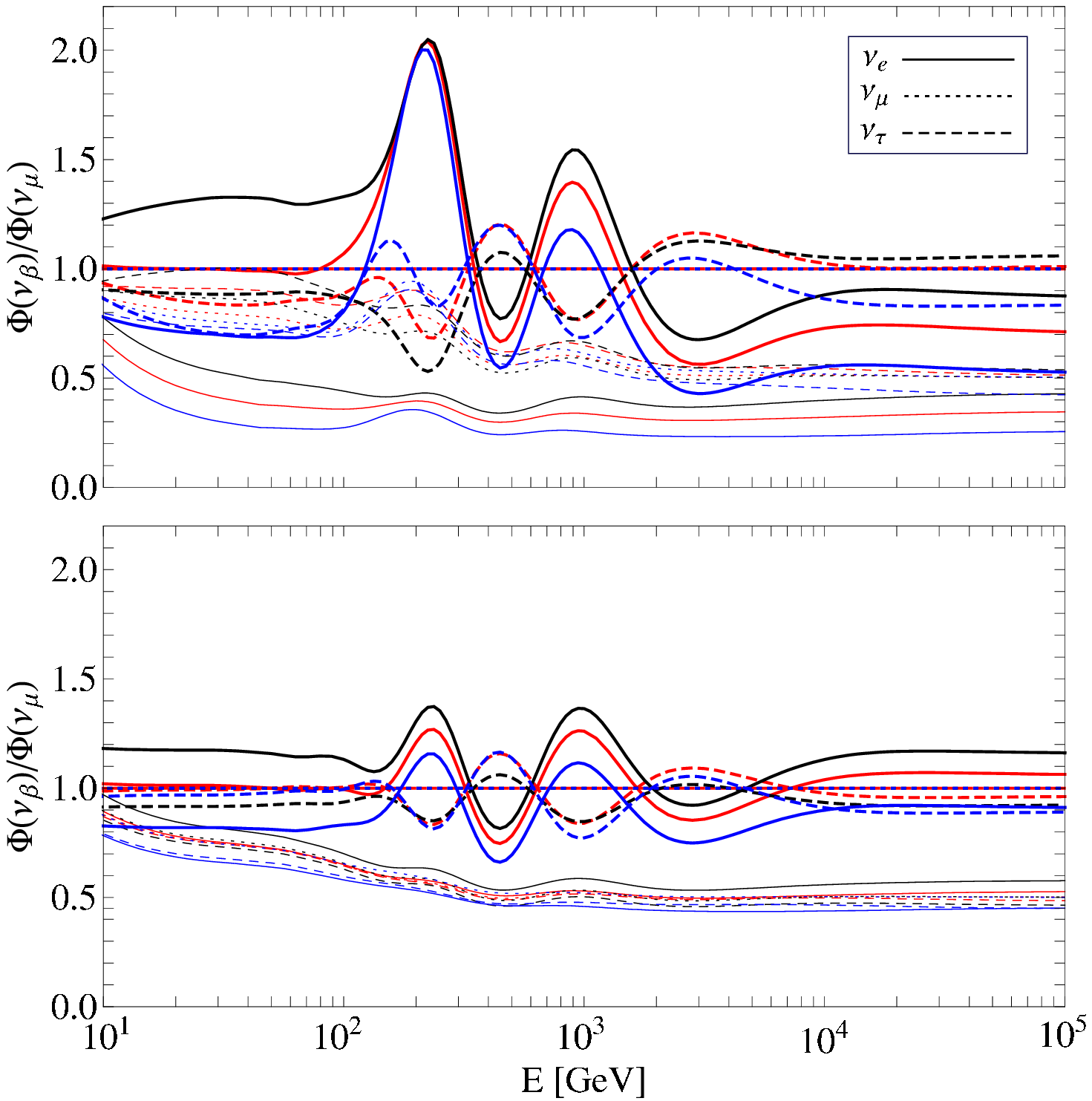,width=10cm}
\caption{ Neutrino and antineutrino flavor ratios as functions of the
neutrino energy for different values of $\sin^2 \theta_{23}$: 0.4
(black lines), 0.5 (red lines), 0.6 (blue lines), and two different
original flavor contents: $\eps : 1 : 0$ (upper panel) and $1 : 2 : 0$
(lower panel).  The thick (thin) lines correspond to the neutrino
(antineutrino) fluxes.  We used profile A, $\sin^2 2\theta_{13} =
0.08$, $\delta_{cp} = 0$ and normal mass hierarchy. }
\label{fr-23}}

Consider dependence of the flavor ratios on the density profile.  With
decrease of density gradient, $k$, (see Fig.~\ref{fr-ab}) the
evolution becomes more adiabatic, the 1-3 peak shifts to smaller
energies and becomes wider; the amplitude of wiggles decreases, the
region of wiggles extends to higher energies.

\FIGURE{\epsfig{file=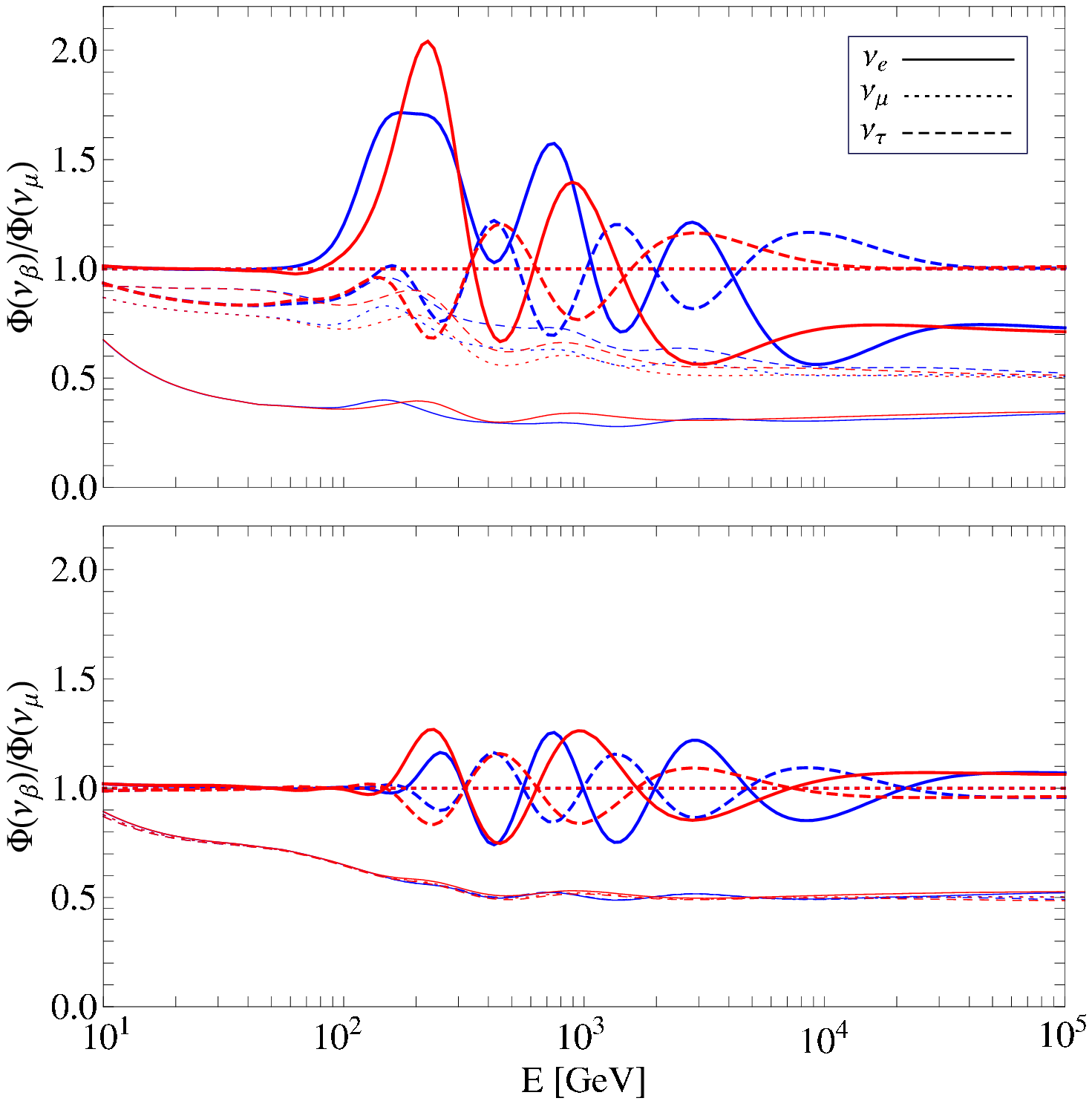,width=10cm}
\caption{ Neutrino and antineutrino flavor ratios as functions of the
neutrino energy for different density profiles: A (red lines) B (blue
lines) and two different original flavor contents: $\eps : 1 : 0$
(upper panel) and $1 : 2 : 0$ (lower panel).  The thick (thin) lines
correspond to the neutrino (antineutrino) fluxes.  We used $\sin^2
2\theta_{13} = 0.08$, $\delta_{cp} = 0$ and normal mass hierarchy. }
\label{fr-ab}}

For the original flavor content $(1:2:0)$, there is substantial
cancellation of contributions from the $\nu_\mu \to \nu_e$ and $\nu_e
\to \nu_e$ transitions, since the transition and the survival
probabilities have opposite dependence on energy.

\FIGURE{\epsfig{file=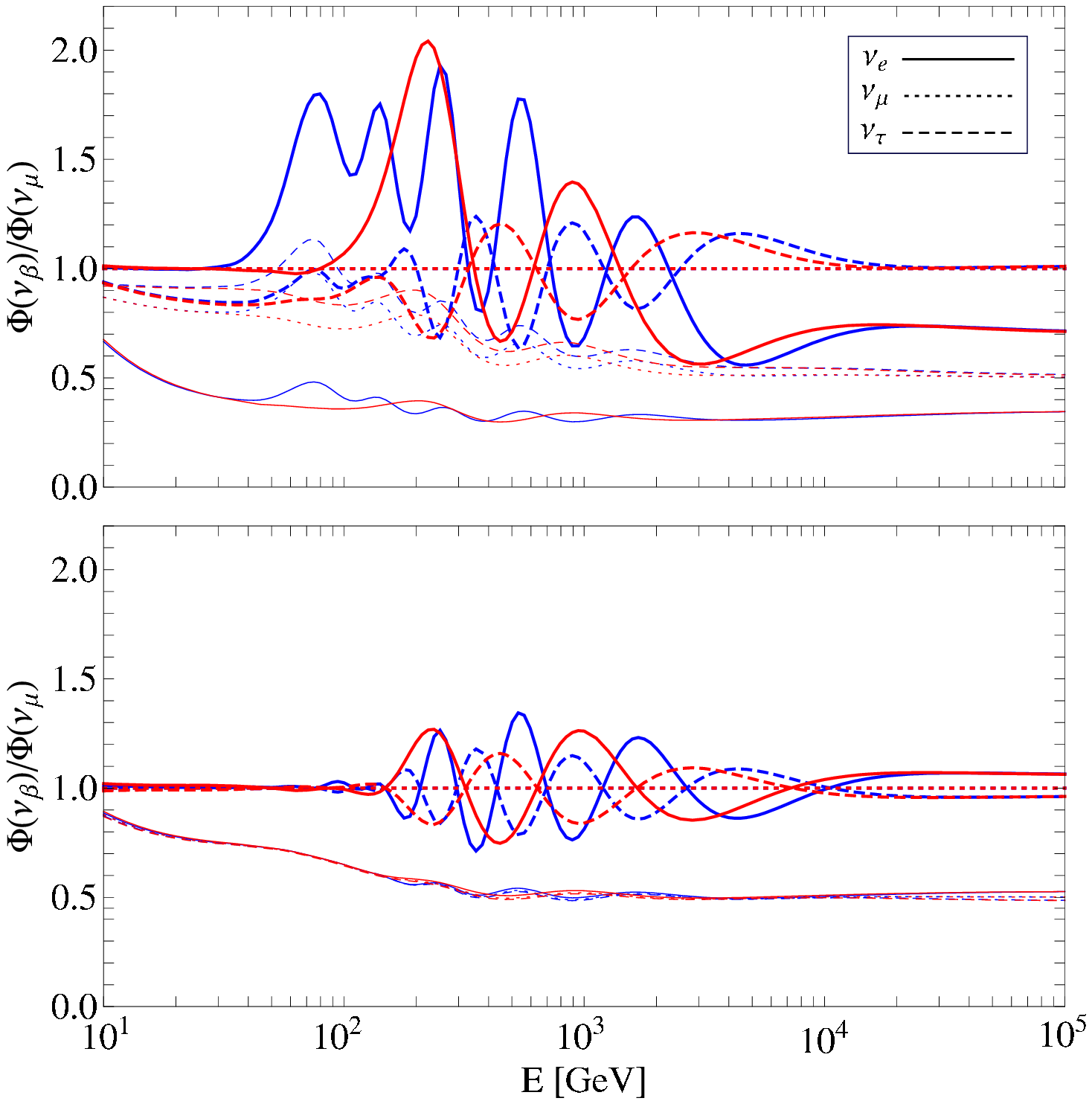,width=10cm}
\caption{ Neutrino and antineutrino flavor ratios as functions of the
neutrino energy for the profile A with different inner number
densities of electrons $n_0 = 10^{23}$ cm$^{-1}$ (red lines) and $n_0
= 2\cdot 10^{23}$ cm$^{-1}$ (blue lines) and two different original
flavor contents: $\eps : 1 : 0$ (upper panel) and $1 : 2 : 0$ (lower
panel).  The thick (thin) lines correspond to the neutrino
(antineutrino) fluxes.  We used $\sin^2 2\theta_{13} = 0.08$,
$\delta_{cp} = 0$ and normal mass hierarchy. }
\label{fr-n0}}

Dependence of the ratios on the initial density $n_0$ is strong (see
Fig.~\ref{fr-n0}): with increase of $n_0$ the resonance peak shifts to
lower energies, the period of wiggles decreases and number of wiggles
increases; the amplitudes of wiggles become slightly smaller.  In the
case of $(1:2:0)$ original content, there is strong cancellation of
contributions of $(\nu_\mu \to \nu_e)$ and $(\nu_e \to \nu_e)$
especially in the region of 1-3 peak. As a consequence, the region of
wiggles starts from about 200 GeV.

In general, in the case of original $(\epsilon:1:0)$ content,
dependences of the flavor ratios on the neutrino parameter as well as
on parameters of the star are much stronger, and this makes hidden
jets to be more prospective than the observable jets as far as
investigation of neutrino parameters is concerned.

In Fig.~\ref{sh-trk} we show the ratio of shower-to-track events on
the neutrino energy defined in Eq. (\ref{sh-trr}).  Qualitatively,
this dependence repeats the corresponding dependence of the flavor
ratio $r_{e/\mu}$ with L-wiggles and asymmetric asymptotics at high
and low energies.  However relative size of the wiggles, and in
general, matter effects (deviations from the averaged oscillation
result) is further suppressed.  As we mentioned before, this
suppression is due to contribution of the $\nu_{\mu} \rightarrow
\nu_{\tau}$ transition, $r_{\tau/\mu}$, as well as the contributions
of antineutrinos which have small matter effect for the normal mass
hierarchy.  According to Fig.~\ref{sh-trk}, the structures in
$r_{sh/tr}$ due to conversion in the matter of the star are up to $40
- 50 \%$. This should be compared with factor of (2 - 2.5) effect in
the ratio $r_{e/\mu}$.  The effect is smaller for the inverted mass
hierarchy. Substantial deviations from the averaged VO result are in
the energy range $(10^2 - 2 \cdot 10^{3})$ GeV which can be extended
to $(30 - 5 \cdot 10^{3})$ GeV for smaller density gradient or larger
density in the inner part of the envelope (bottom panel).  The size of
the matter effect is much smaller: about $\pm 5\%$ in the case of
initial ratio flavor ratio ($1:2:0$).

\FIGURE{\epsfig{file=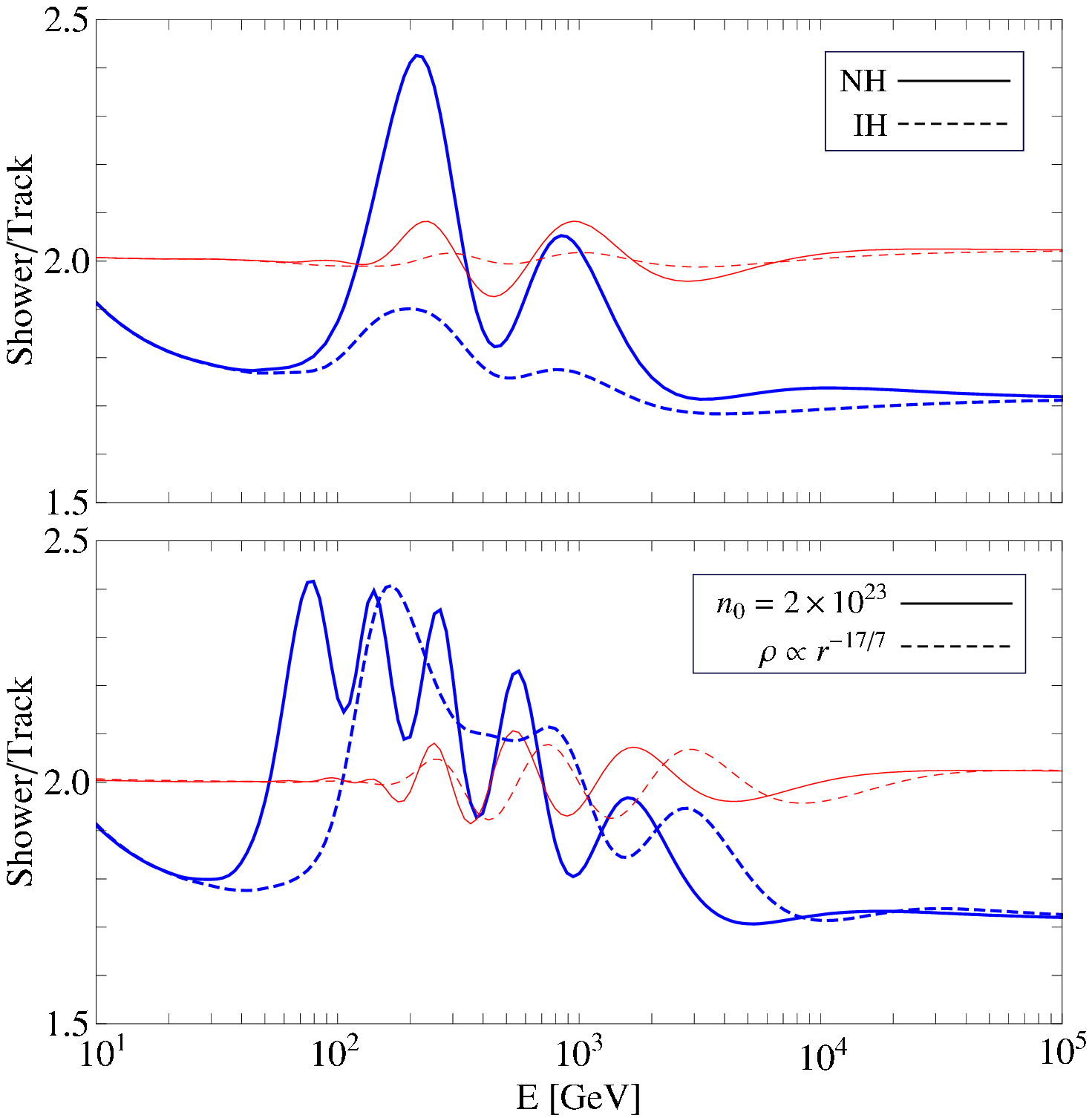,width=10cm}
\caption{Ratio of the shower-to-track events as the functions of the
neutrino energy for two different initial flavor content:
$\epsilon:1:0$ (blue lines) and 1:2:0 (red lines) upper panel: for
normal (solid) and inverted (dashed); we used profile A; lower panel:
for $n_0 = 2\times 10^{23}$ cm$^{-1}$ (solid lines) and for smaller
gradient $k = 17/7$ (dashed lines).  We used $\sin^2 2\theta_{13} =
0.08$, $\delta_{cp} = 0$.}
\label{sh-trk}}

\section{Discussion and conclusion}

Hidden jets realize unique set up (configuration and conditions) in
which at one hand protons can be accelerated up to high energies and
then produce high energy neutrinos, and at the other hand these
neutrinos cross large enough column density of matter on the way from
production region so that flavor conversion in the matter of the star
is important.  We present detailed and comprehensive study of flavor
conversion of neutrinos from hidden sources (jets) - astrophysical
sources invisible in non-thermal $\gamma$ rays or X rays.  Our results
differ from the results of paper \cite{Mena:2006eq}.

\noindent
1. The conversion is affected by matter of a star and the Earth in the
energy interval $\sim (1 - 10^{5})$ GeV.  In this range the
probabilities and flavor ratios substantially deviate from
probabilities and ratios given by the averaged vacuum oscillations.
The borders of this interval are determined by the largest density in
the envelope, $E_L = E_L(n_0)$, and gradient of density in the
envelope $\sim R_\star \Delta m_{31}^2 / 4\pi$ rather than on
$R_\star$, the radius of the stellar envelope.  The interval expands
for larger initial density and smaller gradient.
 For energies below this interval matter effects can be neglected and 
with decrease of energy the probability converges to the 
averaged VO probability. Above the interval the adiabaticity is strongly 
broken and flavor evolution inside a star can be 
neglected. With increase of energy the probability converges to the 
averaged VO probability again.  

\noindent 
2. We discused in details physics involved in the production of
neutrinos.  Neutrinos are produced in $\pi-$, $\mu-$ and $K-$decays in
strong magnetic fields which leads to very short wave packets of
neutrinos.  This, in turn, results in quick separation of the wave
packets and loss of coherence, in spite of high neutrino energy. The
coherence length is comparable with radius of star. The coherence is
not restored in a detector due to finite energy resolution and large
baselines.  Therefore the overal evolution consists of the
flavor-to-mass states transitions inside the star, loss of coherence,
propagation of the nmmass staes in vacuum; so that incoherent fluxes
of mass states arrive at the surface of the Earth.  Detector projects
these mass states back onto flavor states incoherently.

Inelastic scattering becomes important for high energies $\sim 10^4$
GeV. To a good approximation this scattering is flavor invariant, and
therefore factors out and does not influence the flavor conversion
picture. It can be included in determination of the original neutrino
spectra.  We have estimated that some effect of $\nu - \nu$ scattering
may show up at high energies where usual flavor conversion effect
becomes small.

\noindent
3. Mainly the neutrino flavor change is due to adiabatic and partially
adiabatic conversion in the envelope of a star.  Oscillations inside
the Earth lead to an additional distortion of probabilities below 10
GeV.  We show that probabilities as functions of neutrino energy have
several generic features which are well controlled by the density
profile crossed by neutrinos (initial density, density gradient).
This includes
\begin{itemize}
\item
Plateau in the intermediate energy range which is due to adiabatic
conversion in the 1-2 resonance region.  The Earth matter effect
produces an oscillatory dip in the plateau at $E < 10$ GeV.
\item
The dips or peaks (depending on channel) related to the 1-3 resonance
at $E \gtrsim E_H$.
\item  
Non-adiabatic edge of the energy profile is modulated by wiggles.  The
wiggles are new dynamical feature which is not realized in other
objects.  The wiggles are manifestations of interference induced by
adiabaticity violation in the H-resonance (H-wiggles) and in
L-resonance (L-wiggles).
\end{itemize}

Different features of the energy profile of the matter effect such as
low energy border, plateau, the 1-3 peak/dip (its position and size),
wiggles and asymptotics depend on parameters of neutrino and star
differently, so that one will be able to disentangle these dependences
in principle.  Position of these features depends on density profile
of the star: in particular, on the density at the border between jet
and envelope ($n_0$) and gradient of density in an envelope ($k$).
The size of the matter effects depends on values of neutrino
parameters as well as on original flavor content.  In the adiabatic
range there is no dependence on the profile (apart from lower border
of the regions). Therefore plateau as well as the 1-3 peak (for large
enough $\theta_{13}$ or small gradient) do not depend on a
profile. Above the H-resonance in the adiabaticity violation range
(the wiggles region) characteristics of probabilities depend both on
profile and on neutrino parameters. In particular, number of wiggles
is a measure of the density gradient in the envelope.

\noindent
4. At the Earth one can, in principle, determine the flavor ratios
$r_{e/\mu}$, $r_{\tau/\mu}$, $r_{ \bar{e}/\mu}$, $r_{\bar{\tau}/\mu}$,
$r_{\bar{\mu}/\mu}$ as functions of neutrino energy.  To a large
extent these ratios reproduce the energy dependence of probabilities.
Independent measurements of these ratios would give very rich
information on neutrino parameters, density profile of star and
original flavor ratios which encode information about conditions
inside jets. Effect of neutrino flavor conversion on these ratios is
strong, modifying the ratios in certain energy ranges by factor 2 --
3.  With present experimental techniques and detectors, however, one
can determine the ratio of shower-to-track numbers of events only,
without separation of neutrinos and antineutrinos. Sum of
contributions from neutrinos and antineutrios as well as from $\nu_e$
and $\nu_{\tau}$ (which have an opposite phase) lead to significant
damping of the observable conversion effect in $r_{sh/tr}$. The effect
depends strongly on the original flavor content. It can reach 30 -- 40
\% for the content $\epsilon : 1 : 0 $ ($\epsilon \ll 1$) and it is
about $5 -- 10 \%$ for $1 : 2 : 0$. In reality one can expect some
intermediate situation.

\noindent
5. Observability of the neutrino signal from hidden jets is beyond the
scope of this paper, and here we just add some comments. There are two
issues related to observability: (i) number of event in a given
detector, and (ii) a possibility to identify signal and extract it
from a background produced by the atmospheric neutrinos. These issue
have been discussed partly in \cite{Mena:2006eq}.  Detection of signal
from individual source would require large detector and/or a source in
a nearby galaxy such as M82 and NGC253.  Signatures include high
energy neutrino events from certain directions during $\sim 10$ sec.\
which can be repeated.  There should be a correlation between the
jetted neutrino signal ($E \gtrsim 1$~GeV) with burst of low energy
thermal neutrinos ($E \sim 5$ -- 50~MeV) generated during the core
collapse.  Thermal 10--50 MeV neutrinos are also produced in a jet
since the shell density and temperature are initially very high at the
base of the jet.  No significant delay is expected between the thermal
and high-energy neutrinos that are produced in the jet.  There can be
significant delay between the emission of core-collapse thermal
neutrinos and jet thermal/high-energy neutrinos depending on whether
the core collapses directly to a black hole or via an intermediate
neutron star stage. The thermal neutrino emission from core-collapse
preceding the high-energy (and more thermal) neutrino emission and the
time-delay between the two can be used as a probe to learn about the
core-collapse and jet formation processes.

In case of a successful supernova, with or without jet break out,
optical lightcurve can be extrapolated back to the time of explosion
and to check temporal correlation with any neutrino signal detected
from the same direction in the sky.  This can largely reduce the
atmospheric neutrino and muon background.  A neutrino detector setup
outside the Earth's atmosphere, e.g., in space or on the Moon, as the
technology develops, can of course greatly improve the detection
prospect of astrophysical neutrinos.

It is expected that parameters of jets change during jet duration
which leads to time dependence of neutrino signals we have
discussed. In principal, it opens unique possibility to monitor
evolution of jets with neutrinos.

The neutrino oscillation signatures from hidden jets are modified in
the case of isotropic diffuse flux. To identify this flux one can use
distortion of the energy spectrum in the energy range where
atmospheric neutrinos are not affected by oscillations.  In general if
the diffuse flux is dominated by hidden sources as discussed here, the
measured flavor ratio will differ significantly from the usual
($1:1:1$) ratio expected from the optically thin sources in the GeV --
TeV range.

\section*{Acknowledgements}

We thank Cecilia Lunardini for helpful comments on the manuscript.  SR
would like to thank the Abdus Salam International Centre for
Theoretical Physics for warm hospitality where part of this work was
completed.

\appendix

\section{Meson and lepton energy losses}

High energy $\pi$ and $K$ are subject to energy losses due to hadronic
$\pi p$ and $K p$ interactions with cross-sections $\sigma_{\pi
p/Kp}\approx (2/3)\sigma_{pp}$, and due to synchrotron radiation in
the strong magnetic field before they decay.  The cooling time scales
due to these processes are
$$
t^\p_{had} (E^\p_{\pi (K)})\approx 
\frac{E^\p_{\pi (K)}}{n^\p_p \sigma_{\pi (K)p} 
\Delta E^\p_{\pi (K)} }
\sim 10^{-5} ~{\rm s},
$$
for hadronic interactions, where $\Delta E^\p_{\pi (K)} \approx 0.5
E^\p_{\pi (K)}$ is the average inelasticity and we ignored the
logarithmic factor in cross-section, and
$$
t^\p_{em} (E^\p_j) \approx \left( \frac{E^\p_j}{\rm GeV} \right)^{-1}
\left( \frac{B^\p}{10^9~\rm G} \right)^{-2} \times
\cases{
0.34~{\rm s}\, ;~~~~~~~~~ j=K \cr
2.2\times 10^{-3} ~{\rm s}\, ;\, j=\pi \cr
7.0\times 10^{-4} ~{\rm s}\, ;\, j=\mu ~. 
}
$$
for the synchrotron cooling.  As can be seen, from these equations,
the meson energy losses are initially dominated by hadronic and later
by electromagnetic process~\cite{Ando05,Razzaque05}.  With a decay
time scale $t^\p_{j,dec} = \tau_j (E^\p_j/m_j)$, where $\tau_j$ is the
mean lifetime, $\pi$ and $K$ decay without significant energy losses
below $E^\p_{K, b1} \sim 400~{\rm GeV}$ and $E^\p_{\pi, b1} \sim
50~{\rm GeV}$ (which corresponds to $t^\p_{dec} \approx t^\p_{had}$).
The electromagnetic losses become important at energies above
$E^\p_{K,b2} \sim 35~{\rm TeV}$ and $E^\p_{\pi, b2} \sim 220~{\rm
GeV}$ (which corresponds to $t^\p_{em} \approx t^\p_{had}$).  Thus we
can define a suppression factor to be multiplied with the production
fluxes of $\pi$ and $K$ before they decay as
$$ 
\zeta_j (E^\p_{j}) \approx \cases{
1 \,;~~~~~~~~~~~~~~~~~~~~~~~~~~~~~~~~~~~~~~~~~~~~~~~~~~ 
E^\p_{j} \lesssim E^\p_{j,b1} \cr
t^\p_{j,had}/t^\p_{j,dec} \approx (E^\p_j/E^\p_{j,b1})^{-1} \,;\, 
~~~~~~~~~~~~~~~~E^\p_{j,b1} \lesssim E^\p_{j} \lesssim E^\p_{j,b2} \cr
t^\p_{j,em}/t^\p_{j,dec} \approx (E^\p_{j,b1}/E^\p_{j,b2}) 
(E^\p_j/E^\p_{j,b2})^{-2}
\,;~ E^\p_{j} \gtrsim E^\p_{j,b2} ~. 
}
$$ 
Note that muons from $\pi$- or $K$-decays are also subject to
synchrotron energy losses which dominate above $E^\p_{\mu,b} \approx
6$~GeV and the $\mu-$flux before decay will be suppressed by a factor
$\sim (E^\p_\mu/E^\p_{\mu,b})^{-2}$.  Thus, the $\nu$'s from
$\mu-$decay channels contribute little to the total flux at high
energy.  Also, the contributions to the $\nu$ fluxes from $\pi p-$ and
$Kp-$interactions is small and we ignore those.

\section{Comparison of the conversion formulas in
Ref.~\cite{Mena:2006eq} and in this paper}

General formula (\ref{ab-general}) for the conversion probabilities
differ from that used in the paper \cite{Mena:2006eq}. Consequently,
results of this paper differ from the results in \cite{Mena:2006eq}.
According to \cite{Mena:2006eq} the flavor probability equals
\be
P(\nu_\alpha \rightarrow \nu_\beta) =
\sum_\gamma P_* (\nu_\alpha \rightarrow \nu_\gamma) 
\bar{P}_V (\nu_\gamma \rightarrow \nu_\beta),
\label{mena1}
\ee
where $\nu_\gamma$ is a flavor neutrino state, and $\bar{P}_V
(\nu_\gamma \rightarrow \nu_\beta)$ is the averaged oscillation
probability in vacuum.  Taking explicit expression for the latter we
can rewrite the probability (\ref{mena1}) as
\be
P(\nu_\alpha \rightarrow \nu_\beta) =
\sum_\gamma P_* (\nu_\alpha \rightarrow \nu_\gamma)
\sum_i |U_{\beta i}|^2 |U_{\gamma i}|^2 
\label{mena2}
\ee
or 
\be
P(\nu_\alpha \rightarrow \nu_\beta) =  \sum_i |U_{\beta i}|^2 
\sum_\gamma |A_* (\nu_\alpha \rightarrow \nu_\gamma)|^2 |U_{\gamma i}|^2,  
\label{mena3}
\ee
where $A_* (\nu_\alpha \rightarrow \nu_\gamma)$ is the amplitude of
probability of the corresponding flavor transition.  On the other hand,
expression for the probability (\ref{ab-general}) we use in this paper
can be rewritten as
\be
P(\nu_\alpha \rightarrow \nu_\beta) = \sum_i |U_{\beta i}|^2
\left|\sum_\gamma A_* (\nu_\alpha \rightarrow \nu_\gamma) U_{\gamma 
i}\right|^2,
\label{our}
\ee
which clearly differs from (\ref{mena3}).  The difference is that
instead of the flavor states $\nu_\gamma$ at the exit from the star,
we take the mass states $\nu_i$ which are the eigenstates of
propagation in vacuum and these states lose coherence.  In
(\ref{mena3}) the averaging was taken for oscillations outside the
star, and oscillations of mass states are not averaged inside the
star. Therefore Eq.~(\ref{mena3}) leads to fast oscillatory picture
which corresponds to oscillations inside the envelope in the interval
from $r_{jet} - R_\star$. In our consideration the wiggles also
appear.  However they appear in the adiabatisity violation range only,
i.e. above the H-resonance. There is no wiggles in the adiabatic part.
Furthermore, the period of wiggles in the energy scale is much larger:
the phase of wiggles is determined by the oscillation phase on the way
from the production point to the H- resonance region (for H-wiggles)
and to the L-resonance (for the L-wiggles) which is much smaller than
total size of the envelope $r_{jet} - R_\star$.

\section{The Earth matter effect}

The mass-to-flavor transition probabilities inside the Earth can be
written as
\be 
P_E (\nu_i \rightarrow \nu_\beta) = 
|(U_{23} \Gamma_\delta S_E U^{\prime})_{\beta i}|^2 = 
|(U_{23} \Gamma_\delta S_E U_{13} U_{12})_{\beta i}|^2, 
\label{Pearth}
\ee
where according to (\ref{uprime}) $U^{\prime}$ gives projection of the
mass states onto the states of the propagation basis, $S_E$ is the
S-matrix for transitions inside the Earth in the propagation basis,
and $U_{23} \Gamma_\delta$ projects the propagation basis onto the
flavor basis.  (If $S_E = I$, the probability (\ref{Pearth}) is
reduced to $|(U_{PMNS})_{\beta i}|^2$.)

Let us find explicit expression for $S_E$ for $E \gtrsim 1$ GeV where
substantial neutrino flux from jets exists.  For the Earth densities
this energy range is far above the L-resonance and the 1-2 mixing is
strongly suppressed.  So, the problem is reduced to
$2\nu-$oscillations due to the 1-3 mixing.  For definiteness we will
take normal mass hierarchy and consider for simplicity the evolution
in the approximation of constant density. In this case
\be
S_E = U_m^{\prime} D U_m^{\prime \dagger}, 
\label{se}
\ee
where 
\be
D = diag(e^{-i\phi_{12}}, 1, e^{-i\phi_{32}}).  
\label{dia}
\ee
Here $\phi_{12} \equiv \phi_1 - \phi_2$, $\phi_{32} \equiv \phi_3 -
\phi_2$, and $\phi_i = H_i t$ are the oscillation phases of the
eigenstates in matter $\nu_{im}$ ($H_i$ are the eigenvalues in
matter).  The mixing matrix in matter in the propagation basis,
$U_m^{\prime}$, for $E \gg E_L$ equals, according to (\ref{uprimem}),
$$
U^\prime_m  \approx \pmatrix{
0   &  c_{13}^m  &  s_{13}^m \cr
- 1   &  0  &  0 \cr
0   &   - s_{13}^m    & c_{13}^m
}.
$$
Using this expression for $U^\prime_m$ and (\ref{dia}) we find from
(\ref{se}) the evolution matrix in the Earth:
\be
S_E  \approx \pmatrix{
A_{ee}  &  0                   &  A_{e\tau} \cr
0       &  - e^{-i\phi_{12}}   &  0 \cr
A_{e\tau}   &   0    & A_{\tau \tau}
},
\label{se-exp}
\ee
where 
\be
A_{ee} = c_{13}^{m 2} + s_{13}^{m 2} e^{-i\phi_{32}},~~~
A_{e \tau} =  s_{13}^{m} c_{13}^{m} (e^{-i\phi_{32}} - 1),~~~
A_{\tau \tau} =  s_{13}^{m 2} + c_{13}^{m 2} e^{-i\phi_{32}}. 
\ee
Inserting (\ref{se-exp}) into (\ref{Pearth}) we obtain the matrix of
flavor-to-mass transition probabilities.  In particular, we have
\bea
P(\nu_1 \rightarrow \nu_e) & = & 
|c_{13} c_{12} A_{ee} - s_{13} c_{12} A_{e\tau}|^2, 
\nonumber\\
P(\nu_2 \rightarrow \nu_e) & = & |c_{13} s_{12} A_{ee} - 
s_{13} s_{12} A_{e\tau}|^2, 
\nonumber\\
P(\nu_3 \rightarrow \nu_e) & = & |s_{13} A_{ee} + c_{13} A_{e\tau}|^2.  
\eea
Then for the flavor probability $\nu_e \rightarrow \nu_e$ with all
transitions included we obtain using (\ref{tot-m}) and
(\ref{adprobb}):
\be
P(\nu_e \rightarrow \nu_e) = c_{13}^2 s_{12}^2 
|c_{13}  A_{ee} - s_{13} A_{e\tau}|^2
+ s_{13}^2 |s_{13} A_{ee} + c_{13} A_{e\tau}|^2. 
\label{eenn}
\ee
This is reduced to the probability without oscillations inside the
Earth (\ref{ee-norm}) for $A_{ee} = 1$ and $A_{e\tau} = 0$.  In the
first approximation in $s_{13}$ we have from (\ref{eenn})
\be
P(\nu_e \rightarrow \nu_e) \approx s_{12}^2 |A_{ee}|^2 \approx 
s_{12}^2 \left(1 - \sin^2 2 \theta_{13}^m \sin^2 
\frac{\phi_{23}}{2} \right). 
\label{moduli}
\ee

According to this result the plateau is modulated by resonance
oscillations: the oscillatory dip appears in the range $E \sim
E^R_{13} \sim 6$ GeV. Typical width of the dip is given by $\Delta E
\sim 2E_{13}^R \tan 2 \theta_{13}$.  So, for $\sin 2 \theta_{13} =
0.08$ the modulations are in the range $(4 - 8)$ GeV.  Maximal depth
depends on the zenith angle of the neutrino trajectory.  The result
(\ref{moduli}) differs from the Earth matter effect on the solar
neutrinos where both plateau and the Earth effect are due to 1-2
mixing and the Earth matter effect enhances the survival probability.
Here the effect is opposite: the probability becomes smaller.  The
difference stems from the differentce of density profile of the Sun
and an envelope of a star. In particular, density in the central part
of the Sun, where neutrinos are produced are much higher than maximal
density in an envelope.  As a result, in the case of jetted neutrinos,
the Earth matter effect due to 1-2 mixing is at very low energies ($<
0.2$ GeV) - essentially in the low asymptotic region - below $E_L$.
The Earth matter effect in the 1-2 mixing plateau is due to 1-3
mixing. This (to some extend) is similar to the 1-3 dip at $E \gtrsim
10^2$ GeV.

Detection of the Earth matter effect could give an information about
neutrino properties and also about direction to the star.  It is not
clear though, if this effect can be ever extracted from the
atmospheric neutrino background.



\begin{thebibliography}{99}

\bibitem{nu_astro} F.~Halzen, arXiv:0910.0436v1
[astro-ph.HE]; V.~Berezinsky, in Proc. 4th Int. Workshop ``Neutrino
Oscillations in Venice'', ed. Milla Baldo Ceolin, p. 137,
arXiv:0901.1428 [astro-ph]; E.~Waxman, Science, {\bf 315}, 63, (2007).

\bibitem{agn_nu} F.~W.~Stecker, C.~Done, M.~H.~Salamon and P.~Sommers,
Phys.\ Rev.\ Lett.\ {\bf 66}, 2697 (1991); Erratum-ibid.\ {\bf 69},
2738 (1992); L.~Nellen, K.~Mannheim and P.~L.~Biermann, Phys.\ Rev.\ D
{\bf 47}, 5270 (1993); A.~P.~Szabo and R.~J.~Protheroe, Astropart.\
Phys.\ {\bf 2}, 375 (1994); A.~Atoyan and C.~D.~Dermer, Phys.\ Rev.\
Lett.\ {\bf 87}, 221102 (2001); J.~Alvarez-Muniz and P.~Meszaros,
Phys.\ Rev.\ D {\bf 70}, 123001 (2004).

\bibitem{grb_nu} E.~Waxman and J.~N.~Bahcall, Phys.\ Rev.\ Lett.\ {\bf
78}, 2292 (1997); A.~Atoyan and C.~D.~Dermer, Phys.\ Rev.\ Lett.\ {\bf
91} 071102 (2003); S.~Razzaque, P.~Meszaros and E.~Waxman, Phys.\
Rev.\ D {\bf 69}, 023001 (2004); K.~Murase, K.~Ioka, S.~Nagataki,
T.~Nakamura, Astrophys.\ J.\ {\bf 651}, L5 (2006); N.~Gupta and
B.~Zhang, Astropart.\ Phys.\ {\bf 27}, 386 (2007).

\bibitem{sn_nu} E.~Waxman and A.~Loeb, Phys.\ Rev.\ Lett.\ {\bf 87},
071101 (2001); X.-Y.~Wang, S.~Razzaque, P.~Meszaros and Z.-G.~Dai,
Phys.\ Rev.\ D {\bf 76}, 083009 (2007).

\bibitem{snr_nu} J.~Alvarez-Muniz and F.~Halzen, Astrophys.\ J.\ {\bf
576}, L33 (2002); M.~L.~Costantini and F.~Vissani, Astropart.\ Phys.\
{\bf 23}, 477 (2005).

\bibitem{Razzaque04} S.~Razzaque, P.~M\'esz\'aros, and E.~Waxman,
Phys.\ Rev.\ Lett.\ {\bf 93}, 181101 (2004); {\bf 94}, 109903(E)
(2005).

\bibitem{hidden_jet} D.~C.~Leonard, A.~V.~Filippenko, A.~J.~Barth and
T.~Matheson, Astrophys.\ J.\ {\bf 536}, 239 (2000); L.~Wang,
D.~A.~Howell, P.~Hoflich and J.~C.~Wheeler, Astrophys.\ J.\ {\bf 550},
1030 (2001); J.~Granot and E.~Ramirez-Ruiz, Astrophys.\ J.\ {\bf 609},
L9 (2004).

\bibitem{Ando05} S.~Ando and J.~F.~Beacom, Phys.\ Rev.\ Lett.\ {\bf
95}, 061103 (2005).

\bibitem{Razzaque05} S.~Razzaque, P.~M\'esz\'aros, and E.~Waxman,
Mod.\ Phys. Lett. A, {\bf 20}, 2351 (2005).

\bibitem{Ando05b} S.~Ando, J.~F.~Beacom and H.~Yuksel, Phys.\ Rev.\
Lett.\ {\bf 95}, 171101 (2005).

\bibitem{Fermi_starburst} Fermi LAT Collaboration (A.~A.~Abdo, et
al.), arXiv:0911.5327v1 [astro-ph.HE].

\bibitem{IceCube} J.~Ahrens et al.\, Astropart.\ Phys.\ {\bf 20}, 507
(2004).

\bibitem{ANTARES} ANTARES Collaboration (J.~A.~Aguilar et al.),
Astropart.\ Phys.\ {\bf 26}, 314 (2006).

\bibitem{KM3NeT} U.~F.~Katz, Presented at 2nd VLVNT Workshop on Very
Large Neutrino Telescope (VLVNT2), Catania, Italy, 8-11 Nov 2005.
Published in Nucl.\ Instrum.\ Meth.\ A {\bf 567}, 457 (2006).

\bibitem{Kowalski07} M.~Kowalski and A.~Mohr, Astropart.\ Phys.\ {\bf
27}, 533 (2007).

\bibitem{Mena:2006eq}
  O.~Mena, I.~Mocioiu and S.~Razzaque,
  Phys.\ Rev.\  D {\bf 75}, 063003 (2007)
  [arXiv:astro-ph/0612325].

\bibitem{Luna}
  C.~Lunardini and A.~Y.~Smirnov,
  Nucl.\ Phys.\  B {\bf 583}, 260 (2000)
  [arXiv:hep-ph/0002152].

\bibitem{Farzan}
  Y.~Farzan and A.~Y.~Smirnov,
  Nucl.\ Phys.\  B {\bf 805}, 356 (2008)
  [arXiv:0803.0495 [hep-ph]].

\bibitem{MSW}
L. Wolfenstein, Phys. Rev. D {\bf 17} (1978) 2369;
L. Wolfenstein,   in ``Neutrino-78'', Purdue Univ. C3, (1978);
S.~P.~Mikheev and A.~Y.~Smirnov,
 Sov.\ J.\ Nucl.\ Phys.\  {\bf 42}, 913 (1985)
 [Yad.\ Fiz.\  {\bf 42}, 1441 (1985)];
S.~P.~Mikheev and A.~Y.~Smirnov,
 Sov.\ Phys.\ JETP {\bf 64}, 4 (1986)
 [Zh.\ Eksp.\ Teor.\ Fiz.\  {\bf 91}, 7 (1986)]
 [arXiv:0706.0454 [hep-ph]].

\bibitem{McFadyen99} A.~I.~McFadyen and S.~E.~Woosley, Astrophys.\ J.\
{\bf 524}, 262 (1999).

\bibitem{Fermi_GRB} A.~A.~Abdo, et al.\ [Fermi Collaboration],
Science, {\bf 323}, 1688 (2009); A.~A.~Abdo, et al.\ [Fermi
Collaboration \& The Swift Team] ApJL (accepted), arXiv:0909.2470.

\bibitem{McFadyen01} A.~I.~McFadyen, S.~E.~Woosley, and A.~Heger,
Astrophys.\ J.\ {\bf 550}, 410 (2001).

\bibitem{grb_review} T.~Piran, Rev.\ Mod.\
Phys.\ {\bf 76}, 1143 (2004); P.~Meszaros, Rept.\ Prog.\ Phys.\ {\bf
69}, 2259 (2006).

\bibitem{Gaisser} T.~K. Gaisser, ``Cosmic Rays and Particle Physics'',
Cambridge University Press (1990).

\bibitem{Enberg09} R.~Enberg, M.~H.~Reno and I.~Sarcevic, Phys.\ Rev.\
D {\bf 79}, 053006 (2009).

\bibitem{Lipari93} P.~Lipari, Astropart.\ Phys.\ {\bf 1}, 195 (1993).

\bibitem{Koers07} H.~B.~J.~Koers and R.~A.~M.~J.~Wijers,
arXiv:0711.4791v1

\bibitem{Matzner99} C.~D.~Matzner and C.~F.~McKee, Astrophys.\ J.\
{\bf 510}, 379 (1999).

\bibitem{Pont} B. Pontecorvo,
  {\em Zh.\ Eksp.\ Teor.\ Fiz.} {\bf 34} (1958) 247
[{\em Sov.\ Phys.\ JETP} {\bf 7} (1958) 172].
\bibitem{MNS}
  Z. Maki, M. Nakagawa and S. Sakata,
  {\em Prog.\ Theor.\ Phys.}  {\bf 28} (1962) 870.

\bibitem{vacmim} E.~K.~Akhmedov,
  Phys.\ Lett.\  B {\bf 503}, 133 (2001)
  [arXiv:hep-ph/0011136].

\bibitem{sn}
  A.~S.~Dighe and A.~Y.~Smirnov,
  Phys.\ Rev.\  D {\bf 62}, 033007 (2000)
  [arXiv:hep-ph/9907423].

\bibitem{petcov}
  S.~T.~Petcov,
  Phys.\ Lett.\  B {\bf 200}, 373 (1988).

\bibitem{double-bang} J.~G.~Learned and S.~Pakvasa, Astropart.\ Phys.\
{\bf 3}, 267 (1995).

\bibitem{tau-mu} E.~Bugaev, T.~Montaruli, Y.~Shlepin and
I.~Sokalski, Astropart.\ Phys.\ {\bf 21}, 491 (2004); T.~DeYoung,
S.~Razzaque and D.~F.~Cowen, Astropart.\ Phys.\ {\bf 27}, 238 (2007).

\end{thebibliography}
\end{document}